\documentclass[final]{siamltex}
\usepackage[dvips]{graphicx}
\usepackage{amssymb,amsfonts,amsmath}

\def\R{\mathbb{R}}
\def\T{\top}
\def\epsilon{\varepsilon}
\def\bigoh{\mathcal{O}}
\def\phi{\varphi}

\newtheorem{observe}[theorem]{Observation}
\newtheorem{remark1}[theorem]{Remark}

\newenvironment{remark}{\begin{remark1} \rm}{\end{remark1}}

\title{An algorithm for the principal component analysis of large data sets}

\author{Nathan Halko\thanks{Department of Applied Mathematics,
University of Colorado at Boulder, 526 UCB, Boulder, CO 80309-0526
(nathan.halko@colorado.edu)} \and
Per-Gunnar Martinsson\thanks{Department of Applied Mathematics,
University of Colorado at Boulder, 526 UCB, Boulder, CO 80309-0526
(martinss@colorado.edu)} \and
Yoel Shkolnisky\thanks{Department of Applied Mathematics,
School of Mathematical Sciences, Tel Aviv University,
Ramat Aviv, Tel Aviv, 69978, Israel
(yoelsh@post.tau.ac.il)} \ \and
Mark Tygert\thanks{Courant Institute of Mathematical Sciences, NYU,
251 Mercer St., New York, NY 10012}
(tygert@aya.yale.edu)}

\begin{document}

\maketitle

\begin{abstract}
Recently popularized randomized methods for principal component analysis (PCA)
efficiently and reliably produce nearly optimal accuracy ---
even on parallel processors ---
unlike the classical (deterministic) alternatives.
We adapt one of these randomized methods for use with data sets
that are too large to be stored in random-access memory (RAM).
(The traditional terminology is that
our procedure works efficiently {\it out-of-core}.)
We illustrate the performance of the algorithm
via several numerical examples. For example,
we report on the PCA of a data set stored on disk that is so large
that less than a hundredth of it can fit in our computer's RAM.
\end{abstract}



\pagestyle{myheadings}
\thispagestyle{plain}
\markboth{N. HALKO, P.-G. MARTINSSON, Y. SHKOLNISKY, AND M. TYGERT}
         {PRINCIPAL COMPONENT ANALYSIS OF LARGE DATA SETS}

\section{Introduction}

Principal component analysis (PCA) is among the most popular tools
in machine learning, statistics, and data analysis more generally.
PCA is the basis of many techniques in data mining and information retrieval,
including the latent semantic analysis of large databases of text and HTML
documents described in~\cite{LSA}.
In this paper, we compute PCAs of very large data sets
via a randomized version of the block Lanczos method,
summarized in Section~\ref{summary} below.
The proofs in~\cite{halko-martinsson-tropp} and~\cite{rokhlin-szlam-tygert}
show that this method requires only a couple of iterations to produce
nearly optimal accuracy, with overwhelmingly high probability
(the probability is independent of the data being analyzed,
and is typically $1-10^{-15}$ or greater).
The randomized algorithm has many advantages, as shown
in~\cite{halko-martinsson-tropp} and~\cite{rokhlin-szlam-tygert};
the present article adapts the algorithm for use with data sets
that are too large to be stored
in the random-access memory (RAM) of a typical computer system.

Computing a PCA of a data set amounts to constructing
a singular value decomposition (SVD) that accurately approximates
the matrix $A$ containing the data being analyzed
(possibly after suitably ``normalizing'' $A$,
say by subtracting from each column its mean).
That is, if $A$ is $m \times n$,
then we must find a positive integer $k<\min(m,n)$
and construct matrices $U$, $\Sigma$, and $V$ such that
\begin{equation}
\label{svd_approx}
A \approx U \, \Sigma \, V^\T,
\end{equation}
with $U$ being an $m \times k$ matrix whose columns are orthonormal,
$V$ being an $n \times k$ matrix whose columns are orthonormal, and
$\Sigma$ being a diagonal $k \times k$ matrix
whose entries are all nonnegative.
The algorithm summarized in Section~\ref{summary} below is most efficient
when $k$ is substantially less than $\min(m,n)$;
in typical real-world applications, $k \ll \min(m,n)$.
Most often, the relevant measure of the quality
of the approximation in~(\ref{svd_approx})
is the spectral norm of the discrepancy $A - U \, \Sigma \, V^\T$;
see, for example, Section~\ref{summary} below.
The present article focuses on the spectral norm,
though our methods produce similar accuracy
in the Frobenius/Hilbert-Schmidt norm
(see, for example,~\cite{halko-martinsson-tropp}).

The procedure of the present article
works to minimize the total number of times that the algorithm has to access
each entry of the matrix $A$ being approximated.
A related strategy is to minimize the total number of disk seeks
and to maximize the dimensions of the approximation that can be constructed
with a given amount of RAM;
the algorithm in~\cite{martinsson-szlam-tygert} takes this latter approach.

In the present paper, the entries of all matrices are real valued;
our techniques extend trivially to matrices whose entries are complex valued.
The remainder of the article has the following structure:
Section~\ref{informal} explains the motivation behind the algorithm.
Section~\ref{summary} outlines the algorithm.
Section~\ref{outs} details the implementation for very large matrices.
Section~\ref{ccosts} quantifies the main factors influencing the running-time
of the algorithm.
Section~\ref{numexs} illustrates the performance of the algorithm
via several numerical examples.
Section~\ref{application} applies the algorithm to a data set
of interest in biochemical imaging.
Section~\ref{conclusion} draws some conclusions and proposes directions
for further research.

\section{Informal description of the algorithm}
\label{informal}

In this section, we provide a brief, heuristic description.
Section~\ref{summary} below provides more details
on the algorithm described intuitively in the present section.

Suppose that $k$, $m$, and $n$ are positive integers
with $k < m$ and $k < n$, and $A$ is a real $m \times n$ matrix.
We will construct an approximation to $A$ such that
\begin{equation}
\label{sort_of_svd0}
\| A - U \, \Sigma \, V^\T \|_2 \approx \sigma_{k+1},
\end{equation}
where $U$ is a real $m \times k$ matrix whose columns are orthonormal,
$V$ is a real $n \times k$ matrix whose columns are orthonormal,
$\Sigma$ is a diagonal real $k \times k$ matrix
whose entries are all nonnegative,
$\| A - U \, \Sigma \, V^\T \|_2$ is the spectral ($l^2$-operator) norm
of $A - U \, \Sigma \, V^\T$,
and $\sigma_{k+1}$ is the $(k+1)$st greatest singular value of $A$.
To do so, we select nonnegative integers $i$ and $l$ such that
$l \ge k$ and  $(i+2)k \le n$
(for most applications, $l = k+2$ and $i \le 2$ is sufficient;
$\| A - U \, \Sigma \, V^\T \|_2$ will decrease as $i$ and $l$ increase), and
then identify an orthonormal basis for ``most'' of the range of $A$
via the following two steps:
\begin{enumerate}
\item[1.] Using a random number generator,
form a real $n \times l$ matrix $G$ whose entries are
independent and identically distributed Gaussian random variables
of zero mean and unit variance, and compute the $m \times ((i+1)l)$ matrix
\begin{equation}
\label{product20}
H = \left(\begin{array}{c|c|c|c|c}
    A \, G & A \, A^\T \, A \, G & \ldots & (A \, A^\T)^{i-1} \, A \, G
  & (A \, A^\T)^i \, A \, G \end{array}\right).
\end{equation}
\item[2.] Using a pivoted $QR$-decomposition,
form a real $m \times ((i+1)l)$ matrix $Q$ whose columns are orthonormal,
such that there exists a real $((i+1)l) \times ((i+1)l)$ matrix $R$ for which
\begin{equation}
\label{good_approx20}
H = Q \, R.
\end{equation}
(See, for example, Chapter~5 in~\cite{golub-van_loan}
for details concerning the construction of such a matrix $Q$.)
\end{enumerate}
Intuitively, the columns of $Q$ in~(\ref{good_approx20}) constitute
an orthonormal basis for most of the range of $A$.
Moreover, the somewhat simplified algorithm with $i=0$ is sufficient
except when the singular values of $A$ decay slowly; see, for example,
\cite{halko-martinsson-tropp}.

Notice that $Q$ may have many fewer columns than $A$,
that is, $k$ may be substantially less than $n$
(this is the case for most applications of principal component analysis).
This is the key to the efficiency of the algorithm.

Having identified a good approximation to the range of $A$,
we perform some simple linear algebraic manipulations in order to obtain
a good approximation to $A$, via the following four steps:
\begin{enumerate}
\item[3.] Compute the $n \times ((i+1)l)$ product matrix
\begin{equation}
\label{product_t0}
T = A^\T \, Q.
\end{equation}
\item[4.] Form an SVD of $T$,
\begin{equation}
\label{svd_small0}
T = \tilde{V} \, \tilde{\Sigma} \, W^\T,
\end{equation}
where $\tilde{V}$ is a real $n \times ((i+1)l)$ matrix
whose columns are orthonormal,
$W$ is a real $((i+1)l) \times ((i+1)l)$ matrix whose columns are orthonormal,
and $\tilde{\Sigma}$ is a real diagonal $((i+1)l) \times ((i+1)l)$ matrix
such that $\tilde{\Sigma}_{1,1} \ge \tilde{\Sigma}_{2,2} \ge \dots \ge
\tilde{\Sigma}_{(i+1)l-1,(i+1)l-1} \ge \tilde{\Sigma}_{(i+1)l,(i+1)l} \ge 0$.
(See, for example, Chapter~8 in~\cite{golub-van_loan} for details
concerning the construction of such an SVD.)
\item[5.] Compute the $m \times ((i+1)l)$ product matrix
\begin{equation}
\label{product30}
\tilde{U} = Q \, W.
\end{equation}
\item[6.] Retrieve the leftmost $m \times k$ block $U$ of $\tilde{U}$,
the leftmost $n \times k$ block $V$ of $\tilde{V}$,
and the leftmost uppermost $k \times k$ block $\Sigma$ of $\tilde{\Sigma}$.
\end{enumerate}
The matrices $U$, $\Sigma$, and $V$ obtained
via Steps~1--6 above satisfy~(\ref{sort_of_svd0});
in fact, they satisfy the more detailed bound~(\ref{sort_of_svd})
described below.

\section{Summary of the algorithm}
\label{summary}

In this section, we will construct
a low-rank (say, rank $k$) approximation $U \, \Sigma \, V^\T$
to any given real matrix $A$, such that
\begin{equation}
\label{sort_of_svd}
\| A - U \, \Sigma \, V^\T \|_2 \le
\sqrt{ (C k n)^{1/(2i+1)} + \min(1,C/n) } \; \sigma_{k+1}
\end{equation}
with high probability (independent of $A$),
where $m$ and $n$ are the dimensions of the given $m \times n$ matrix $A$,
$U$ is a real $m \times k$ matrix whose columns are orthonormal,
$V$ is a real $n \times k$ matrix whose columns are orthonormal,
$\Sigma$ is a real diagonal $k \times k$ matrix
whose entries are all nonnegative,
$\sigma_{k+1}$ is the $(k+1)$st greatest singular value of $A$,
and $C$ is a constant determining the probability of failure
(the probability of failure is small when $C = 10$,
negligible when $C = 100$).
In~(\ref{sort_of_svd}), $i$ is any nonnegative integer such that $(i+2)k \le n$
(for most applications, $i = 1$ or $i = 2$ is sufficient;
the algorithm becomes less efficient as $i$ increases), and
$\| A - U \, \Sigma \, V^\T \|_2$ is the spectral ($l^2$-operator) norm
of $A - U \, \Sigma \, V^\T$, that is,
\begin{equation}
\| A - U \, \Sigma \, V^\T \|_2
= \max_{x \in \R^n : \|x\|_2 \ne 0}
  \frac{\| (A - U \, \Sigma \, V^\T)x \|_2}{\| x \|_2},
\end{equation}
\begin{equation}
\| x \|_2 = \sqrt{\sum_{j=1}^n (x_j)^2}.
\end{equation}
To simplify the presentation, we will assume that $n \le m$
(if $n > m$, then the user can apply the algorithm to $A^\T$).
In this section, we summarize the algorithm;
see~\cite{halko-martinsson-tropp} and~\cite{rokhlin-szlam-tygert}
for an in-depth discussion,
including proofs of more detailed variants of~(\ref{sort_of_svd}).

The minimal value of the spectral norm $\| A - B \|_2$,
minimized over all rank-$k$ matrices $B$, is $\sigma_{k+1}$
(see, for example, Theorem~2.5.3 in~\cite{golub-van_loan}).
Hence, (\ref{sort_of_svd})
guarantees that the algorithm summarized below produces approximations
of nearly optimal accuracy.

To construct a rank-$k$ approximation to $A$,
we could apply $A$ to about $k$ random vectors,
in order to identify the part of its range corresponding
to the larger singular values.
To help suppress the smaller singular values,
we apply $A \, (A^\T \, A)^i$, too.
Once we have identified ``most'' of the range of $A$,
we perform some linear-algebraic manipulations in order to recover
an approximation satisfying~(\ref{sort_of_svd}).

A numerically stable realization of the scheme outlined
in the preceding paragraph is the following.
We choose an integer $l \ge k$ such that $(i+1)l \le n-k$
(it is generally sufficient to choose $l = k+2$;
increasing $l$ can improve the accuracy marginally,
but increases computational costs),
and make the following six steps:

\begin{enumerate}
\item[1.] Using a random number generator,
form a real $n \times l$ matrix $G$ whose entries are
independent and identically distributed Gaussian random variables
of zero mean and unit variance, and compute the $m \times l$ matrices
$H^{(0)}$, $H^{(1)}$, \dots, $H^{(i-1)}$, $H^{(i)}$
defined via the formulae
\begin{equation}
\label{first_prod}
H^{(0)} = A \, G,
\end{equation}
\begin{equation}
\label{second_prod}
H^{(1)} = A \, (A^\T H^{(0)}),
\end{equation}
\begin{equation}
\label{third_prod}
H^{(2)} = A \, (A^\T H^{(1)}),
\end{equation}
\begin{equation*}
\vdots
\end{equation*}
\begin{equation}
\label{last_prod}
H^{(i)} = A \, (A^\T H^{(i-1)}).
\end{equation}
Form the $m \times ((i+1)l)$ matrix
\begin{equation}
\label{product23}
H = \left(\begin{array}{c|c|c|c|c}
    H^{(0)} & H^{(1)} & \ldots & H^{(i-1)} & H^{(i)}
    \end{array}\right).
\end{equation}
\item[2.] Using a pivoted $QR$-decomposition,
form a real $m \times ((i+1)l)$ matrix $Q$ whose columns are orthonormal,
such that there exists a real $((i+1)l) \times ((i+1)l)$ matrix $R$ for which
\begin{equation}
\label{good_approx23}
H = Q \, R.
\end{equation}
(See, for example, Chapter~5 in~\cite{golub-van_loan}
for details concerning the construction of such a matrix $Q$.)
\item[3.] Compute the $n \times ((i+1)l)$ product matrix
\begin{equation}
\label{product_t3}
T = A^\T \, Q.
\end{equation}
\item[4.] Form an SVD of $T$,
\begin{equation}
\label{svd_small3}
T = \tilde{V} \, \tilde{\Sigma} \, W^\T,
\end{equation}
where $\tilde{V}$ is a real $n \times ((i+1)l)$ matrix
whose columns are orthonormal,
$W$ is a real $((i+1)l) \times ((i+1)l)$ matrix whose columns are orthonormal,
and $\tilde{\Sigma}$ is a real diagonal $((i+1)l) \times ((i+1)l)$ matrix
such that $\tilde{\Sigma}_{1,1} \ge \tilde{\Sigma}_{2,2} \ge \dots \ge
\tilde{\Sigma}_{(i+1)l-1,(i+1)l-1} \ge \tilde{\Sigma}_{(i+1)l,(i+1)l} \ge 0$.
(See, for example, Chapter~8 in~\cite{golub-van_loan} for details
concerning the construction of such an SVD.)
\item[5.] Compute the $m \times ((i+1)l)$ product matrix
\begin{equation}
\label{product33}
\tilde{U} = Q \, W.
\end{equation}
\item[6.] Retrieve the leftmost $m \times k$ block $U$ of $\tilde{U}$,
the leftmost $n \times k$ block $V$ of $\tilde{V}$,
and the leftmost uppermost $k \times k$ block $\Sigma$ of $\tilde{\Sigma}$.
The product $U \, \Sigma \, V^\T$ then approximates $A$
as in~(\ref{sort_of_svd}) (we omit the proof; see~\cite{halko-martinsson-tropp}
for proofs of similar, more general bounds).
\end{enumerate}

\begin{remark}
In the present paper, we assume that the user specifies the rank $k$
of the approximation $U \, \Sigma \, V^\T$ being constructed.
See~\cite{halko-martinsson-tropp} for techniques
for determining the rank $k$ adaptively, such that the accuracy
$\| A - U \, \Sigma \, V^\T \|_2$ satisfying~(\ref{sort_of_svd})
also meets a user-specified threshold.
\end{remark}

\begin{remark}
Variants of the fast Fourier transform (FFT) permit additional accelerations;
see~\cite{halko-martinsson-tropp},
\cite{liberty-woolfe-martinsson-rokhlin-tygert},
and~\cite{woolfe-liberty-rokhlin-tygert}.
However, these accelerations have negligible effect on the algorithm
running out-of-core.
For out-of-core computations, the simpler techniques of the present paper
are preferable.
\end{remark}

\begin{remark}
The algorithm described in the present section can underflow or overflow
when the range of the floating-point exponent is inadequate
for representing simultaneously both the spectral norm $\|A\|_2$
and its $(2i+1)$st power $(\|A\|_2)^{2i+1}$.
A convenient alternative is the algorithm described
in~\cite{martinsson-szlam-tygert};
another solution is to process $A/\|A\|_2$ rather than $A$.
\end{remark}

\section{Out-of-core computations}
\label{outs}

With suitably large matrices, some steps in Section~\ref{summary} above
require either storage on disk, or on-the-fly computations obviating the need
for storing all the entries of the $m \times n$ matrix $A$ being approximated.
Conveniently, Steps~2, 4, 5, and 6 involve only matrices
having $\bigoh((i+1) \, l \, (m+n))$ entries;
we perform these steps using only storage in random-access memory (RAM).
However, Steps~1 and~3 involve $A$, which has $mn$ entries;
we perform Steps~1 and~3 differently depending on how $A$ is provided,
as detailed below in Subsections~\ref{on-the-fly} and~\ref{on-disk}.

\subsection{Computations with on-the-fly evaluation of matrix entries}
\label{on-the-fly}

If $A$ does not fit in memory, but we have access to a computational routine
that can evaluate each entry (or row or column) of $A$ individually,
then obviously we can perform Steps~1 and~3 using only storage in RAM.
Every time we evaluate an entry (or row or column) of $A$ in order
to compute part of a matrix product involving $A$ or $A^\T$,
we immediately perform all computations associated
with this particular entry (or row or column)
that contribute to the matrix product.

\subsection{Computations with storage on disk}
\label{on-disk}

If $A$ does not fit in memory, but is provided as a file on disk,
then Steps~1 and~3 require access to the disk.
We assume for definiteness that $A$ is provided in row-major format on disk
(if $A$ is provided in column-major format,
then we apply the algorithm to $A^\T$ instead).
To construct the matrix product in~(\ref{first_prod}),
we retrieve as many rows of $A$ from disk as will fit in memory,
form their inner products with the appropriate columns of $G$,
store the results in $H^{(0)}$, and then repeat with the remaining rows of $A$.
To construct the matrix product in~(\ref{product_t3}),
we initialize all entries of $T$ to zeros,
retrieve as many rows of $A$ from disk as will fit in memory,
add to $T$ the transposes of these rows, weighted by the appropriate entries
of $Q$, and then repeat with the remaining rows of $A$.
We construct the matrix product in~(\ref{second_prod}) similarly,
forming $F = A^\T \, H^{(0)}$ first, and $H^{(1)} = A \, F$ second.
Constructing the matrix products
in~(\ref{third_prod})--(\ref{last_prod}) is analogous.

\section{Computational costs}
\label{ccosts}

In this section, we tabulate the computational costs
of the algorithm described in Section~\ref{summary},
for the particular out-of-core implementations described
in Subsections~\ref{on-the-fly} and~\ref{on-disk}.
We will be using the notation from Section~\ref{summary},
including the integers $i$, $k$, $l$, $m$, and $n$,
and the $m \times n$ matrix $A$.

\begin{remark}
For most applications, $i \le 2$ suffices.
In contrast, the classical Lanczos algorithm generally requires
many iterations in order to yield adequate accuracy,
making the computational costs of the classical algorithm prohibitive
for out-of-core (or parallel) computations
(see, for example, Chapter~9 in~\cite{golub-van_loan}).
\end{remark}

\subsection{Costs with on-the-fly evaluation of matrix entries}

We denote by $C_A$ the number of floating-point operations (flops) required
to evaluate all nonzero entries in $A$.
We denote by $N_A$ the number of nonzero entries in $A$.
With on-the-fly evaluation of the entries of $A$,
the six steps of the algorithm described in Section~\ref{summary}
have the following costs:
\begin{enumerate}
\item Forming~$H^{(0)}$ in~(\ref{first_prod}) costs
$C_A + \bigoh(l \, N_A)$ flops.
Forming any of the matrix products
in~(\ref{second_prod})--(\ref{last_prod}) costs
$2 C_A + \bigoh(l \, N_A)$ flops.
Forming~$H$ in~(\ref{product23}) costs $\bigoh(ilm)$ flops.
All together, Step~1 costs $(2i+1) \, C_A + \bigoh(il(m+N_A))$ flops.
\item Forming~$Q$ in~(\ref{good_approx23}) costs
$\bigoh(i^2 l^2 m)$ flops.
\item Forming~$T$ in~(\ref{product_t3}) costs
$C_A + \bigoh(il \, N_A)$ flops.
\item Forming the SVD of $T$ in~(\ref{svd_small3}) costs
$\bigoh(i^2 l^2 n)$ flops.
\item Forming $\tilde{U}$ in~(\ref{product33}) costs
$\bigoh(i^2 l^2 m)$ flops.
\item Forming $U$, $\Sigma$, and $V$ in Step~6 costs
$\bigoh(k(m+n))$ flops.
\end{enumerate}

Summing up the costs for the six steps above,
and using the fact that $k \le l \le n \le m$,
we see that the full algorithm requires
\begin{equation}
\label{on-the-fly_cost}
C_{\hbox{\footnotesize{on-the-fly}}}
= 2(i+1) \, C_A + \bigoh(il \, N_A + i^2 l^2 m)
\end{equation}
flops, where $C_A$ is the number of flops required to evaluate
all nonzero entries in $A$, and $N_A$ is the number of nonzero entries in $A$.
In practice, we choose $l \approx k$ (usually a good choice is $l = k+2$).

\subsection{Costs with storage on disk}

We denote by $j$ the number of floating-point words
of random-access memory (RAM) available to the algorithm.
With $A$ stored on disk,
the six steps of the algorithm described in Section~\ref{summary} have
the following costs
(assuming for convenience that $j > 2 \, (i+1) \, l \, (m+n)$):
\begin{enumerate}
\item Forming~$H^{(0)}$ in~(\ref{first_prod}) requires at most
$\bigoh(lmn)$ floating-point operations (flops),
$\bigoh(mn/j)$ disk accesses/seeks,
and a total data transfer of $\bigoh(mn)$ floating-point words.
Forming any of the matrix products
in~(\ref{second_prod})--(\ref{last_prod}) also requires
$\bigoh(lmn)$ flops, $\bigoh(mn/j)$ disk accesses/seeks,
and a total data transfer of $\bigoh(mn)$ floating-point words.
Forming~$H$ in~(\ref{product23}) costs $\bigoh(ilm)$ flops.
All together, Step~1 requires $\bigoh(ilmn)$ flops,
$\bigoh(imn/j)$ disk accesses/seeks, and a total data transfer of
$\bigoh(imn)$ floating-point words.
\item Forming~$Q$ in~(\ref{good_approx23}) costs
$\bigoh(i^2 l^2 m)$ flops.
\item Forming~$T$ in~(\ref{product_t3}) requires
$\bigoh(ilmn)$ floating-point operations,
$\bigoh(mn/j)$ disk accesses/seeks,
and a total data transfer of $\bigoh(mn)$ floating-point words.
\item Forming the SVD of $T$ in~(\ref{svd_small3}) costs
$\bigoh(i^2 l^2 n)$ flops.
\item Forming $\tilde{U}$ in~(\ref{product33}) costs
$\bigoh(i^2 l^2 m)$ flops.
\item Forming $U$, $\Sigma$, and $V$ in Step~6 costs
$\bigoh(k(m+n))$ flops.
\end{enumerate}

In practice, we choose $l \approx k$ (usually a good choice is $l = k+2$).
Summing up the costs for the six steps above,
and using the fact that $k \le l \le n \le m$,
we see that the full algorithm requires
\begin{equation}
C_{\rm flops} = \bigoh(ilmn + i^2 l^2 m)
\end{equation}
flops,
\begin{equation}
C_{\rm accesses} = \bigoh(imn/j)
\end{equation}
disk accesses/seeks
(where $j$ is the number of floating-point words of RAM available
to the algorithm), and a total data transfer of
\begin{equation}
C_{\rm words} = \bigoh(imn)
\end{equation}
floating-point words (more specifically, $C_{\rm words} \approx 2(i+1)mn$).

\section{Numerical examples}
\label{numexs}

In this section, we describe the results of several numerical tests
of the algorithm of the present paper.

We set $l = k+2$ for all examples,
setting $i = 3$ for the first two examples, and $i = 1$ for the last two,
where $i$, $k$, and $l$ are the parameters from Section~\ref{summary} above.
We ran all examples on a laptop with 1.5~GB of random-access memory (RAM),
connected to an external hard drive via USB 2.0.
The processor was a single-core 32-bit 2-GHz Intel Pentium M,
with 2~MB of L2 cache.
We ran all examples in Matlab~7.4.0,
storing floating-point numbers in RAM using IEEE standard double-precision
variables (requiring 8 bytes per real number),
and on disk using IEEE standard single-precision variables
(requiring 4 bytes per real number).

All our numerical experiments indicate that
the quality and distribution of the pseudorandom numbers have little effect
on the accuracy of the algorithm of the present paper.
We used Matlab's built-in pseudorandom number generator
for all results reported below.

\subsection{Synthetic data}

In this subsection, we illustrate the performance of the algorithm
with the principal component analysis of three examples,
including a computational simulation.

For the first example, we apply the algorithm to the $m \times n$ matrix
\begin{equation}
A = E \, S \, F,
\end{equation}
where $E$ and $F$ are $m \times m$ and $n \times n$
unitary discrete cosine transforms of the second type (DCT-II),
and $S$ is an $m \times n$ matrix whose entries are zero
off the main diagonal, with
\begin{equation}
S_{j,j} = \left\{ \begin{array}{ll}
          10^{-4(j-1)/19}, & j = 1, 2, \dots, 19,\hbox{ or }20 \\
          10^{-4}/(j-20)^{1/10}, & j = 21, 22, \dots, n-1,\hbox{ or }n.
          \end{array} \right.
\end{equation}
Clearly, $S_{1,1}$,~$S_{2,2}$, \dots, $S_{n-1,n-1}$,~$S_{n,n}$
are the singular values of $A$.

For the second example, we apply the algorithm to the $m \times n$ matrix
\begin{equation}
A = E \, S \, F,
\end{equation}
where $E$ and $F$ are $m \times m$ and $n \times n$
unitary discrete cosine transforms of the second type (DCT-II),
and $S$ is an $m \times n$ matrix whose entries are zero
off the main diagonal, with
\begin{equation}
S_{j,j} = \left\{ \begin{array}{ll}
          1.00, & j = 1, 2,\hbox{ or }3 \\
          0.67, & j = 4, 5,\hbox{ or }6 \\
          0.34, & j = 7, 8,\hbox{ or }9 \\
          0.01, & j = 10, 11,\hbox{ or }12 \\
          0.01 \cdot \frac{n-j}{n-13}, & j = 13, 14, \dots, n-1,\hbox{ or }n.
          \end{array} \right.
\end{equation}
Clearly, $S_{1,1}$,~$S_{2,2}$, \dots, $S_{n-1,n-1}$,~$S_{n,n}$
are the singular values of $A$.

Table~1a summarizes results of applying the algorithm
to the first example, storing on disk the matrix being approximated.
Table~1b summarizes results of applying the algorithm
to the first example, generating on-the-fly the columns
of the matrix being approximated.

Table~2a summarizes results of applying the algorithm
to the second example, storing on disk the matrix being approximated.
Table~2b summarizes results of applying the algorithm
to the second example, generating on-the-fly the columns
of the matrix being approximated.

The following list describes the headings of the tables:
\begin{itemize}
\item $m$ is the number of rows in the matrix $A$ being approximated.
\item $n$ is the number of columns in the matrix $A$ being approximated.
\item $k$ is the parameter from Section~\ref{summary} above;
      $k$ is the rank of the approximation being constructed.
\item $t_{\rm gen}$ is the time in seconds required to generate and store
      on disk the matrix $A$ being approximated.
\item $t_{\rm PCA}$ is the time in seconds required to compute the rank-$k$
      approximation (the PCA) provided by the algorithm of the present paper.
\item $\epsilon_0$ is the spectral norm of the difference between
      the matrix $A$ being approximated and its best rank-$k$ approximation.
\item $\epsilon$ is an estimate of the spectral norm
      of the difference between the matrix $A$ being approximated
      and the rank-$k$ approximation produced by the algorithm
      of the present paper.
      The estimate $\epsilon$ of the error is accurate to within a factor
      of two with extraordinarily high probability;
      the expected accuracy of the estimate $\epsilon$ of the error
      is about 10\%, relative to the best possible error $\epsilon_0$
      (see~\cite{kuczynski-wozniakowski}).
      The appendix below details the construction of the estimate $\epsilon$
      of the spectral norm of $D = A-U \Sigma V^\T$,
      where $A$ is the matrix being approximated,
      and $U \Sigma V^\T$ is the rank-$k$ approximation
      produced by the algorithm of the present paper.
\end{itemize}

\begin{table}
\begin{center}
\footnotesize{\sc Table 1a} \\
\footnotesize{\it On-disk storage of the first example.} \\\vspace{.5em}
\begin{tabular}{ccccccc}
$m$ & $n$ & $k$ & $t_{\rm gen}$ & $t_{\rm PCA}$ & $\epsilon_0$ & $\epsilon$
\\\hline
2E5 & 2E5 &  16 &         2.7E4 &         6.6E4 &       4.3E-4 &     4.3E-4
\\
2E5 & 2E5 &  20 &         2.7E4 &         6.6E4 &       1.0E-4 &     1.0E-4
\\
2E5 & 2E5 &  24 &         2.7E4 &         6.9E4 &       1.0E-4 &     1.0E-4
\end{tabular}
\end{center}
\end{table}

\begin{table}
\begin{center}
\footnotesize{\sc Table 1b} \\
\footnotesize{\it On-the-fly generation of the first example.} \\\vspace{.5em}
\begin{tabular}{cccccc}
$m$ & $n$ & $k$ & $t_{\rm PCA}$ & $\epsilon_0$ & $\epsilon$
\\\hline
2E5 & 2E5 &  16 &         7.7E1 &       4.3E-4 &     4.3E-4
\\
2E5 & 2E5 &  20 &         1.0E2 &       1.0E-4 &     1.0E-4
\\
2E5 & 2E5 &  24 &         1.3E2 &       1.0E-4 &     1.0E-4
\end{tabular}
\end{center}
\end{table}

\begin{table}
\begin{center}
\footnotesize{\sc Table 2a} \\
\footnotesize{\it On-disk storage of the second example.} \\\vspace{.5em}
\begin{tabular}{ccccccc}
$m$ & $n$ & $k$ & $t_{\rm gen}$ & $t_{\rm PCA}$ & $\epsilon_0$ & $\epsilon$
\\\hline
2E5 & 2E5 &  12 &         2.7E4 &         6.3E4 &       1.0E-2 &     1.0E-2
\\
2E5 & 2E4 &  12 &         1.9E3 &         6.1E3 &       1.0E-2 &     1.0E-2
\\
5E5 & 8E4 &  12 &         2.2E4 &         6.5E4 &       1.0E-2 &     1.0E-2
\end{tabular}
\end{center}
\end{table}

\begin{table}
\begin{center}
\footnotesize{\sc Table 2b} \\
\footnotesize{\it On-the-fly generation of the second example.} \\\vspace{.5em}
\begin{tabular}{cccccc}
$m$ & $n$ & $k$ & $t_{\rm PCA}$ & $\epsilon_0$ & $\epsilon$
\\\hline
2E5 & 2E5 &  12 &         5.5E1 &       1.0E-2 &     1.0E-2
\\
2E5 & 2E4 &  12 &         2.7E1 &       1.0E-2 &     1.0E-2
\\
5E5 & 8E4 &  12 &         7.9E1 &       1.0E-2 &     1.0E-2
\end{tabular}
\end{center}
\end{table}

For the third example, we apply the algorithm with $k = 3$
to an $m \times 1000$ matrix whose rows are
independent and identically distributed (i.i.d.)\ realizations
of the random vector
\begin{equation}
\label{simulation}
\alpha \, w_1 + \beta \, w_2 + \gamma \, w_3 + \delta,
\end{equation}
where $w_1$,~$w_2$, and~$w_3$ are orthonormal $1 \times 1000$ vectors,
$\delta$ is a $1 \times 1000$ vector whose entries
are i.i.d.\ Gaussian random variables
of mean zero and standard deviation $0.1$,
and $(\alpha, \beta, \gamma)$ is drawn at random
from inside an ellipsoid with axes of lengths $a=1.5$,~$b=1$, and~$c=0.5$,
specifically,
\begin{equation}
\alpha = a \, r \, (\cos \phi) \, \sin \theta,
\end{equation}
\begin{equation}
\beta = b \, r \, (\sin \phi) \, \sin \theta,
\end{equation}
\begin{equation}
\gamma = c \, r \cos \theta,
\end{equation}
with $r$ drawn uniformly at random from $[0,1]$,
$\phi$ drawn uniformly at random from $[0,2\pi]$,
and $\theta$ drawn uniformly at random from $[0,\pi]$.
We obtained $w_1$, $w_2$, and~$w_3$ by applying the Gram-Schmidt process
to three vectors whose entries were i.i.d.\ centered Gaussian random variables;
$w_1$, $w_2$, and~$w_3$ are exactly the same in every row,
whereas the realizations of $\alpha$, $\beta$, $\gamma$, and $\delta$
in the various rows are independent.
We generated all the random numbers on-the-fly
using a high-quality pseudorandom number generator;
whenever we had to regenerate exactly the same matrix
(as the algorithm requires with $i > 0$),
we restarted the pseudorandom number generator with the original seed.

Figure~1a plots the inner product ({\it i.e.}, correlation)
of~$w_1$ in~(\ref{simulation}) and the (normalized) right singular vector
associated with the greatest singular value
produced by the algorithm of the present article.
Figure~1a also plots the inner product of~$w_2$ in~(\ref{simulation})
and the (normalized) right singular vector
associated with the second greatest singular value,
as well as the inner product of~$w_3$
and the (normalized) right singular vector
associated with the third greatest singular value.
Needless to say, the inner products ({\it i.e.}, correlations)
all tend to 1, as $m$ increases --- as they should.
Figure~1b plots the time required to run the algorithm of the present paper,
generating on-the-fly the entries of the matrix being processed.
The running-time is roughly proportional to $m$,
in accordance with~(\ref{on-the-fly_cost}).

\begin{figure}[p]
\begin{center}
\scalebox{1}{\includegraphics{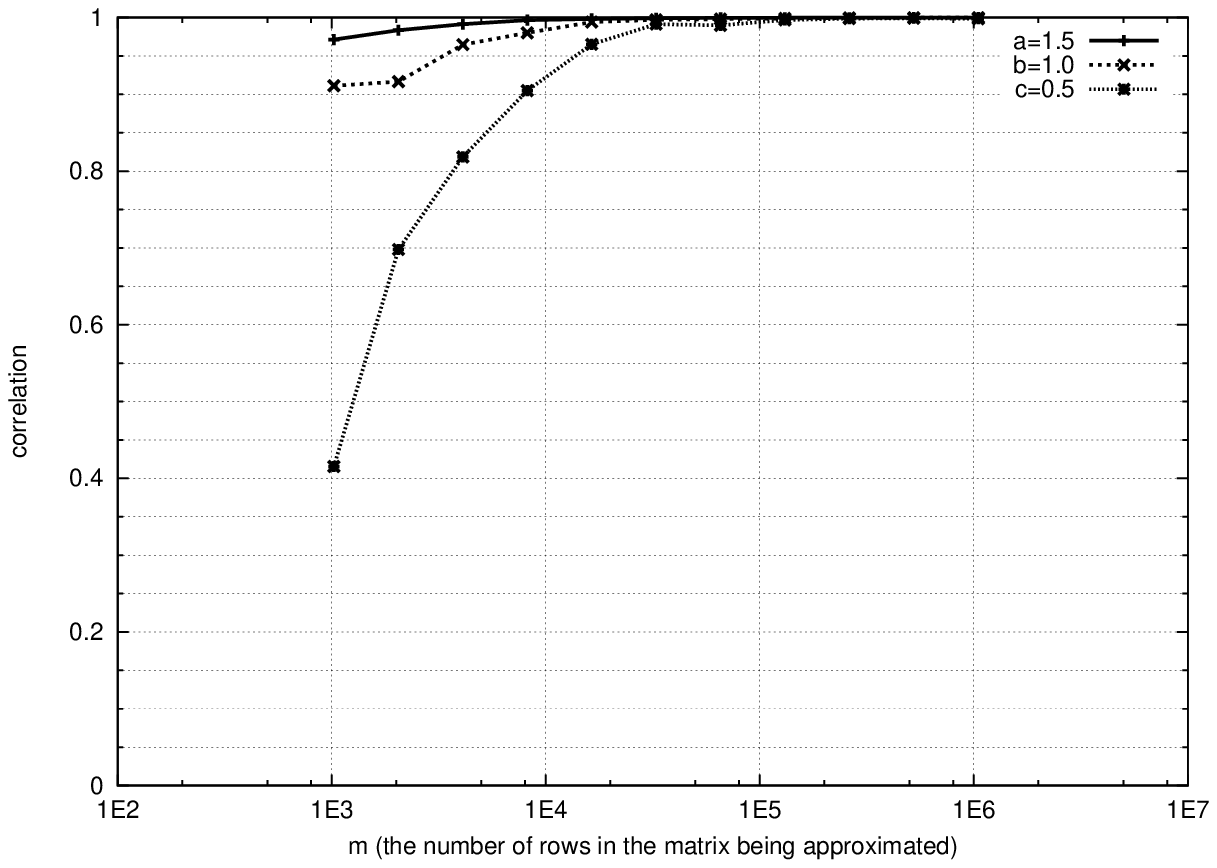}}
\\\vspace{.25em}
\footnotesize{
{\sc Fig. 1a.} {\it Convergence for the third example
                (the computational simulation).}
}
\end{center}
\end{figure}

\begin{figure}[p]
\begin{center}
\scalebox{1}{\includegraphics{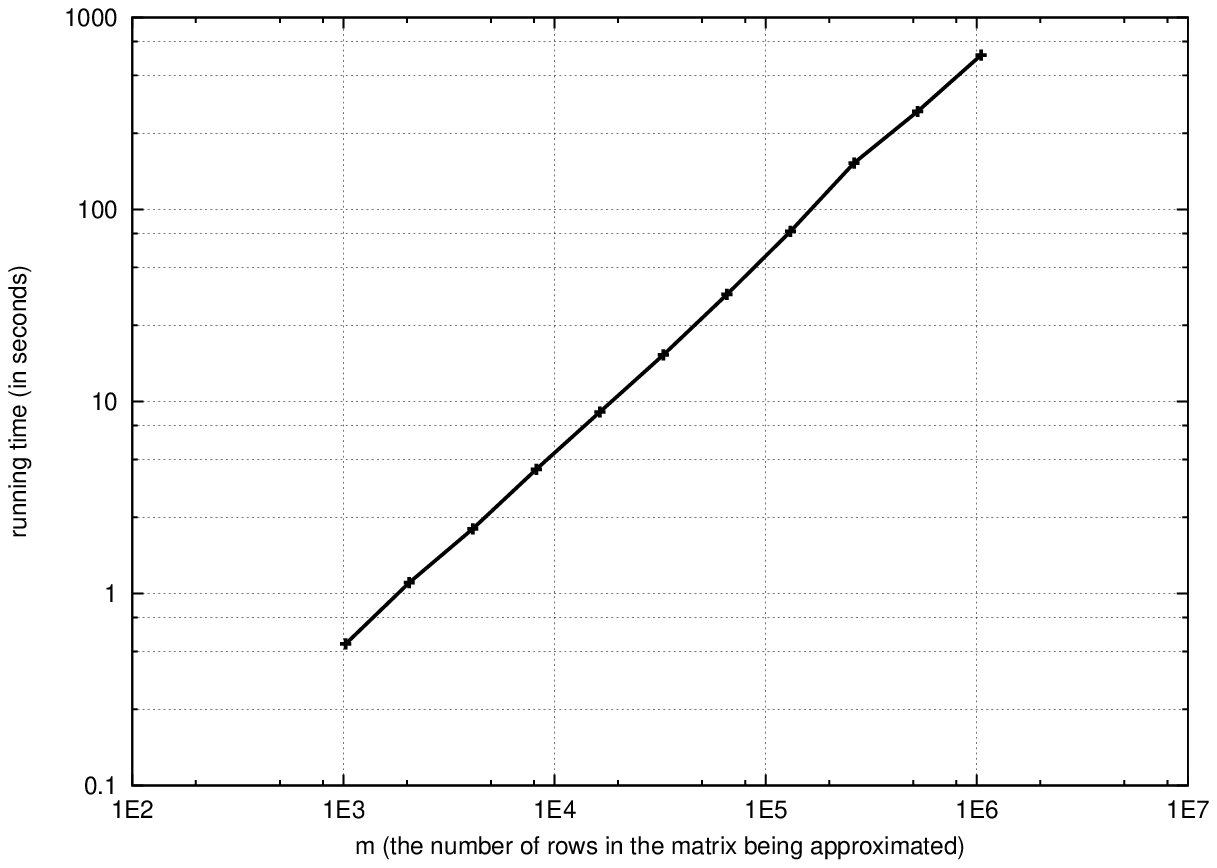}}
\\\vspace{.25em}
\footnotesize{
{\sc Fig. 1b.} {\it Timing for the third example
                (the computational simulation).}
}
\end{center}
\end{figure}

\subsection{Measured data}

In this subsection, we illustrate the performance of the algorithm
with the principal component analysis of images of faces.

We apply the algorithm with $k = 50$ to the 393,216 $\times$ 102,042 matrix
whose columns consist of images from the FERET database
of faces described in~\cite{feret1} and~\cite{feret2},
with each image duplicated three times.
For each duplicate, we set the values of a random choice of 10\% of the pixels
to numbers chosen uniformly at random from the integers 0,~1, \dots, 254,~255;
all pixel values are integers from 0,~1, \dots, 254,~255.
Before processing with the algorithm of the present article,
we ``normalized'' the matrix by subtracting from each column its mean,
then dividing the resulting column by its Euclidean norm.
The algorithm of the present paper required 12.3 hours to process
all 150~GB of this data set stored on disk,
using the laptop computer with 1.5~GB of RAM described earlier
(at the beginning of Section~\ref{numexs}).

Figure~2a plots the computed singular values.
Figure~2b displays the computed ``eigenfaces''
(that is, the left singular vectors)
corresponding to the five greatest singular values.

While this example does not directly provide a reasonable means
for performing face recognition or any other task of image processing,
it does indicate that the sheer brute force of linear algebra
(that is, computing a low-rank approximation) can be used directly
for processing (or preprocessing) a very large data set.
When used alone, this kind of brute force is inadequate
for face recognition and other tasks of image processing;
most tasks of image processing can benefit from more specialized methods
(see, for example, \cite{moon-phillips}, \cite{feret1}, and~\cite{feret2}).
Nonetheless, the ability to compute principal component analyses
of very large data sets could prove helpful, or at least convenient.

\begin{figure}
\begin{center}
\scalebox{.72}{\includegraphics{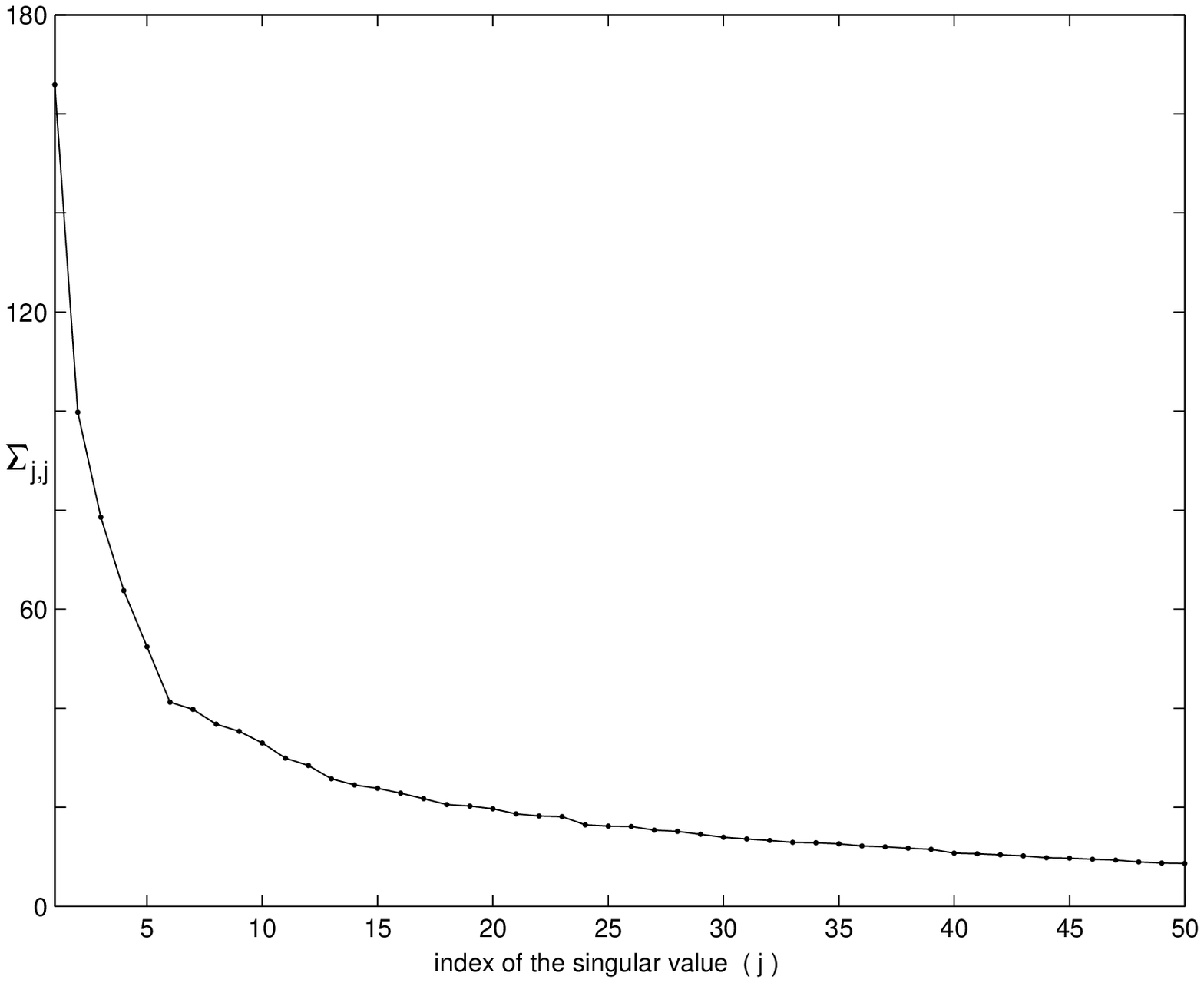}}
\\\vspace{.25em}
\footnotesize{
{\sc Fig. 2a.} {\it Singular values computed for the fourth example
                (the database of images).}
}
\end{center}
\end{figure}

\begin{figure*}
\begin{center}
\scalebox{.29}{\includegraphics{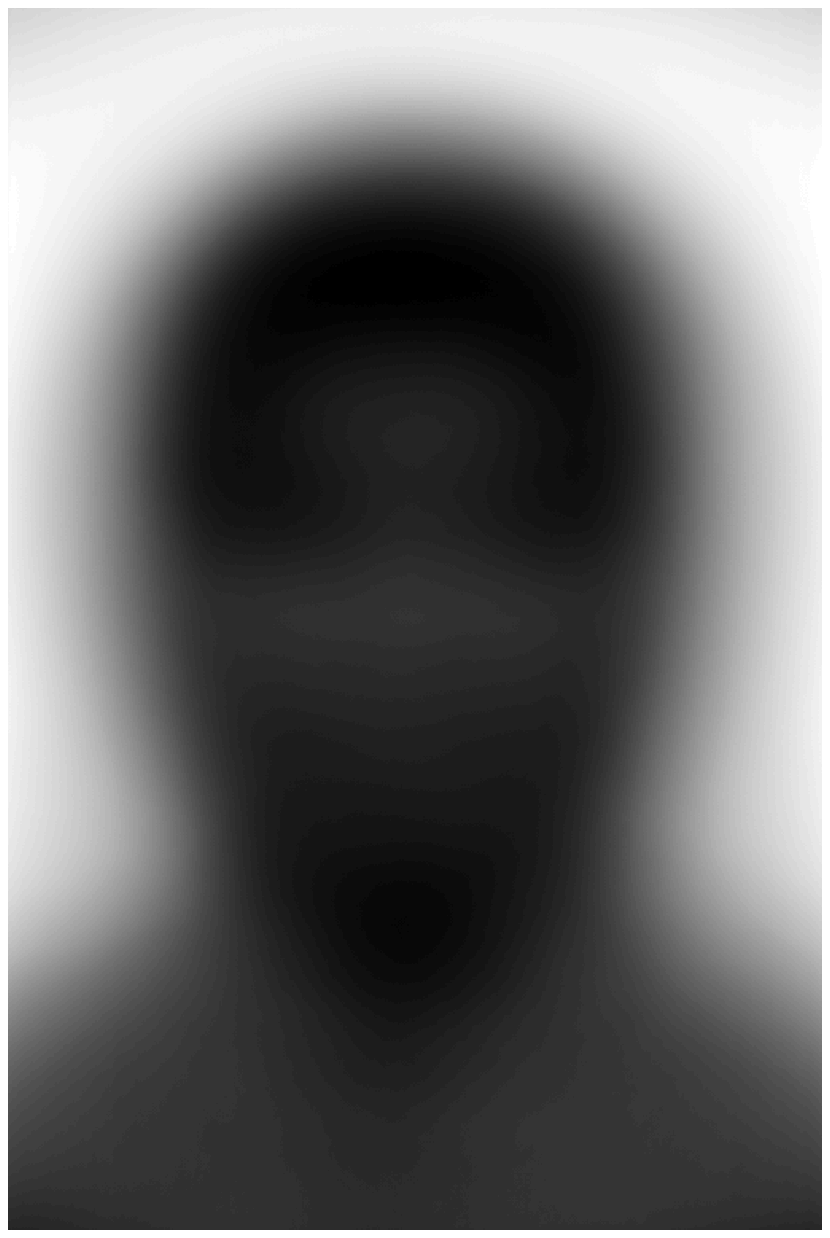}}
\scalebox{.29}{\includegraphics{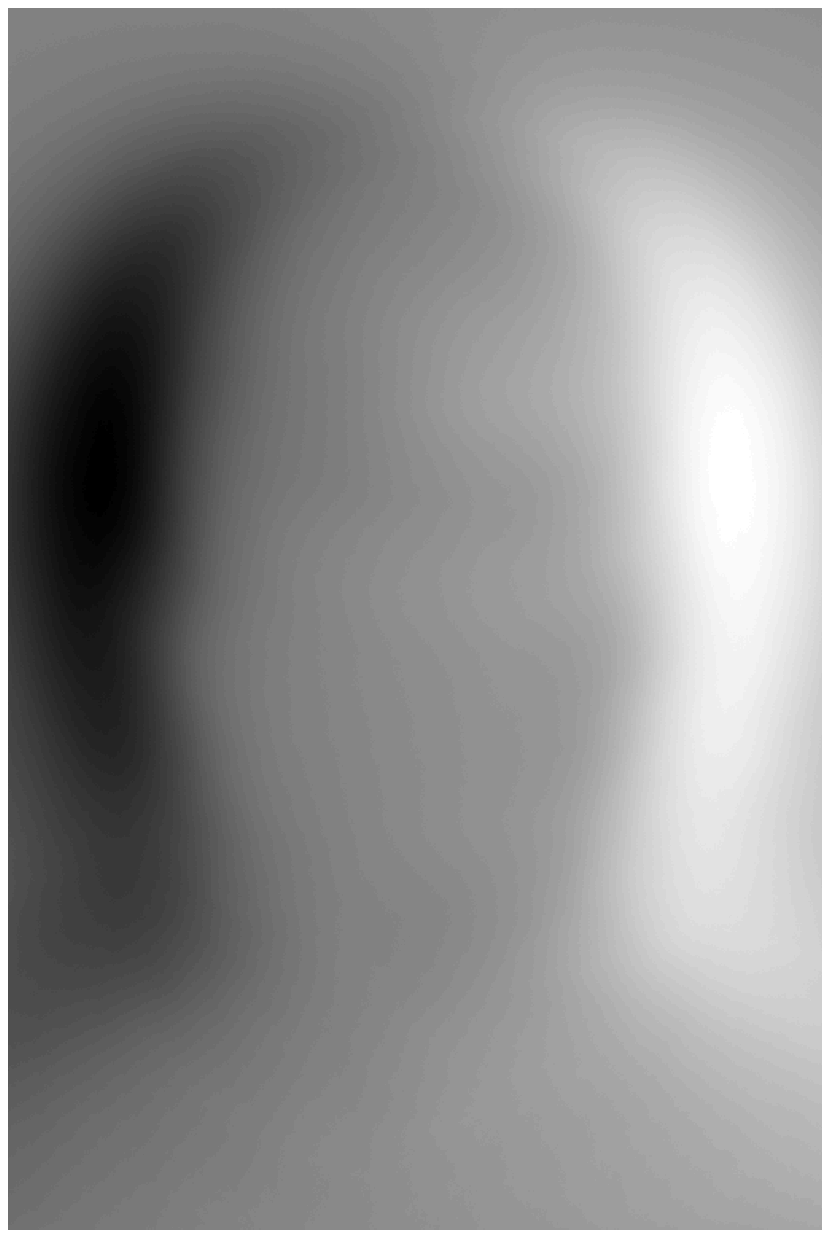}}
\scalebox{.29}{\includegraphics{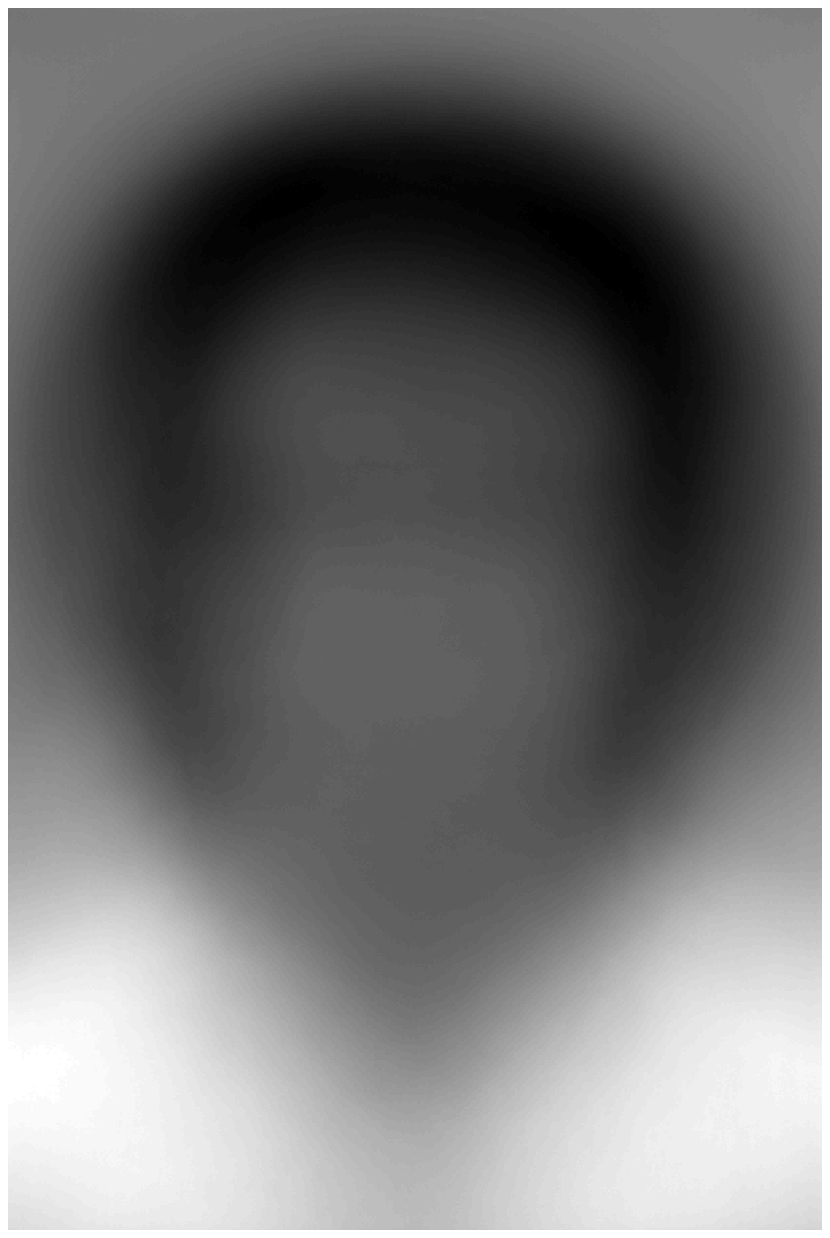}}
\scalebox{.29}{\includegraphics{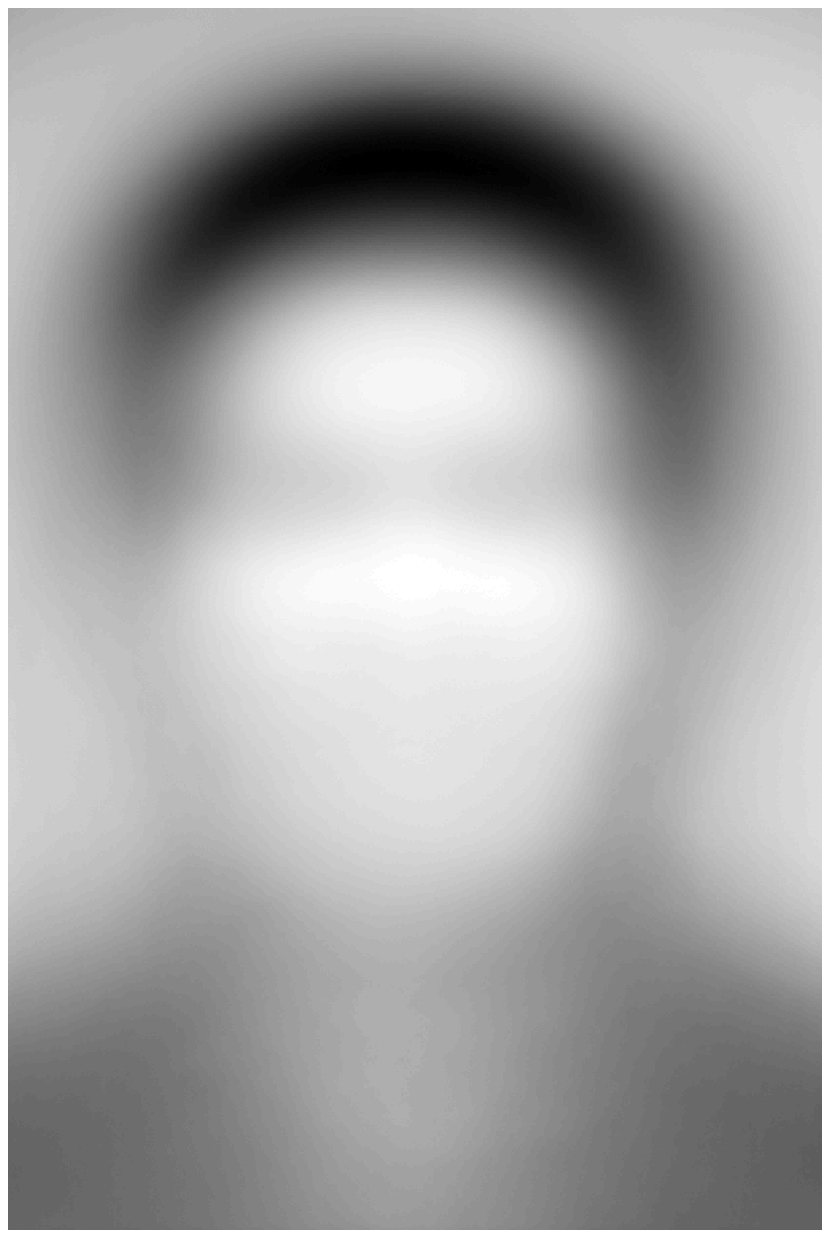}}
\scalebox{.29}{\includegraphics{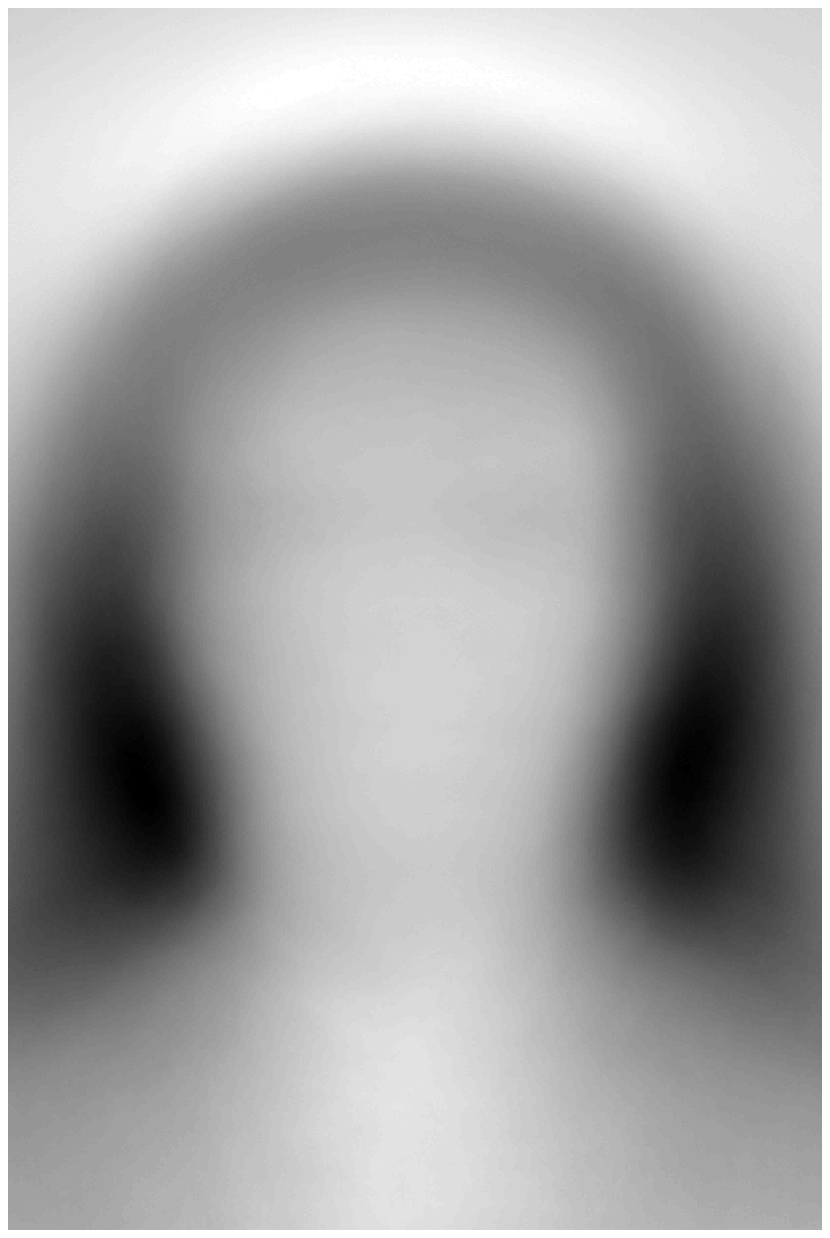}}
\\\vspace{.25em}
\footnotesize{
{\sc Fig. 2b.} {\it Dominant singular vectors computed for the fourth example
                (the database of images).}
}
\end{center}
\end{figure*}

\section{An application}
\label{application}

In this section, we apply the algorithm of the present paper
to a data set of interest in a currently developing imaging modality
known as single-particle cryo-electron microscopy.
For an overview of the field, see~\cite{frank}, \cite{ponce-singer},
and their compilations of references.

The data set consists of 10,000 two-dimensional images
of the (three-dimensional) charge density map
of the E.~coli 50S ribosomal subunit,
projected from uniformly random orientations, then added
to white Gaussian noise whose magnitude is 32 times larger
than the original images',
and finally rotated by 0,~1,~2, \dots, 358,~359 degrees.
The entire data set thus consists of 3,600,000 images,
each 129 pixels wide and 129 pixels high;
the matrix being processed is 3,600,000 $\times$ $129^2$.
We set $i = 1$, $k = 250$, and $l = k+2$,
where $i$, $k$, and $l$ are the parameters from Section~\ref{summary} above.
Processing the data set required 5.5~hours
on two 2.8~GHz quad-core Intel Xeon x5560 microprocessors
with 48~GB of random-access memory.

Figure~3a displays the 250 computed singular values.
Figure~3b displays the computed right singular vectors
corresponding to the 25 greatest computed singular values.
Figure~3c displays several noisy projections,
their versions before adding the white Gaussian noise,
and their denoised versions.
Each denoised image is the projection
of the corresponding noisy image on the computed right singular vectors
associated with the 150 greatest computed singular values.
The denoising is clearly satisfactory.

\begin{figure}
\begin{center}
\includegraphics[width=.9\textwidth]{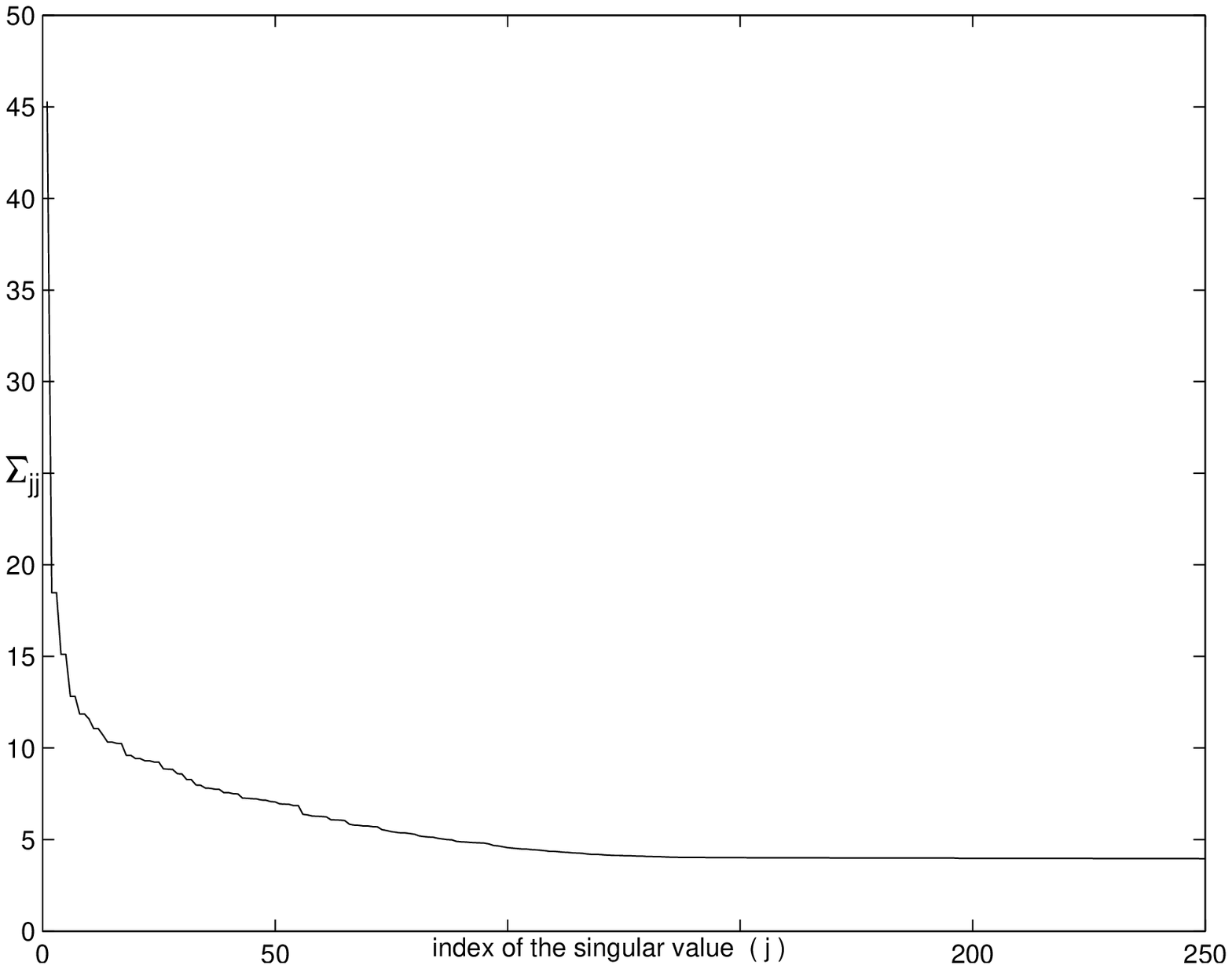}
\label{fig:singular values}
\\\vspace{.25em}
\footnotesize{
{\sc Fig. 3a.} {\it Singular values computed for the E. coli data set.}
}
\end{center}
\end{figure}

\begin{figure}
\begin{center}
\includegraphics[width=0.18\textwidth]{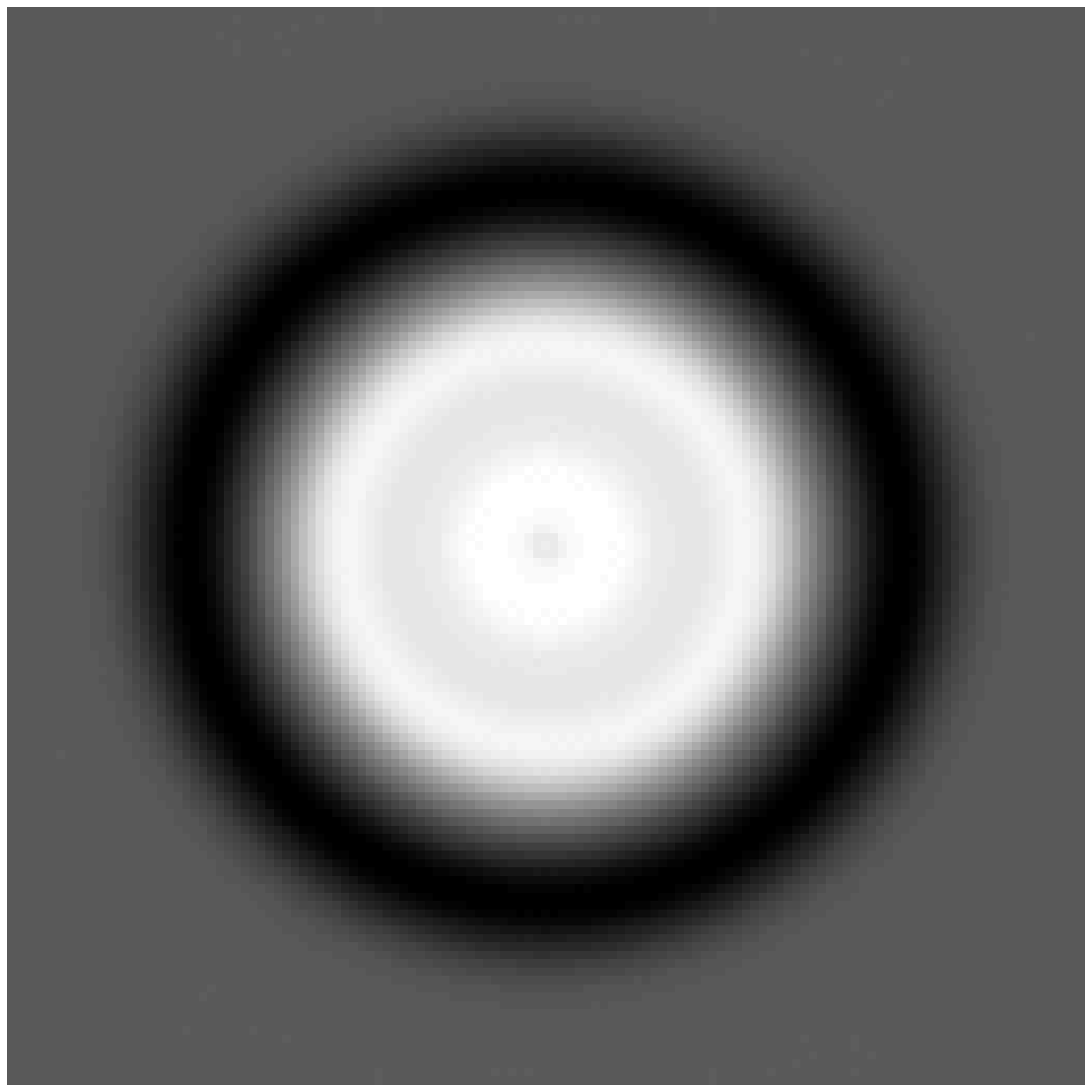}
\includegraphics[width=0.18\textwidth]{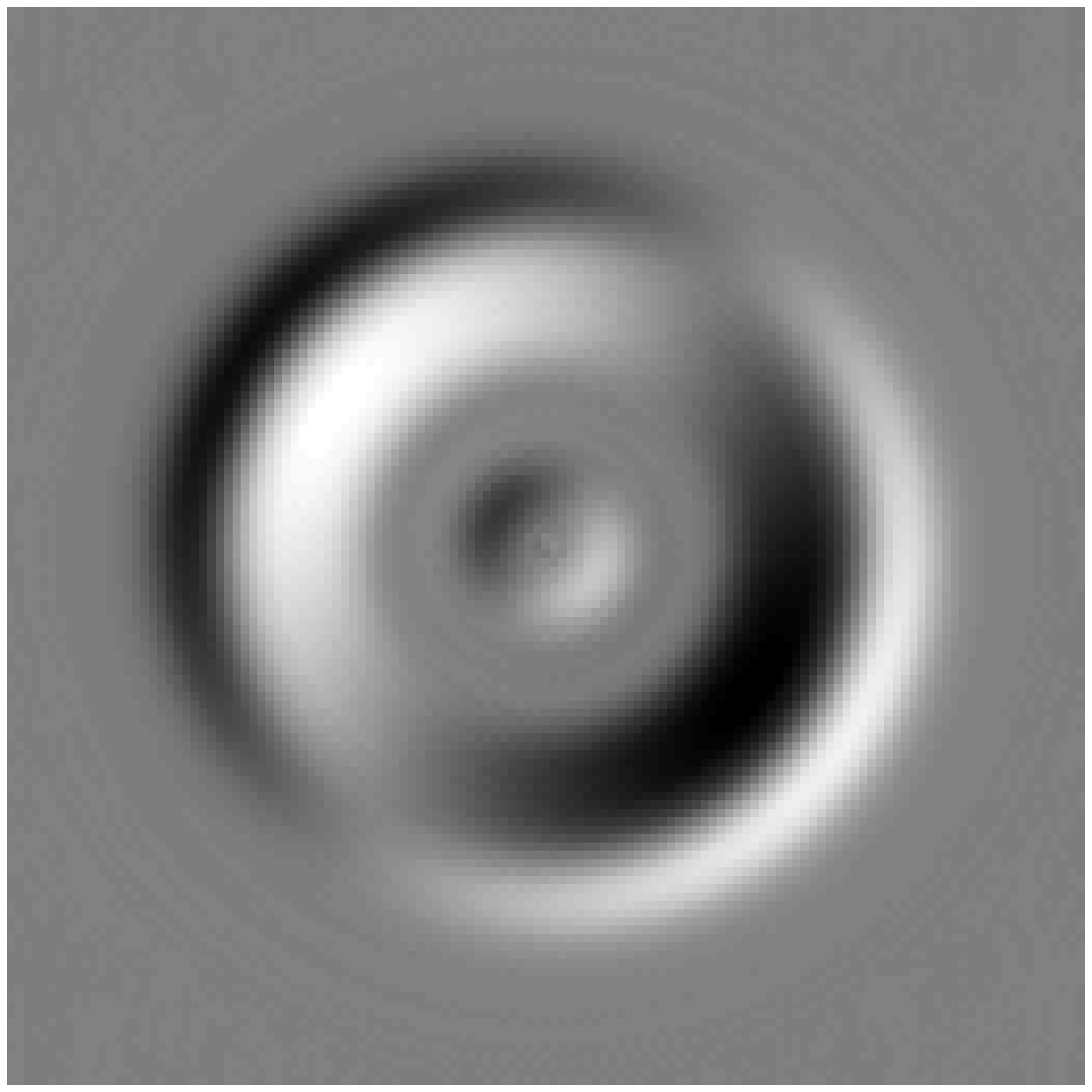}
\includegraphics[width=0.18\textwidth]{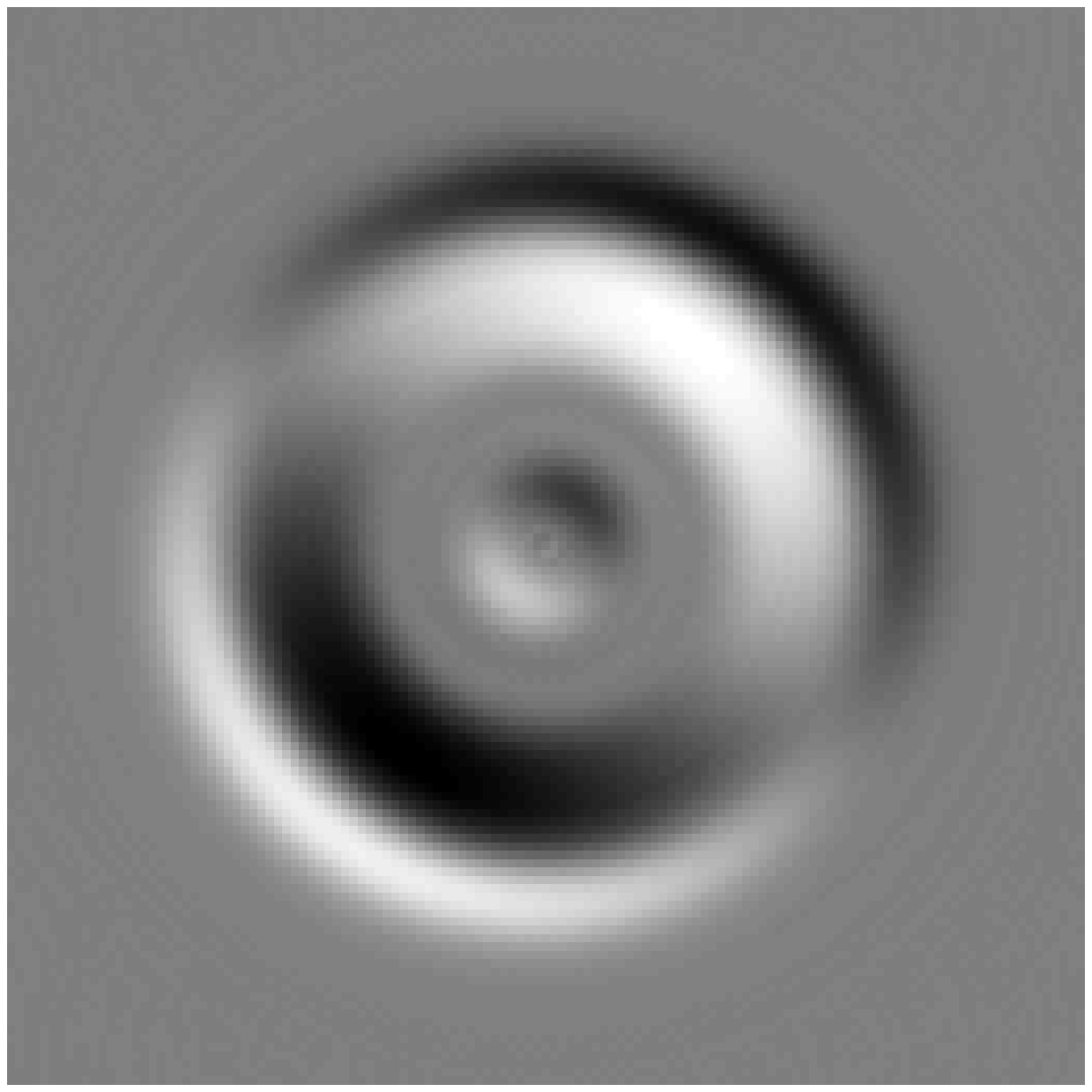}
\includegraphics[width=0.18\textwidth]{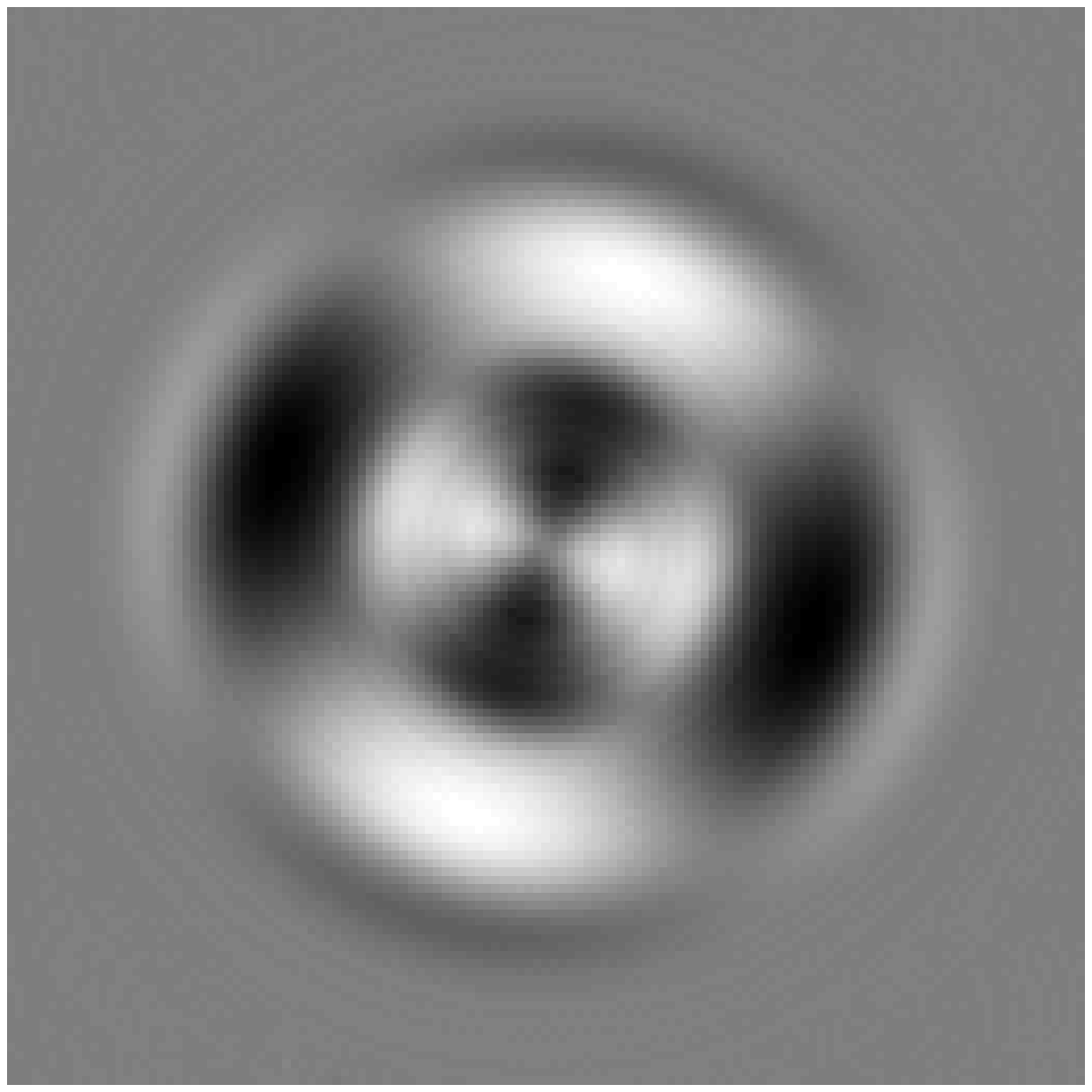}
\includegraphics[width=0.18\textwidth]{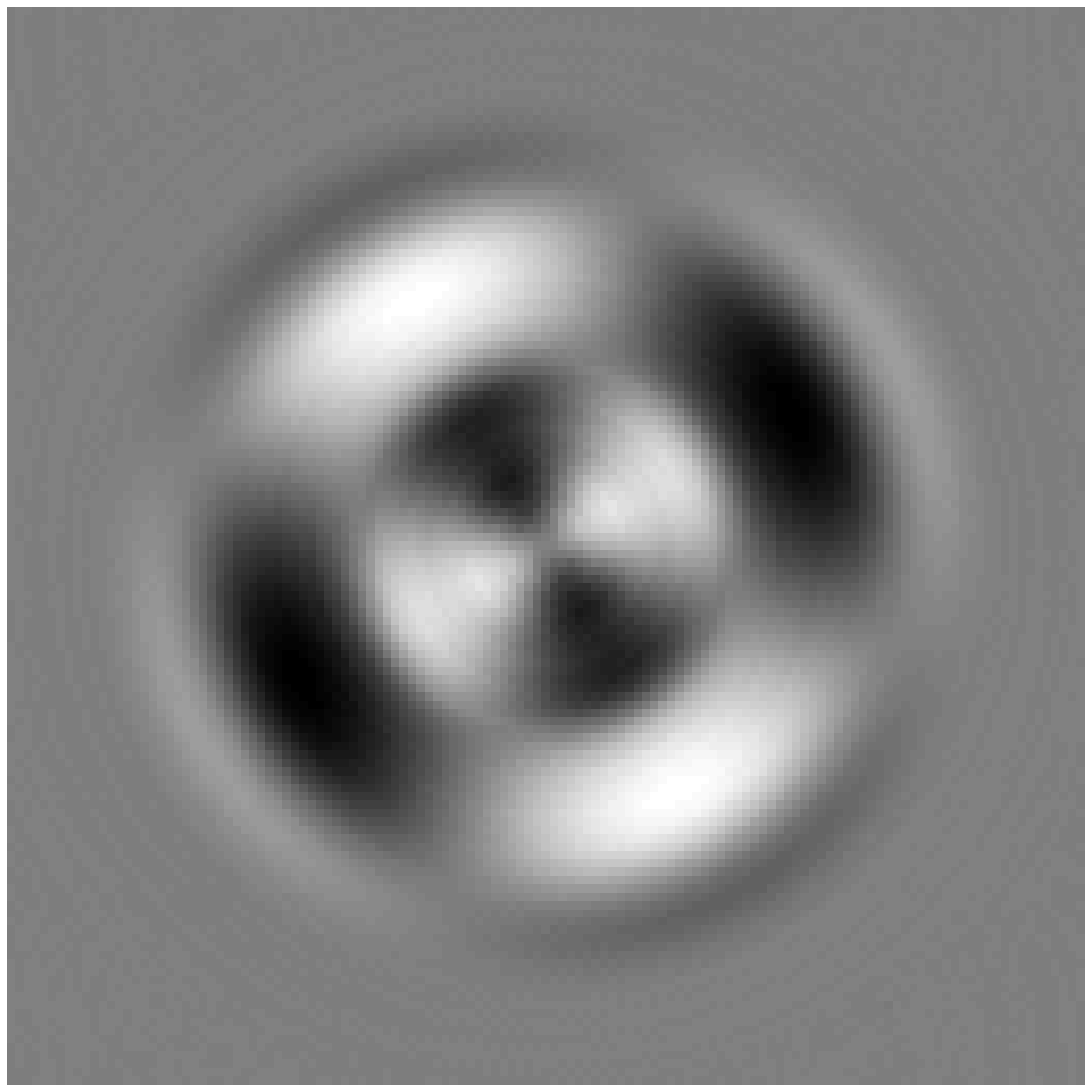} \\
\includegraphics[width=0.18\textwidth]{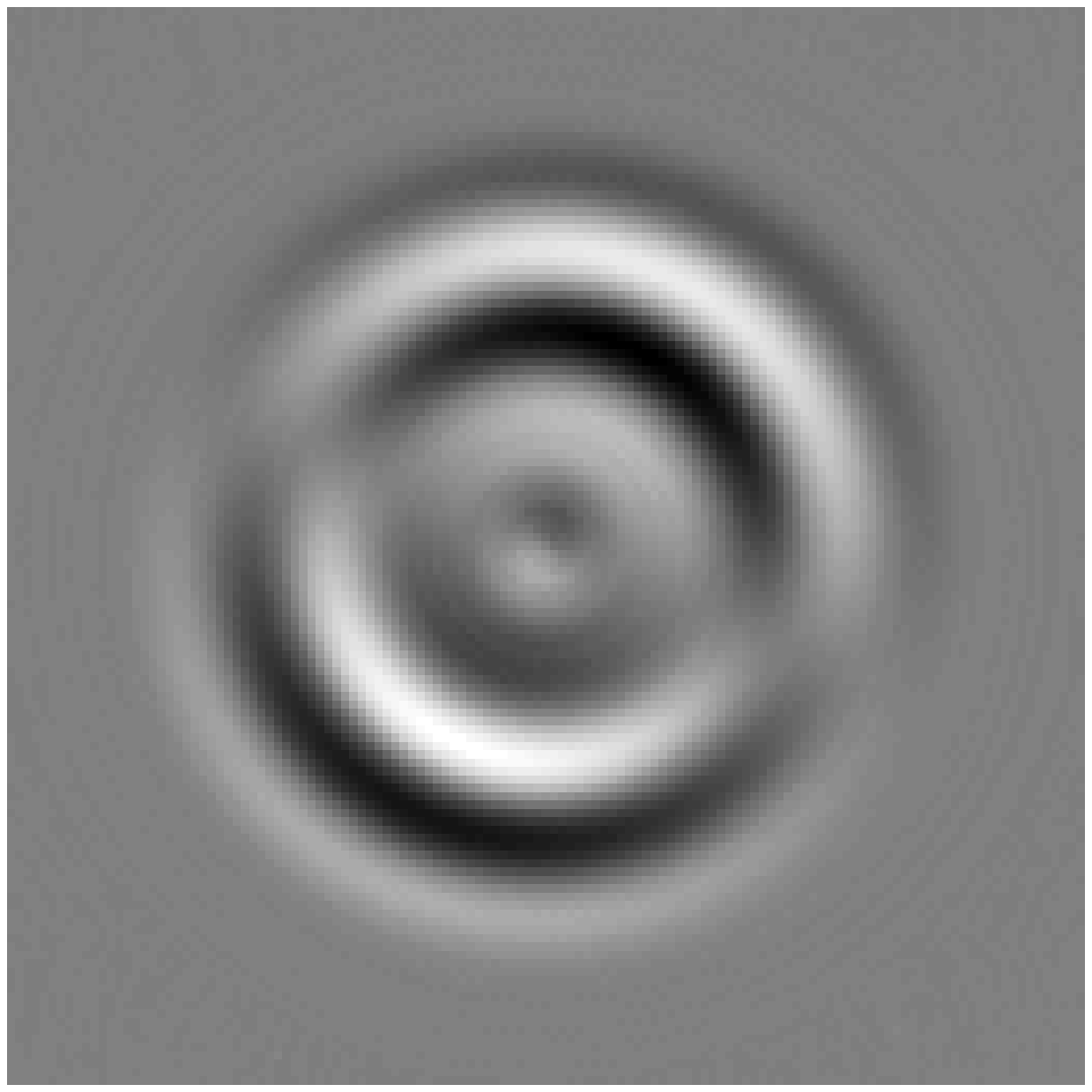}
\includegraphics[width=0.18\textwidth]{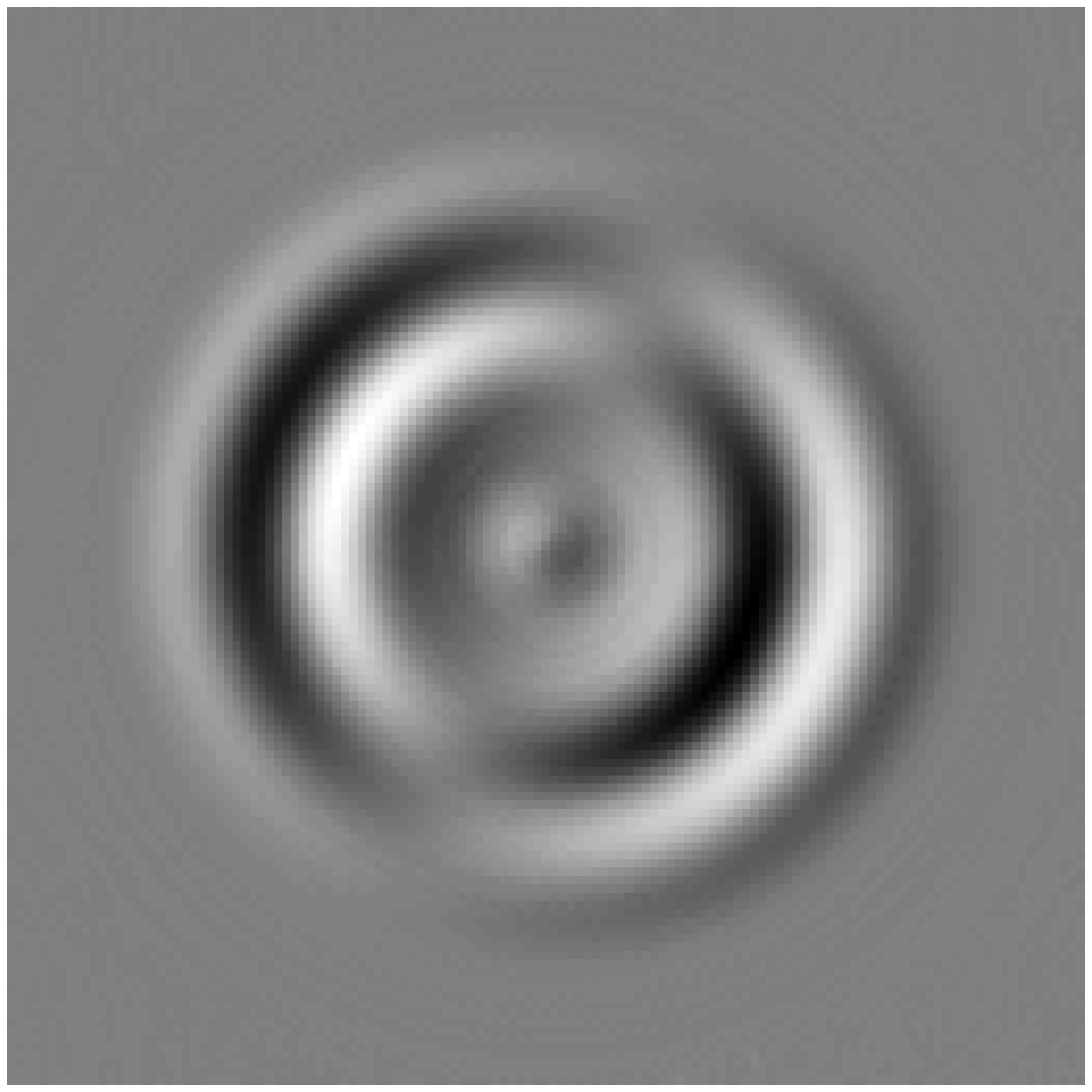}
\includegraphics[width=0.18\textwidth]{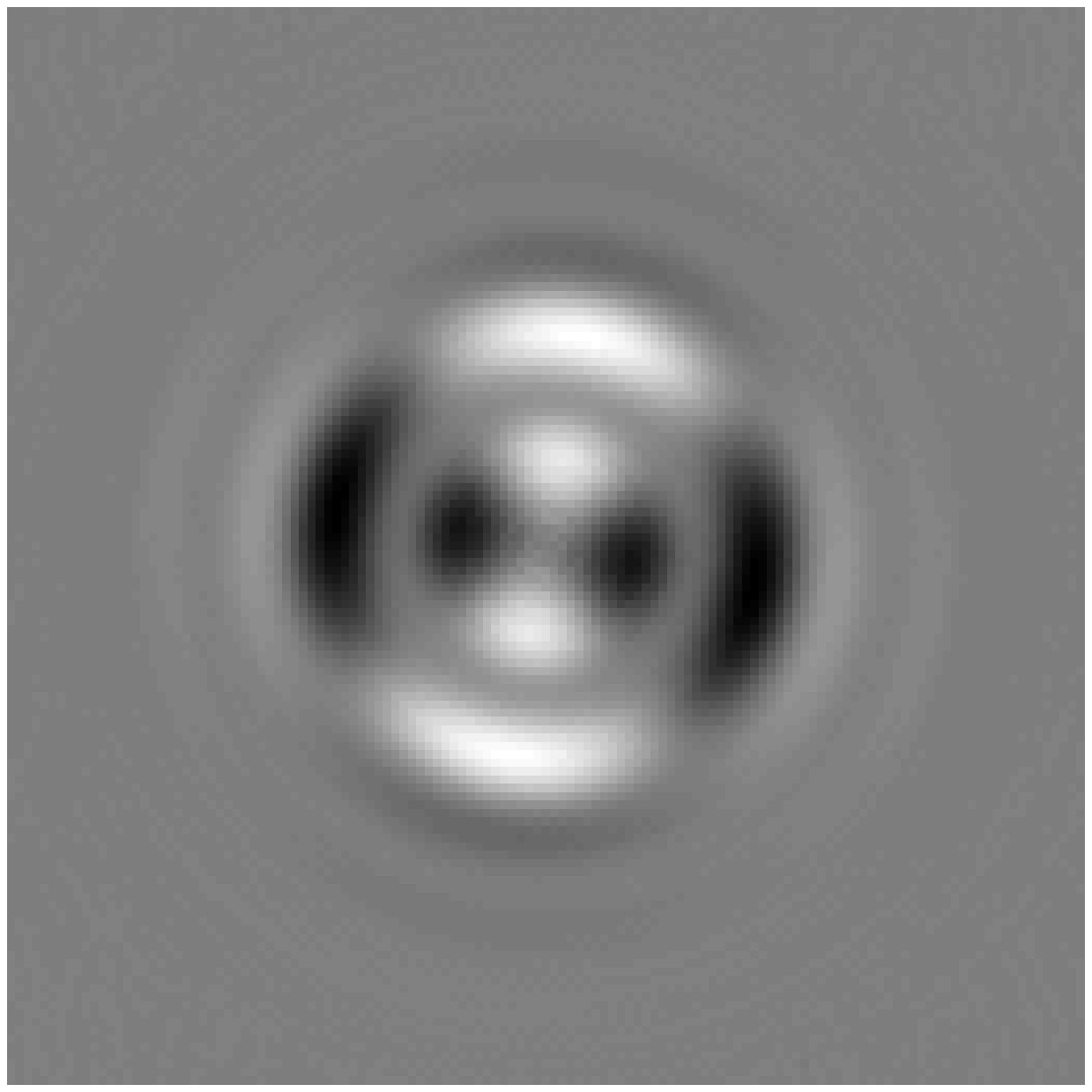}
\includegraphics[width=0.18\textwidth]{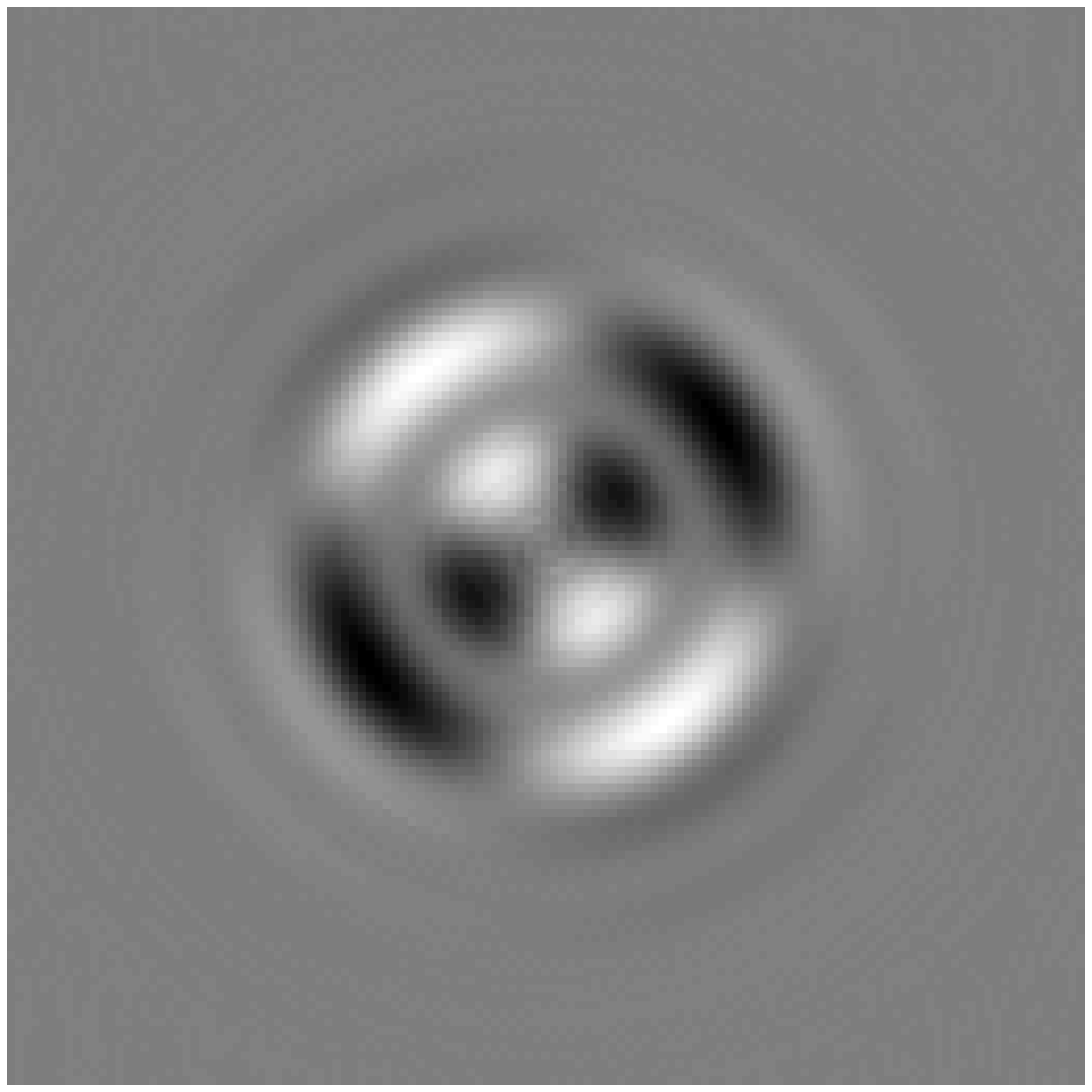}
\includegraphics[width=0.18\textwidth]{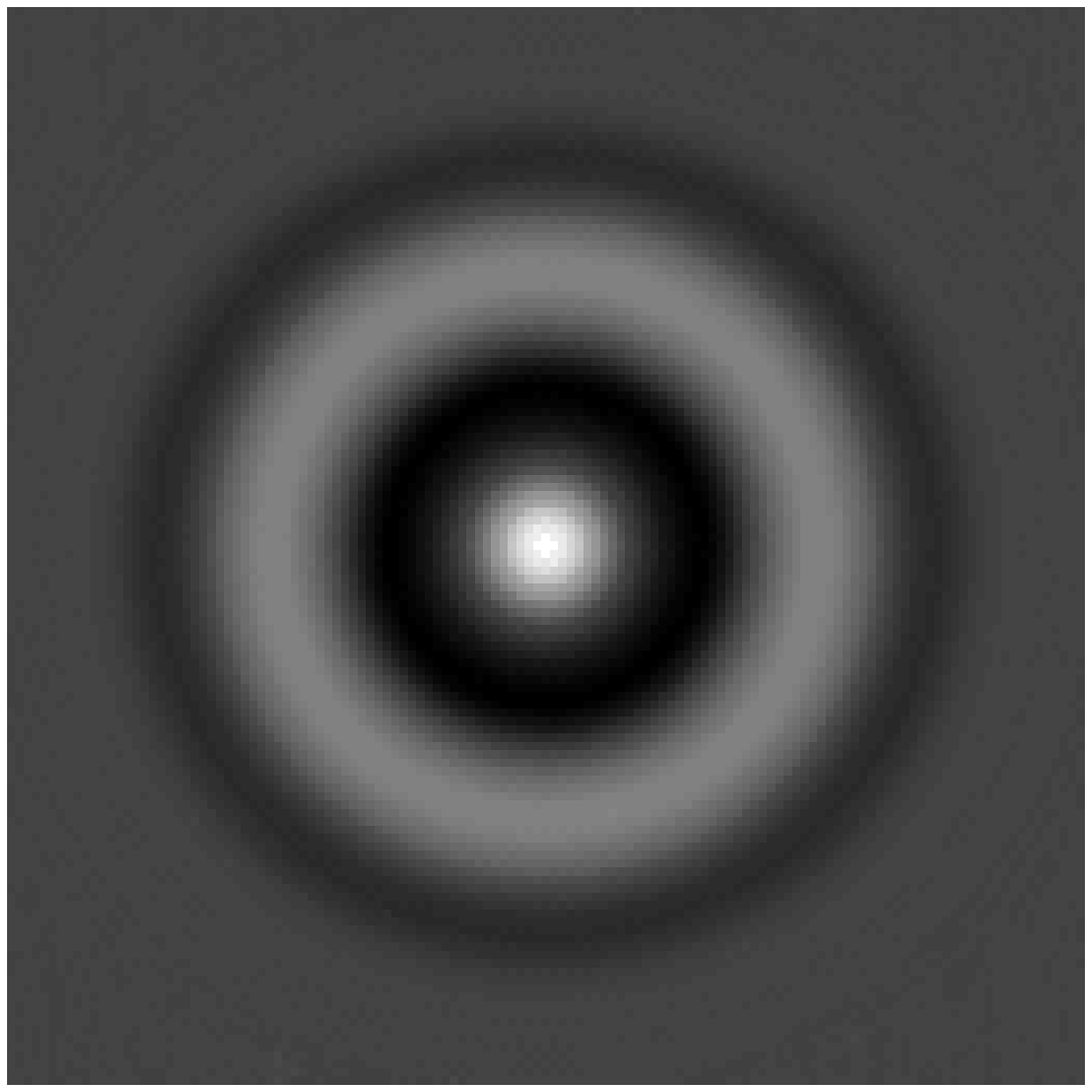} \\
\includegraphics[width=0.18\textwidth]{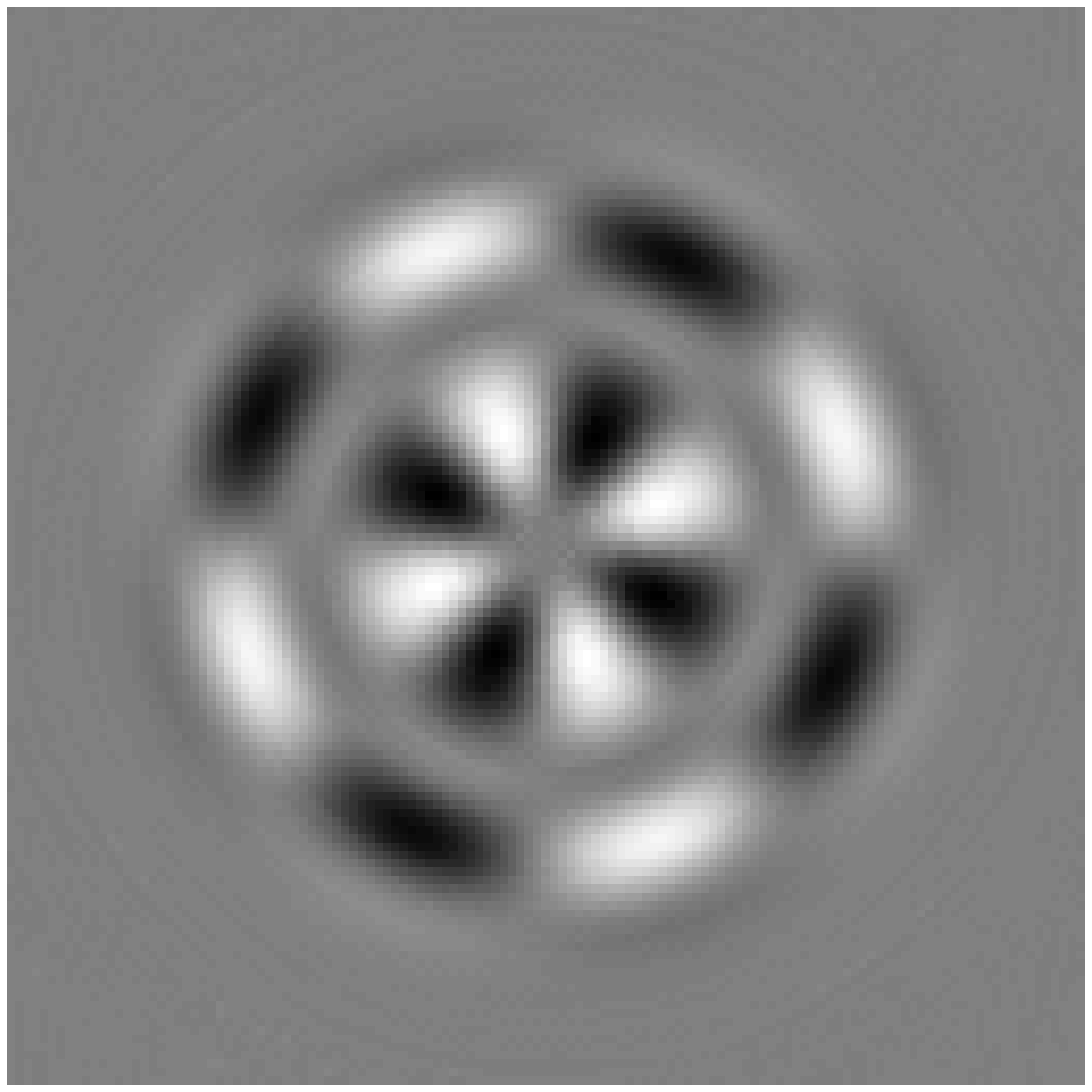}
\includegraphics[width=0.18\textwidth]{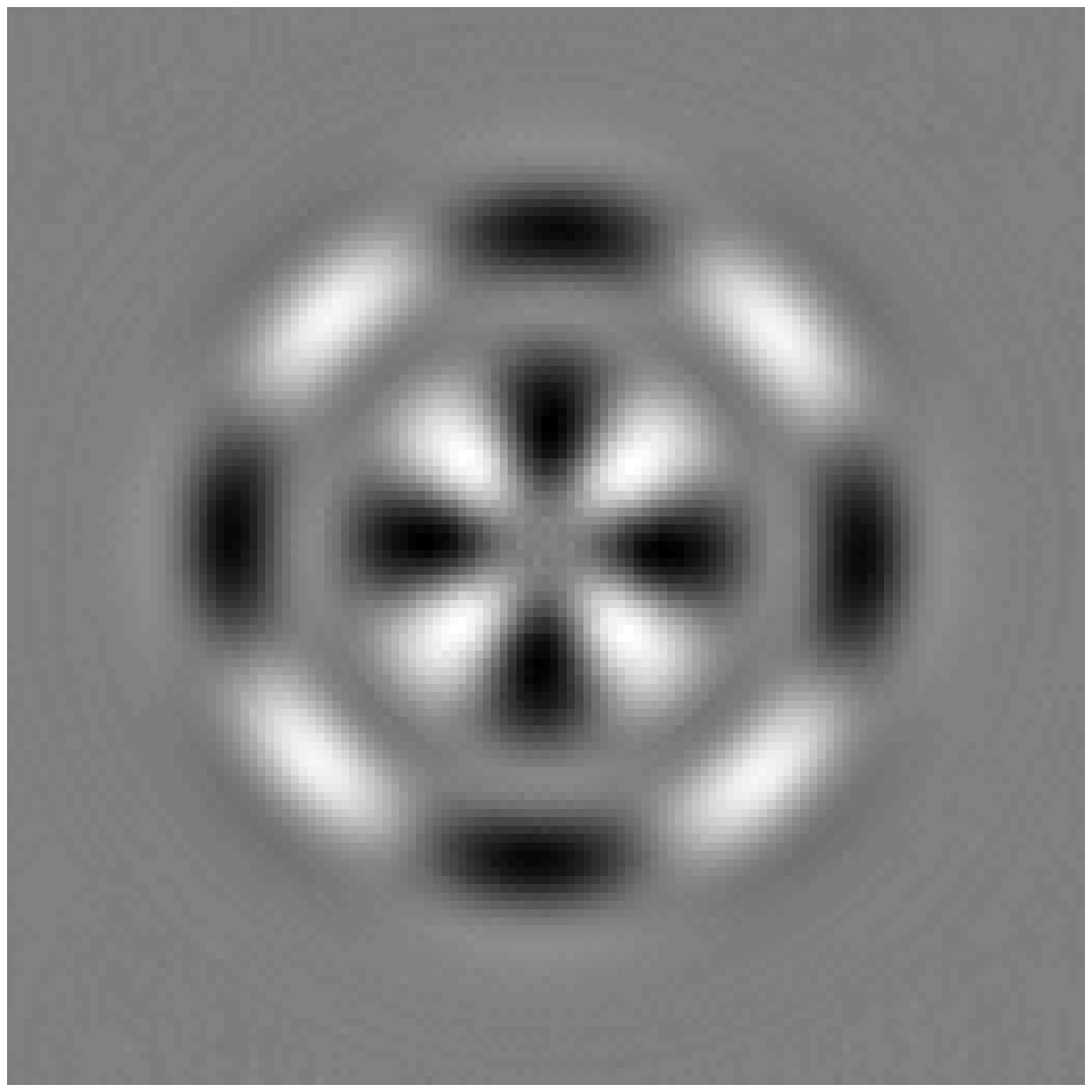}
\includegraphics[width=0.18\textwidth]{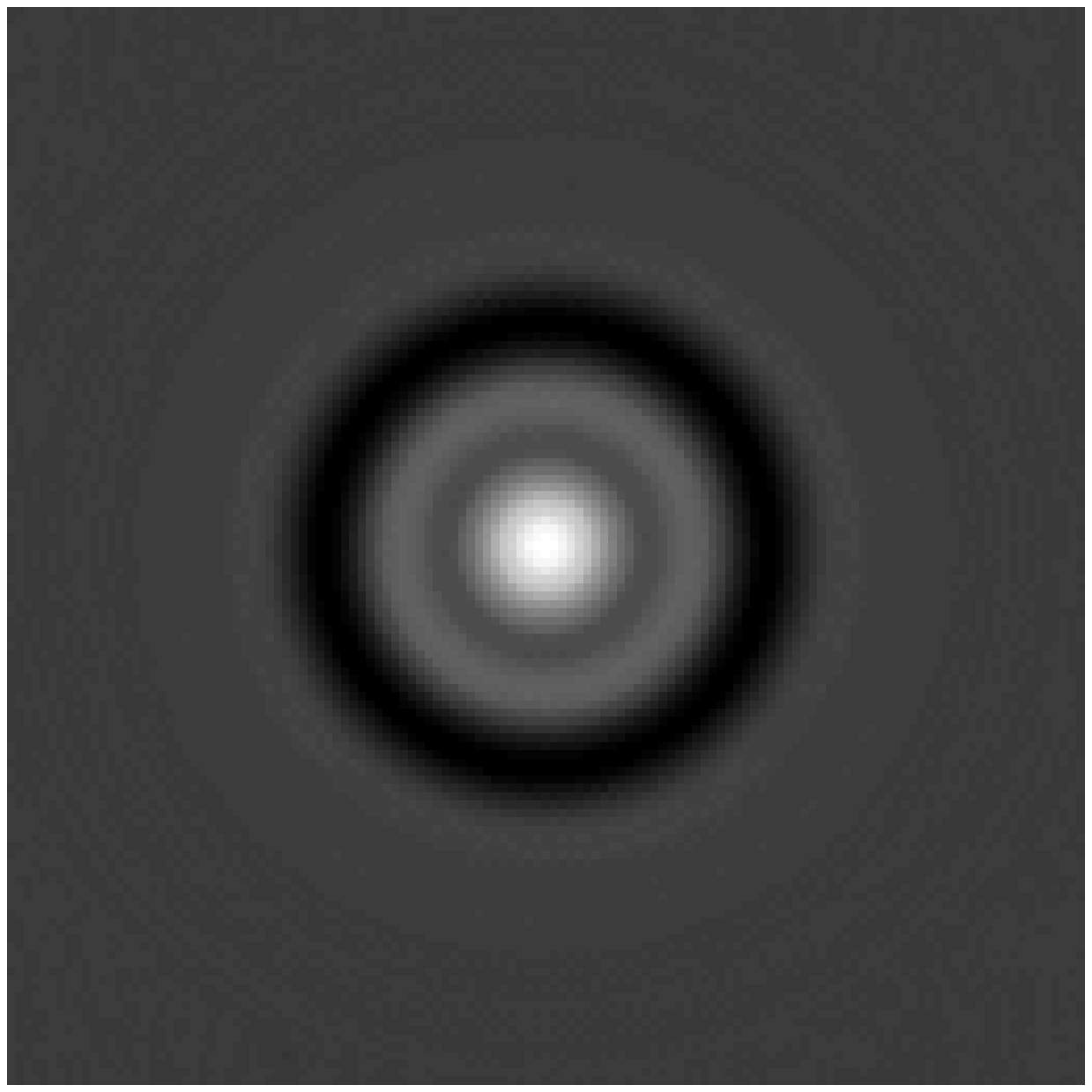}
\includegraphics[width=0.18\textwidth]{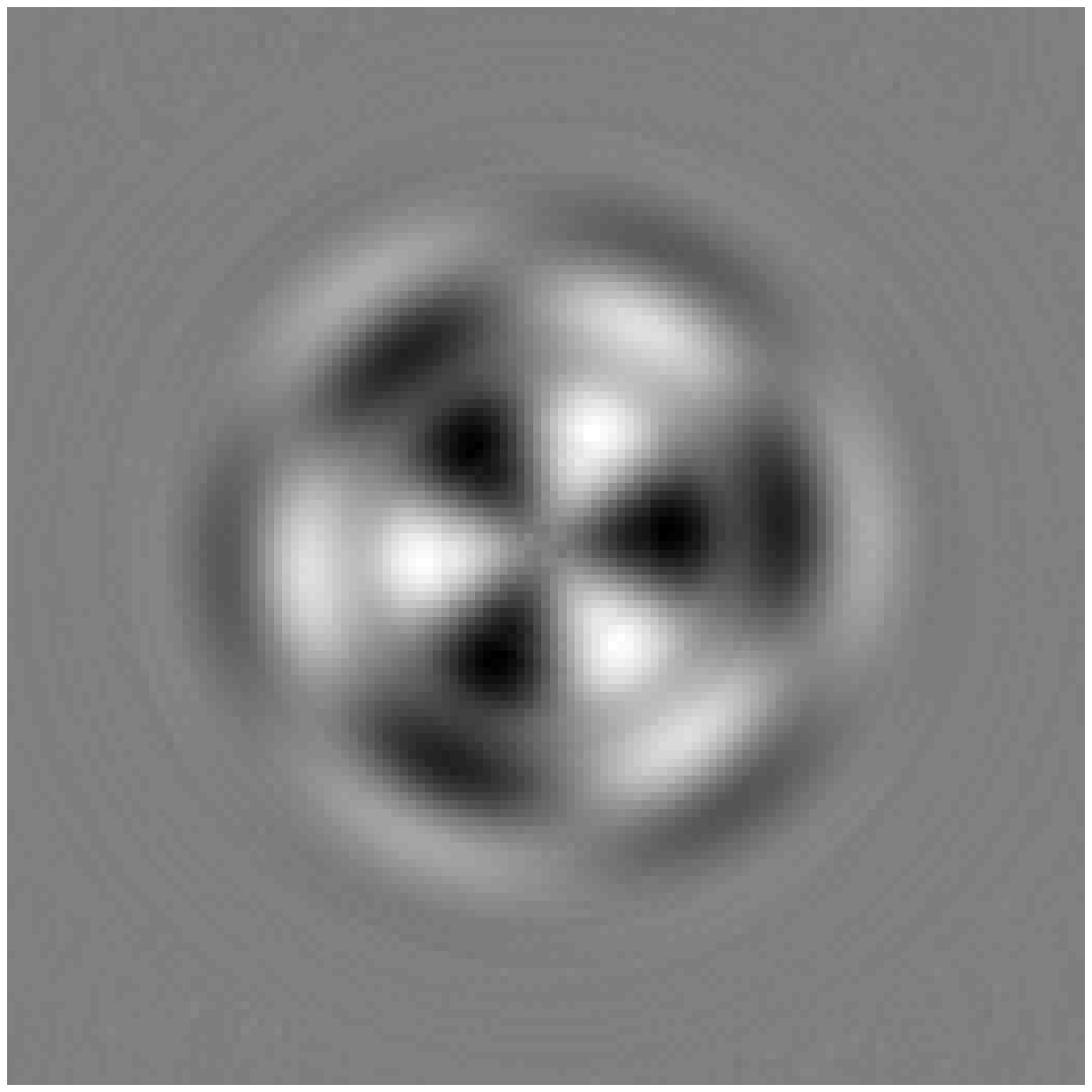}
\includegraphics[width=0.18\textwidth]{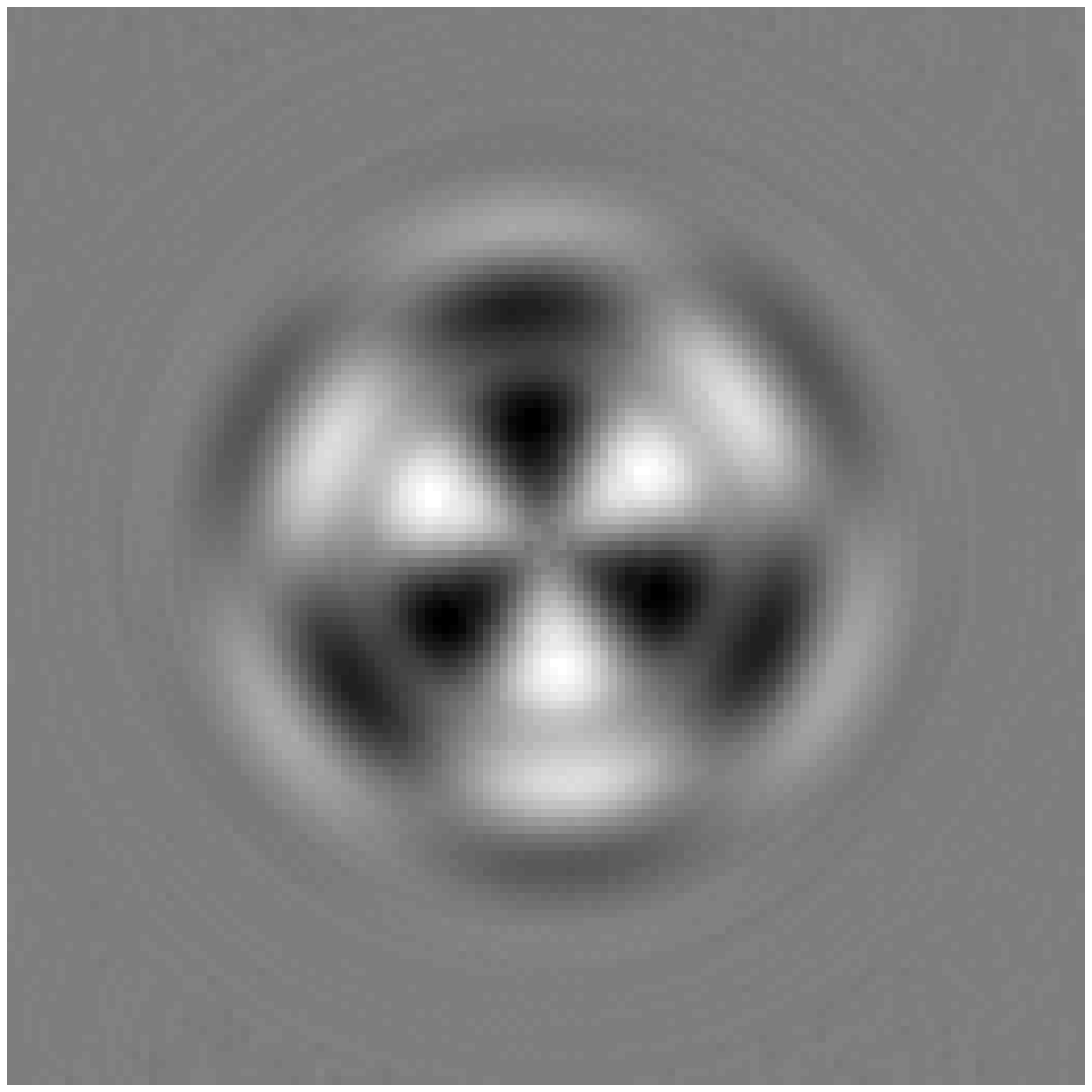}\\
\includegraphics[width=0.18\textwidth]{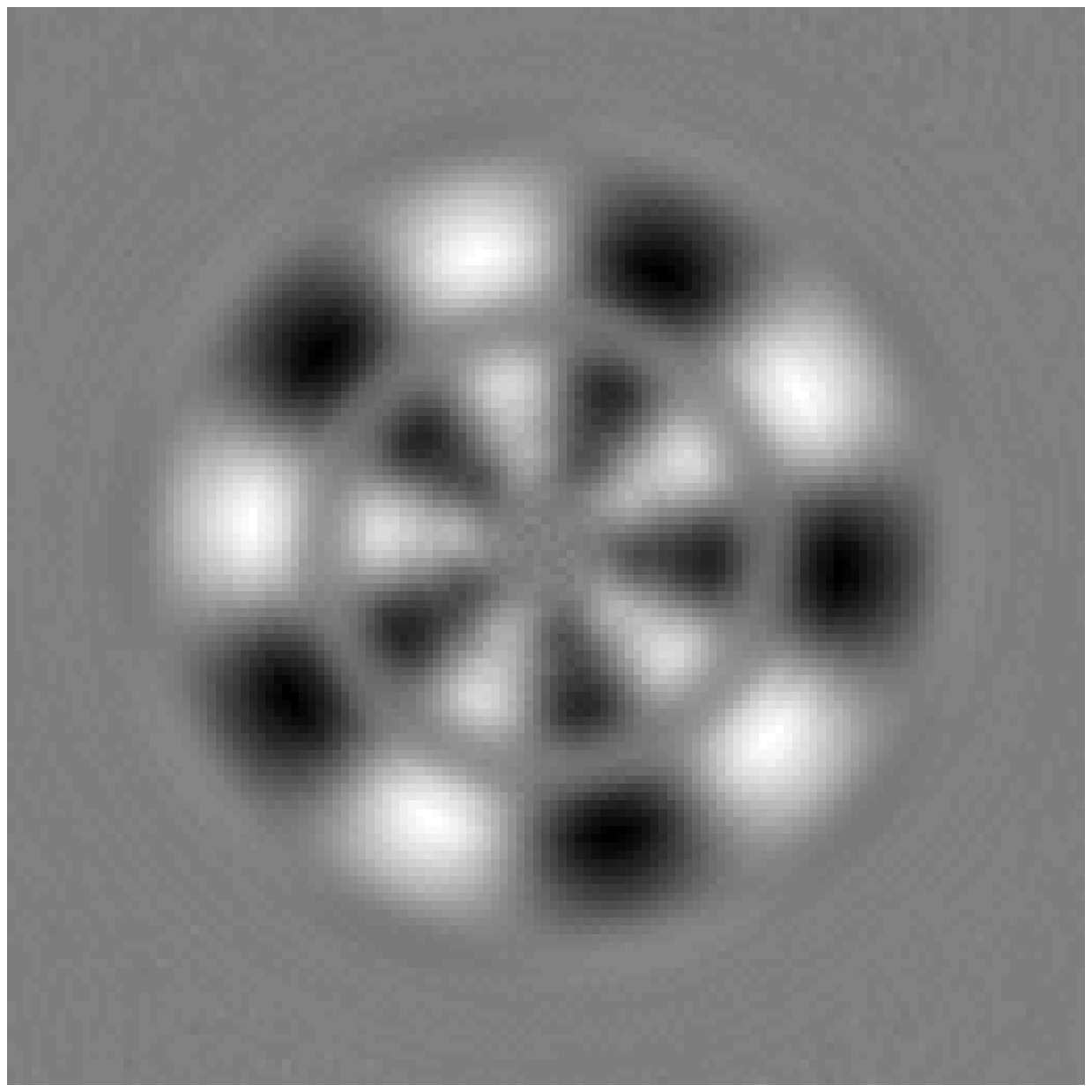}
\includegraphics[width=0.18\textwidth]{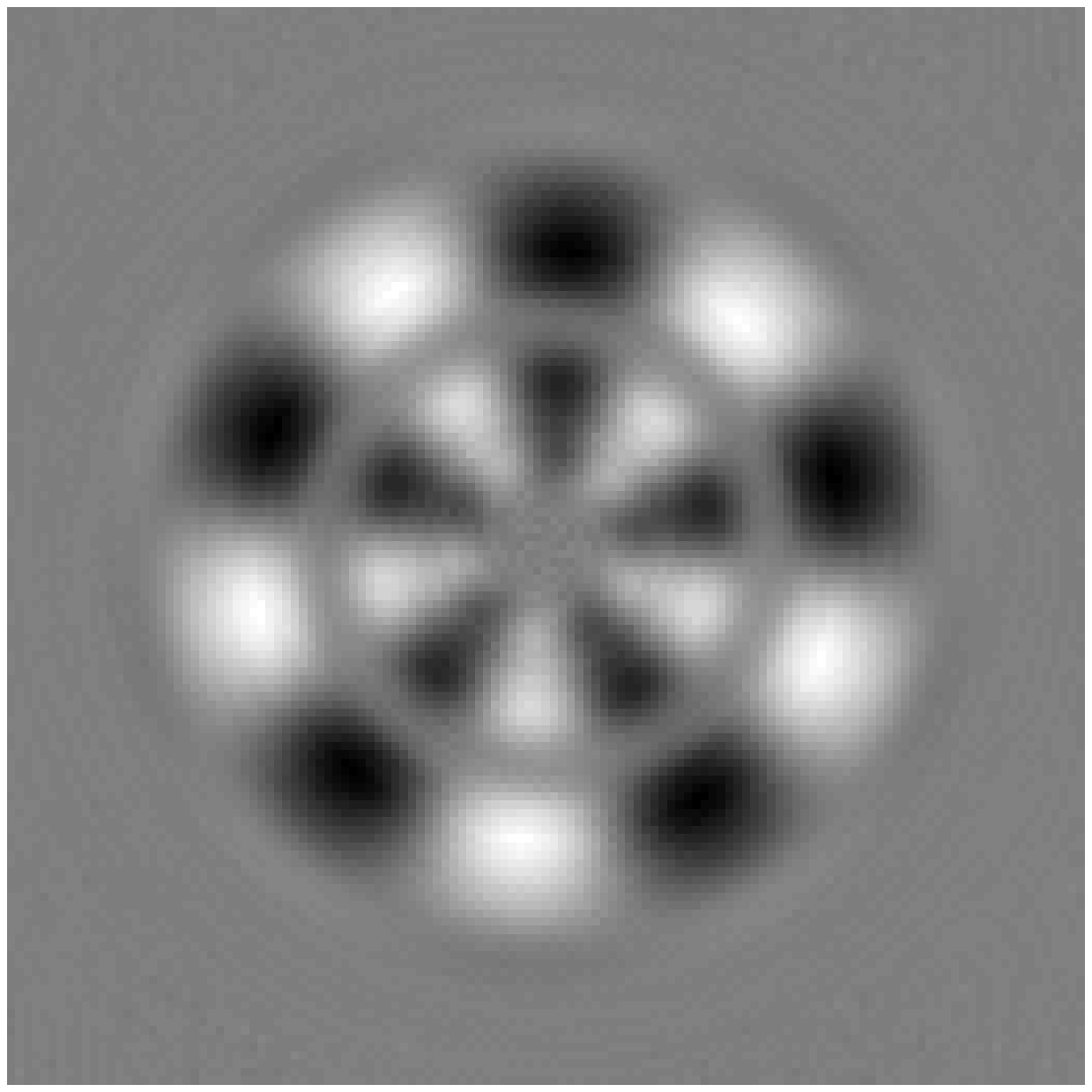}
\includegraphics[width=0.18\textwidth]{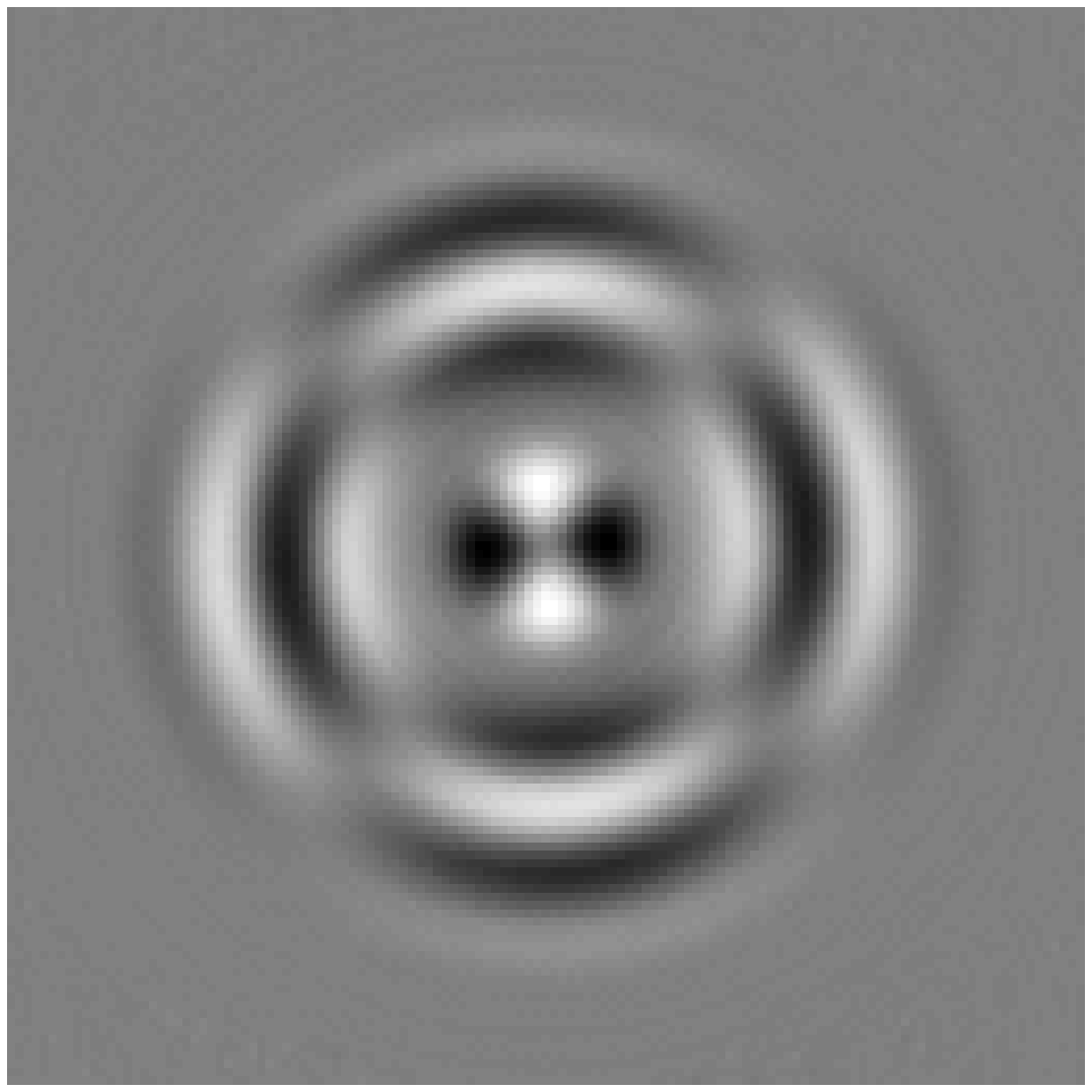}
\includegraphics[width=0.18\textwidth]{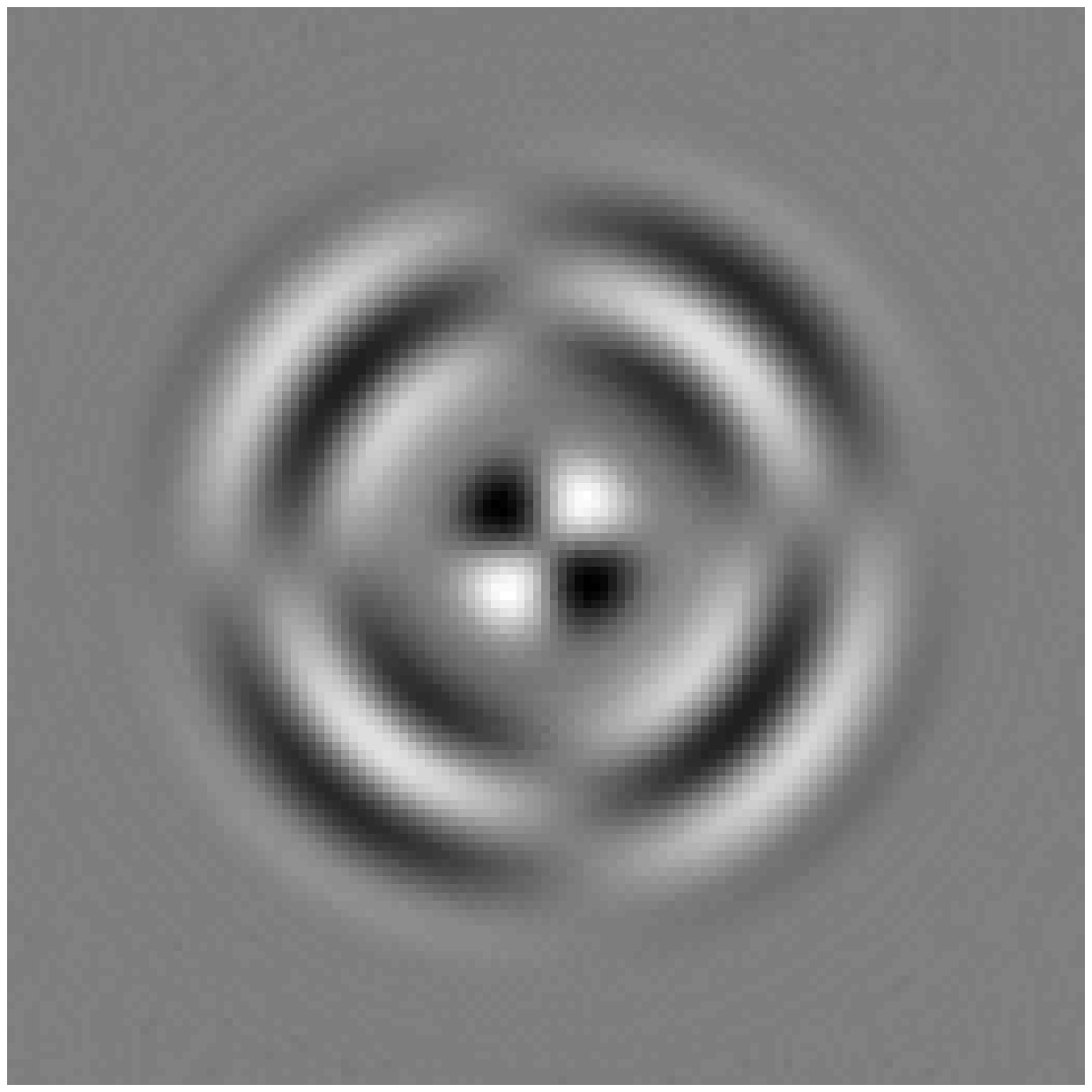}
\includegraphics[width=0.18\textwidth]{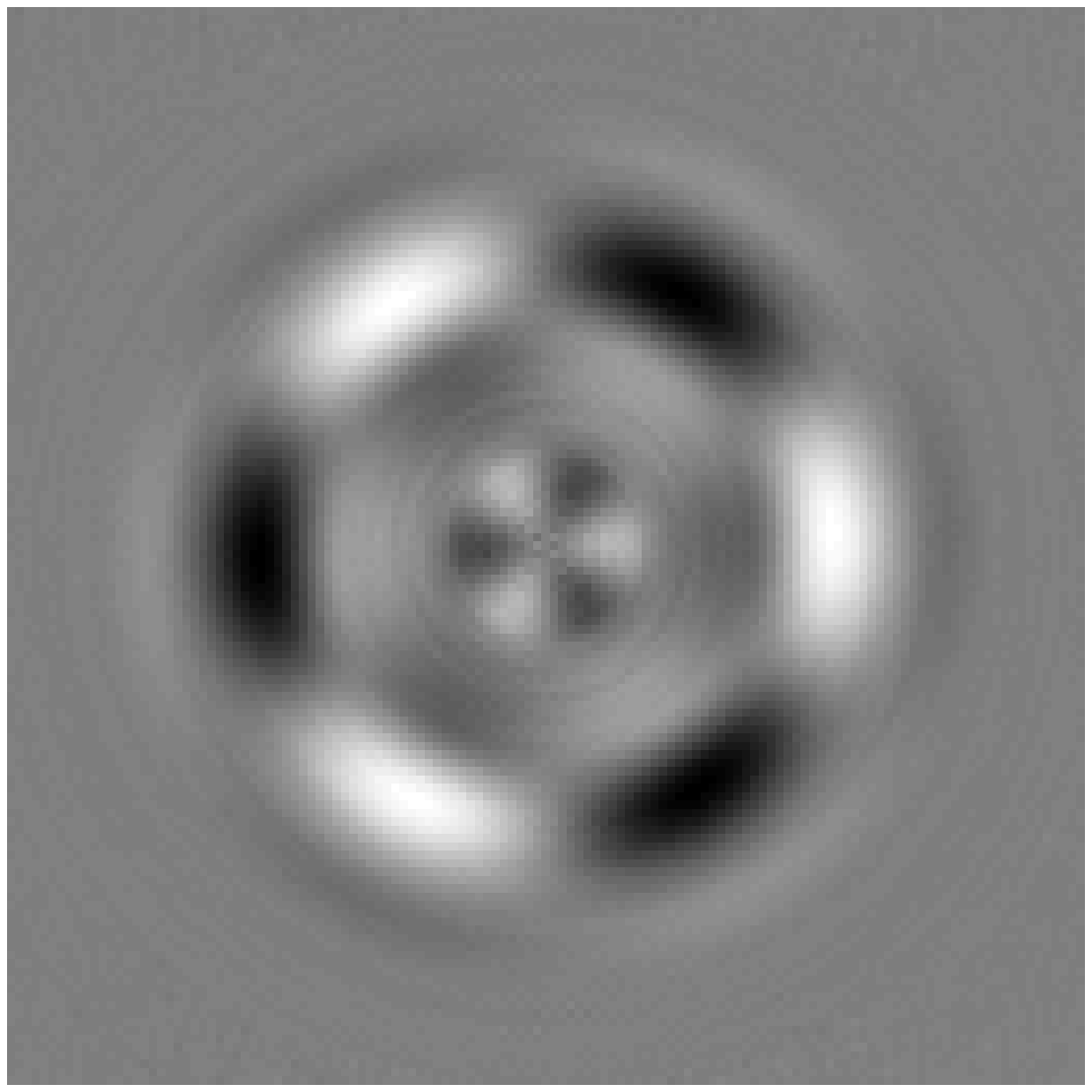}\\
\includegraphics[width=0.18\textwidth]{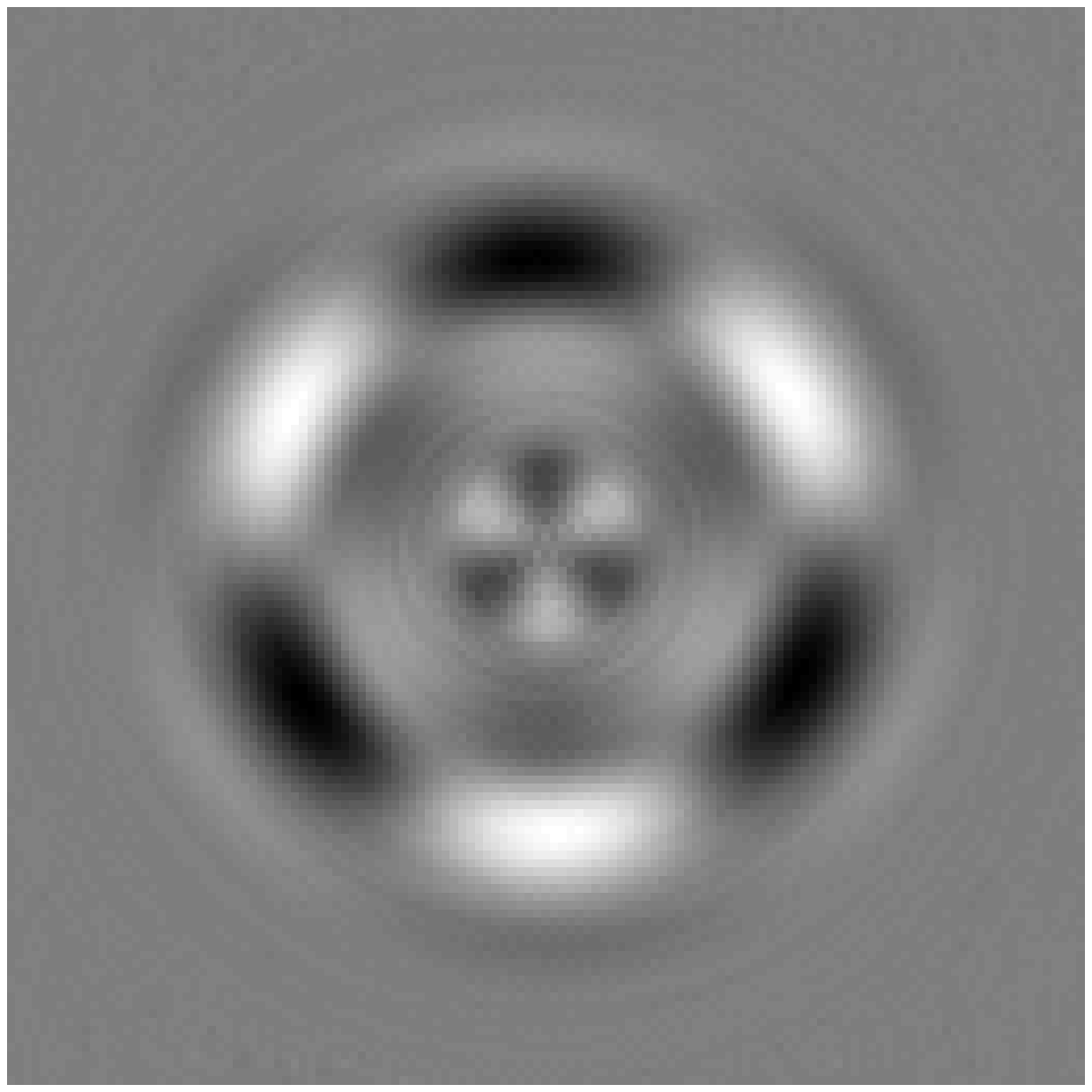}
\includegraphics[width=0.18\textwidth]{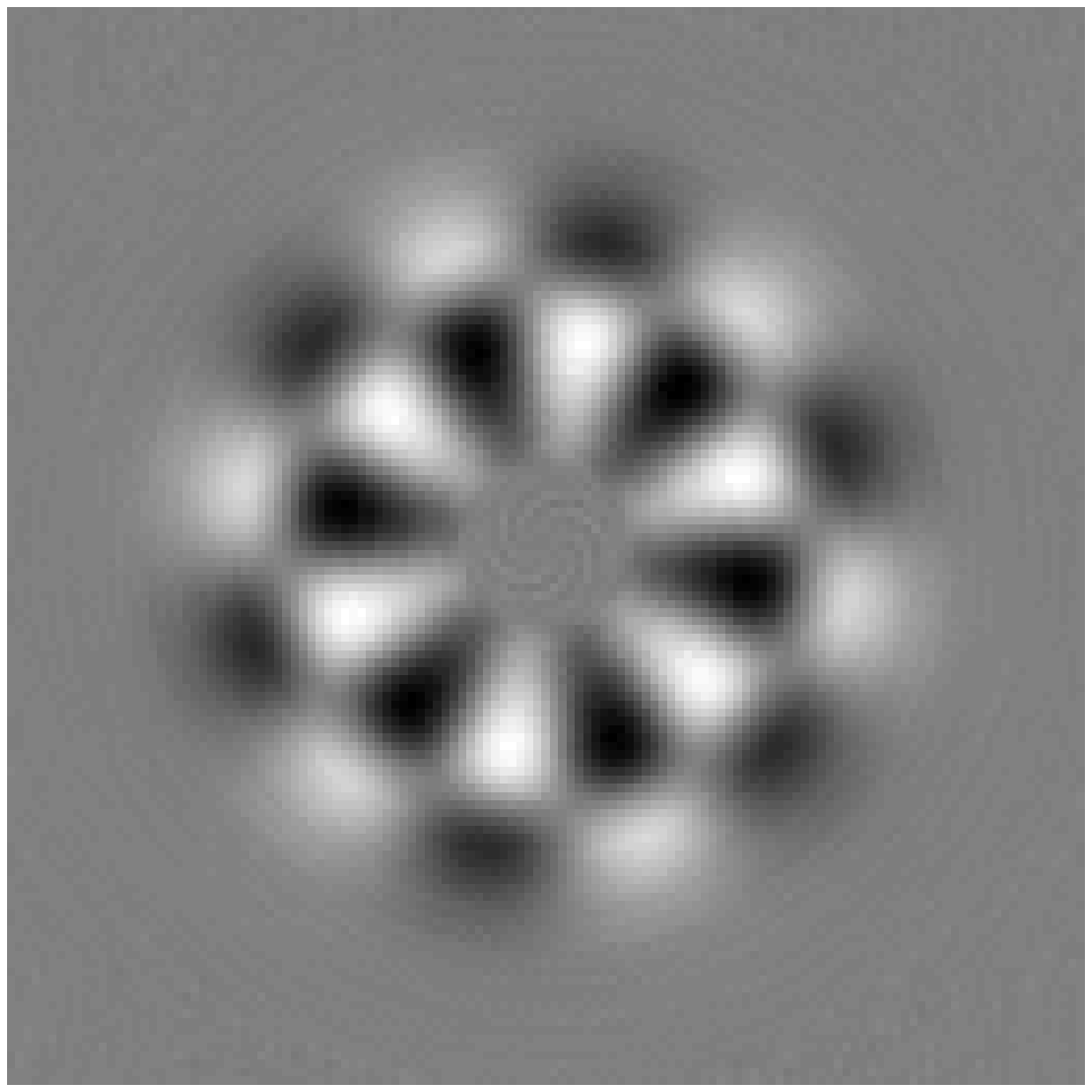}
\includegraphics[width=0.18\textwidth]{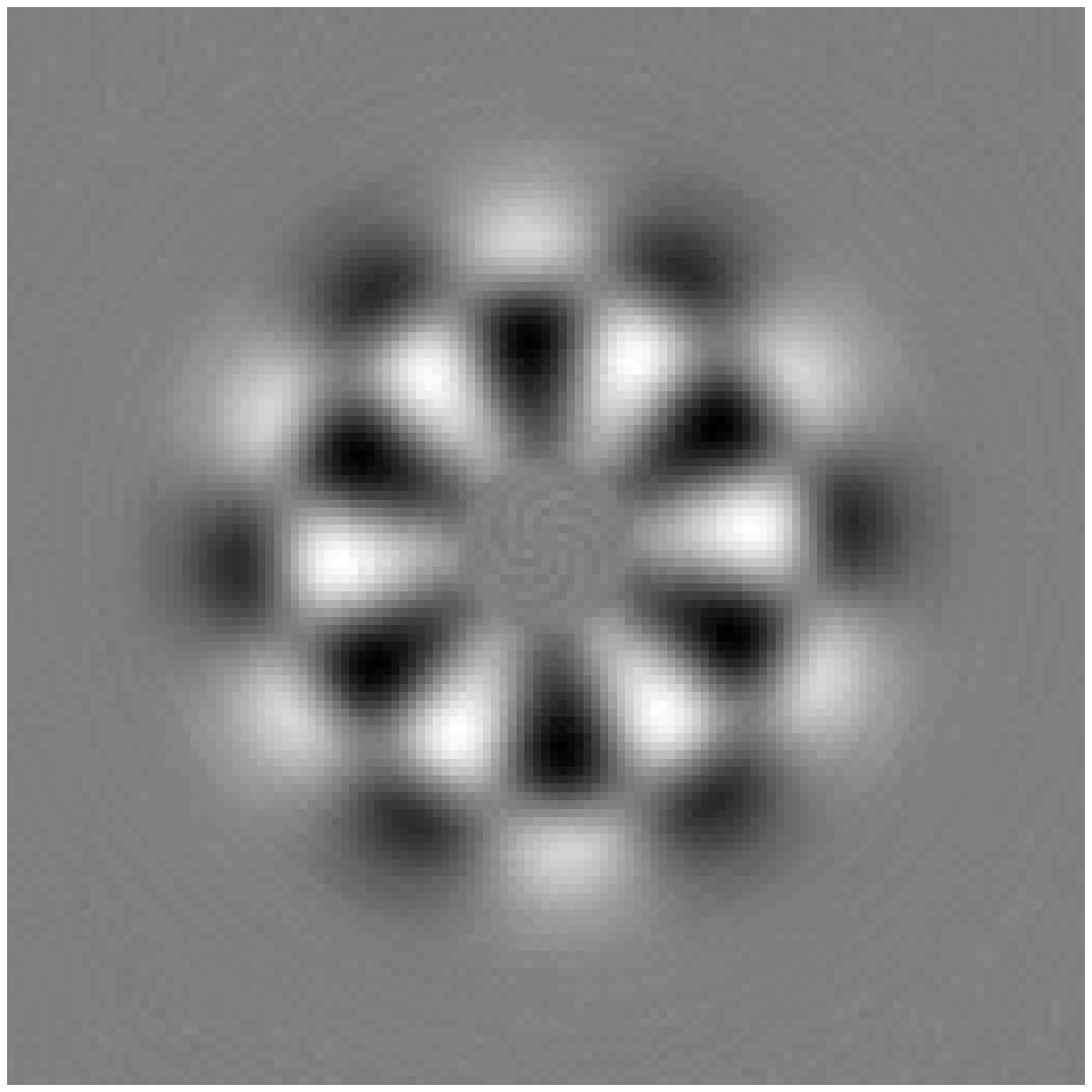}
\includegraphics[width=0.18\textwidth]{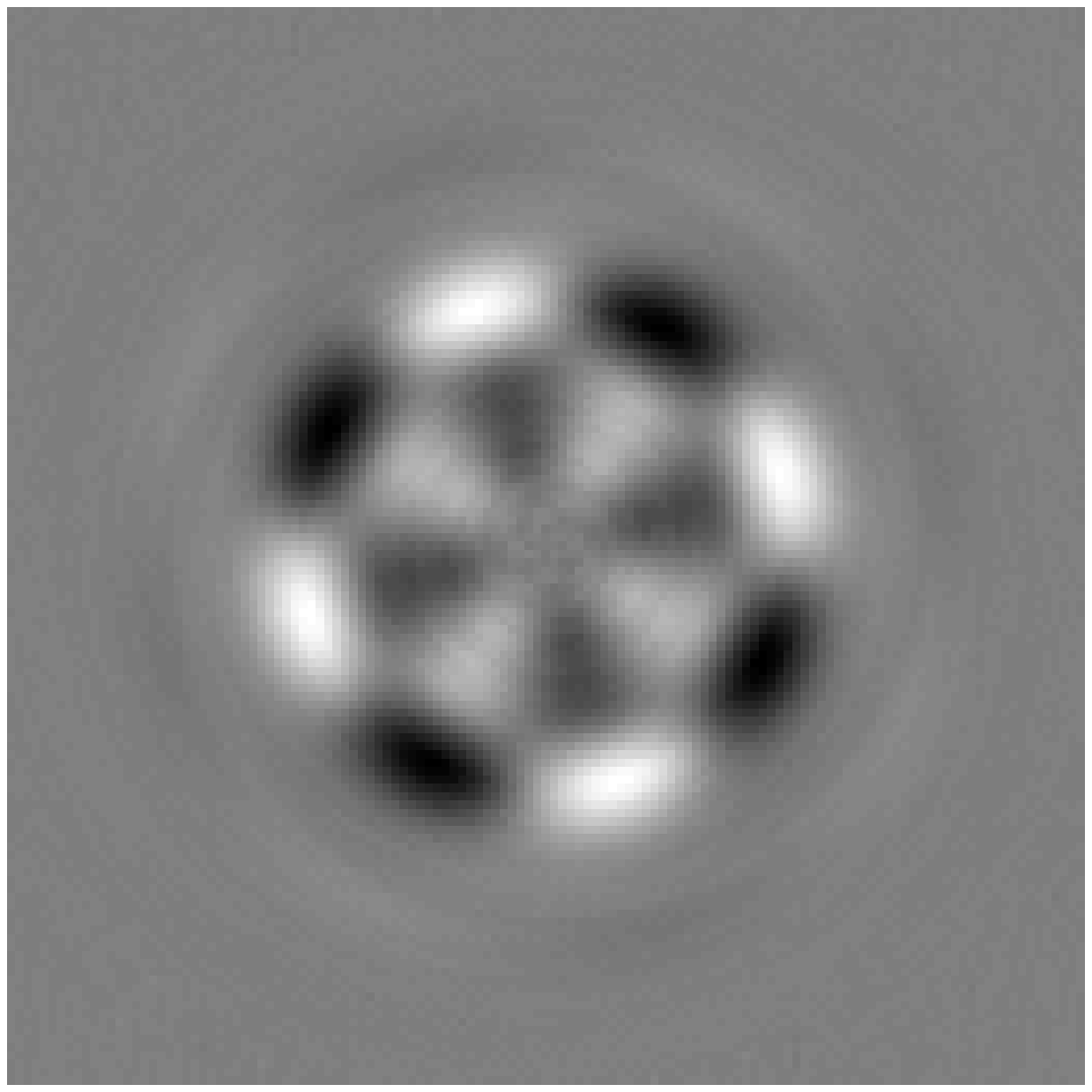}
\includegraphics[width=0.18\textwidth]{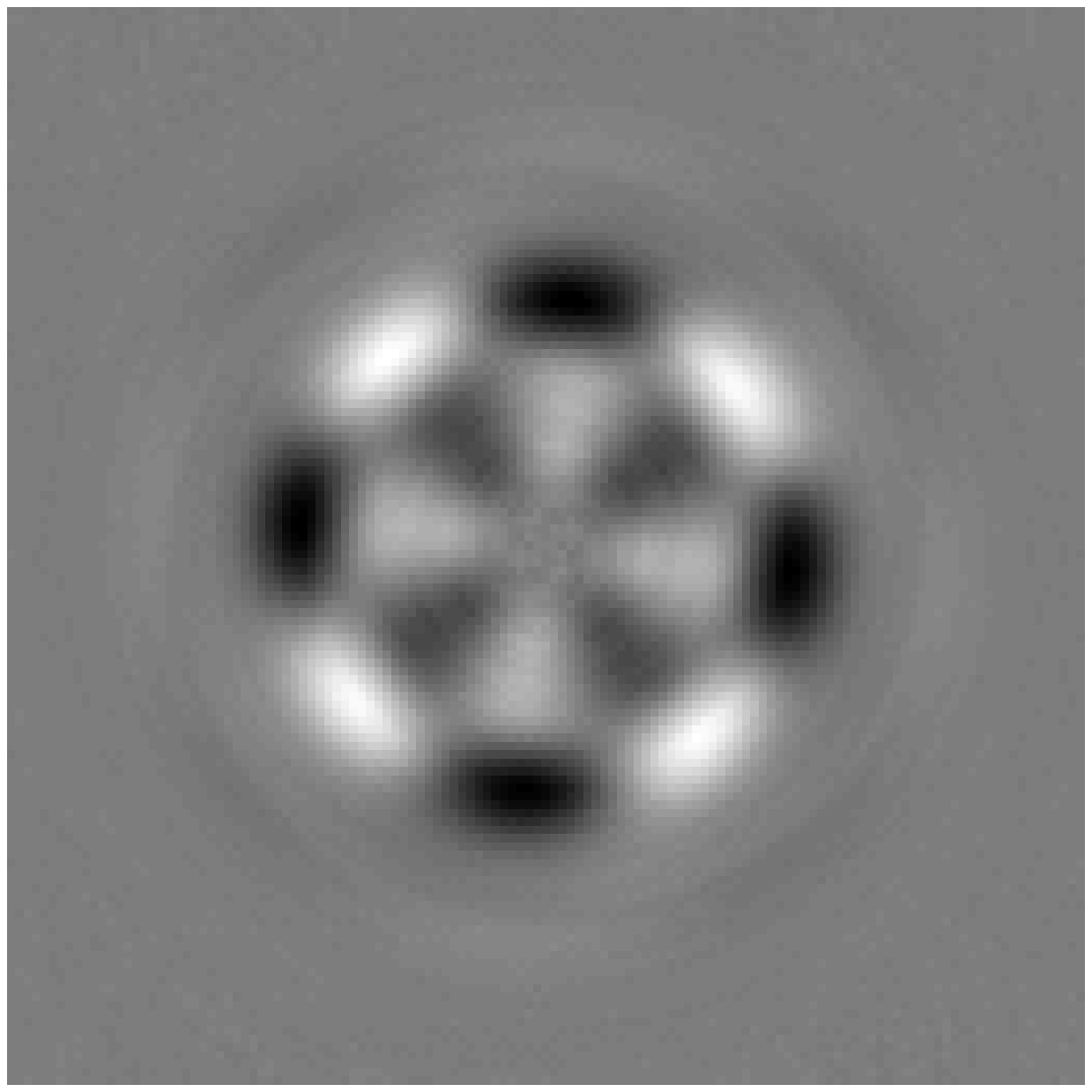}
\\\vspace{.25em}
\footnotesize{
{\sc Fig. 3b.} {\it Dominant singular vectors computed for the E. coli data set.}
}
\end{center}
\end{figure}

\begin{figure}
\begin{center}
\begin{tabular}{ccc}
     Noisy & Clean & Denoised \\\\
       \includegraphics[width=0.2\textwidth]{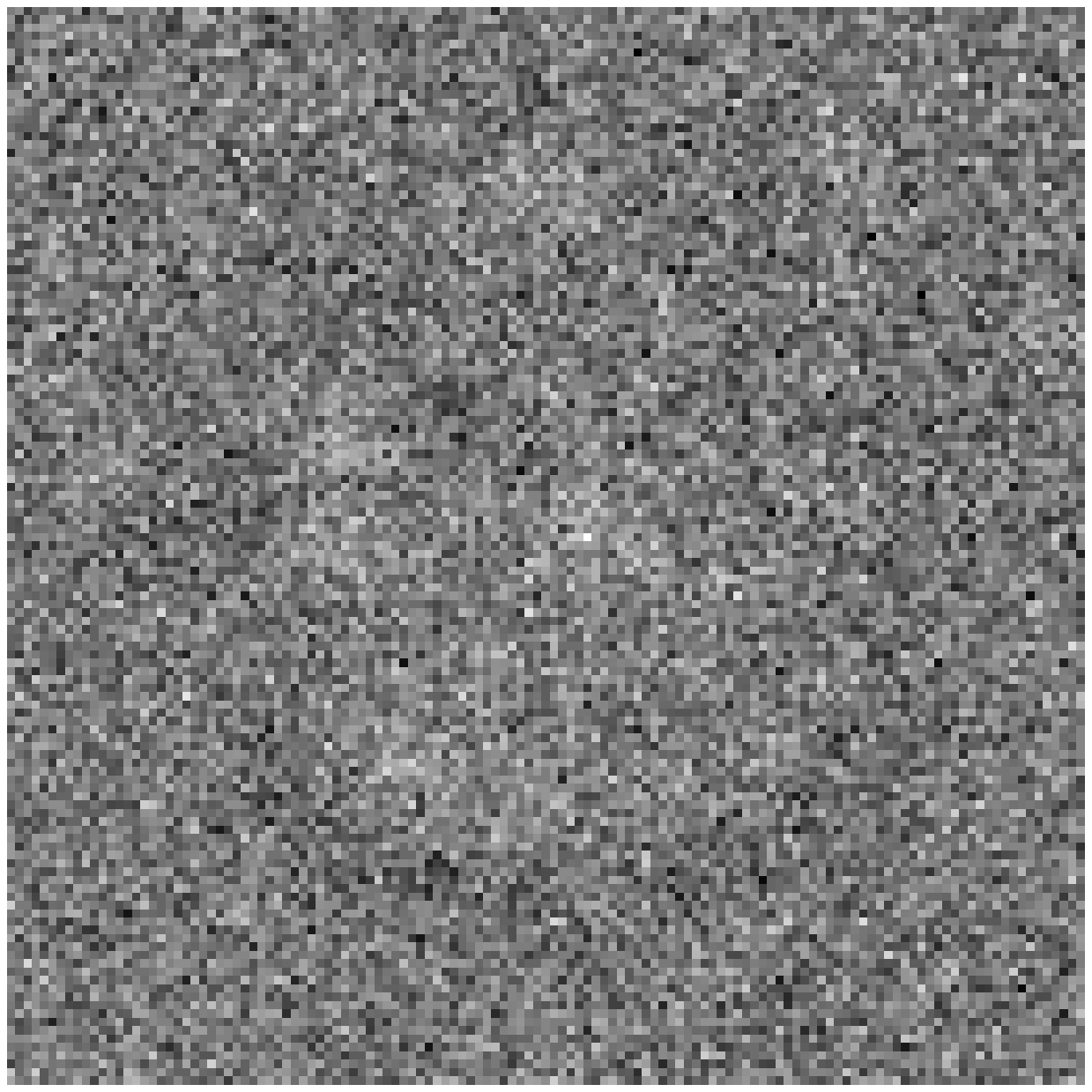}
     & \includegraphics[width=0.2\textwidth]{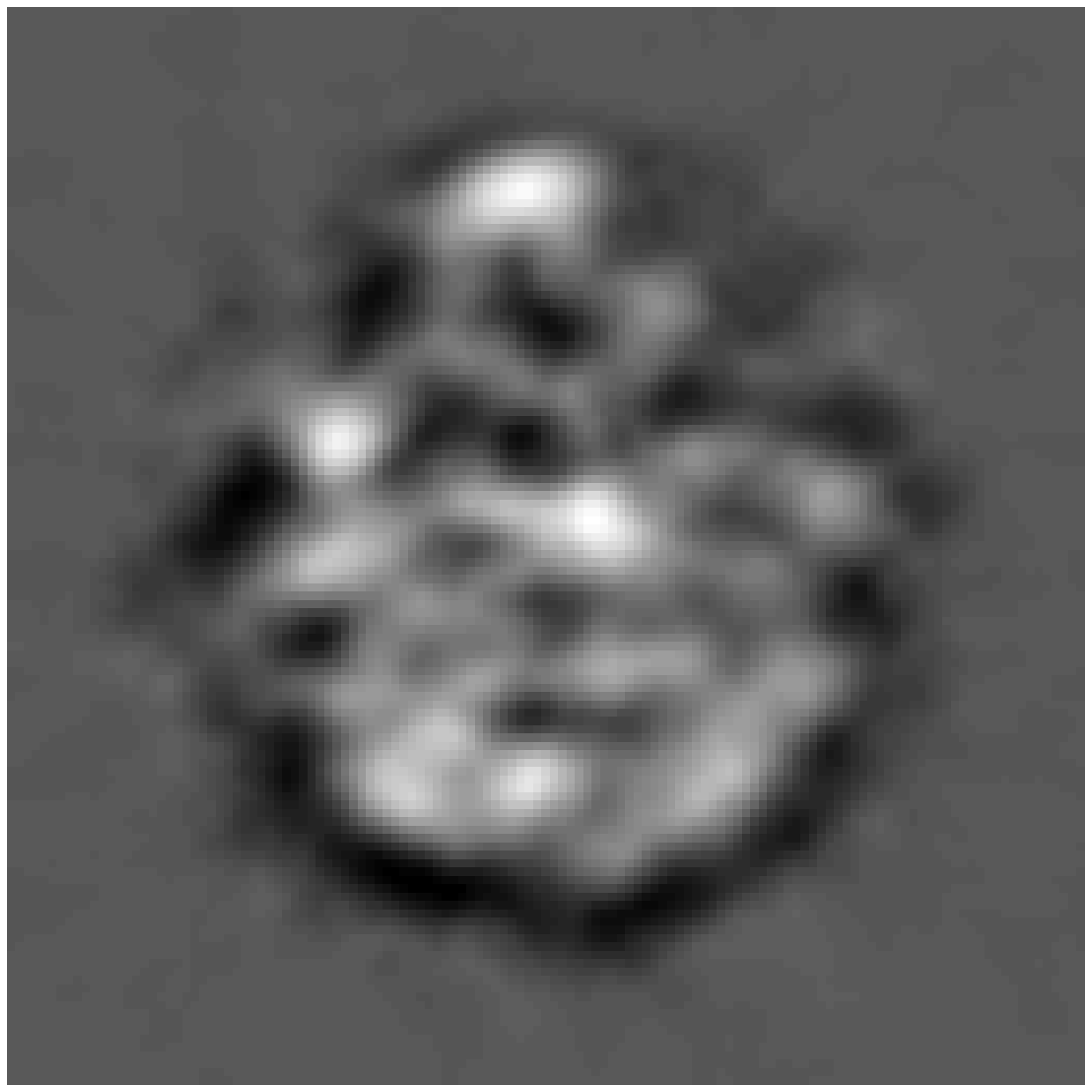}
     & \includegraphics[width=0.2\textwidth]{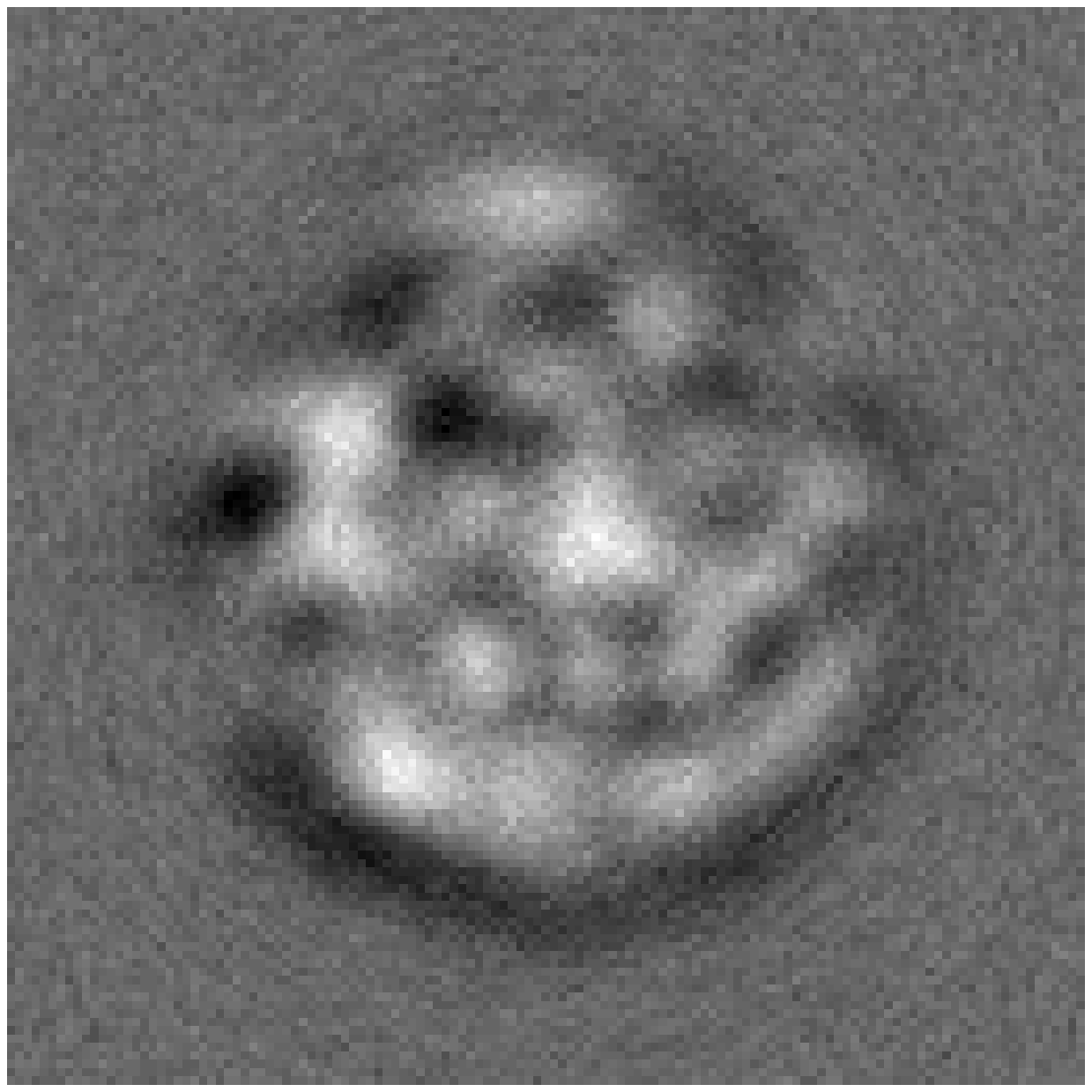} \\\\\\
       \includegraphics[width=0.2\textwidth]{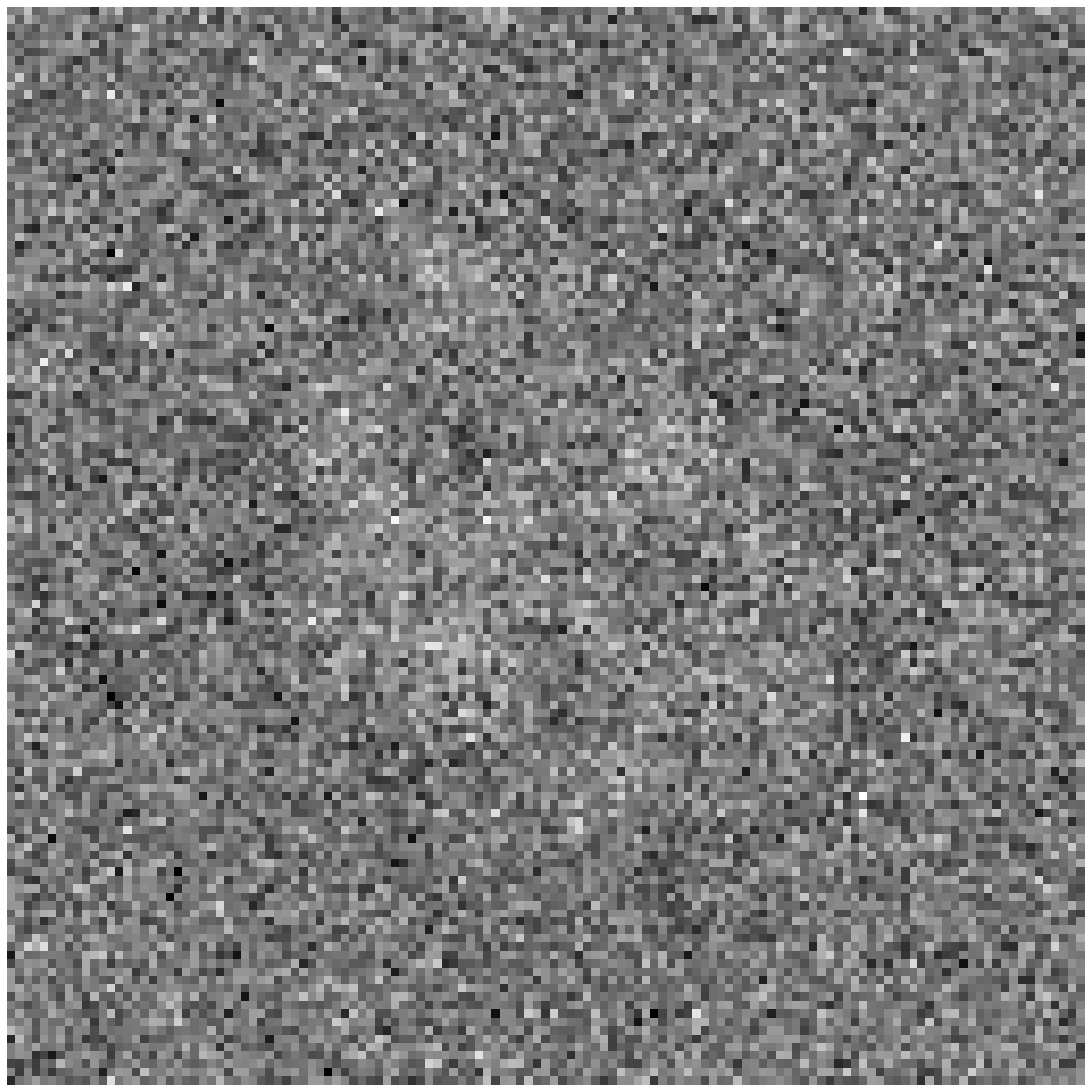}
     & \includegraphics[width=0.2\textwidth]{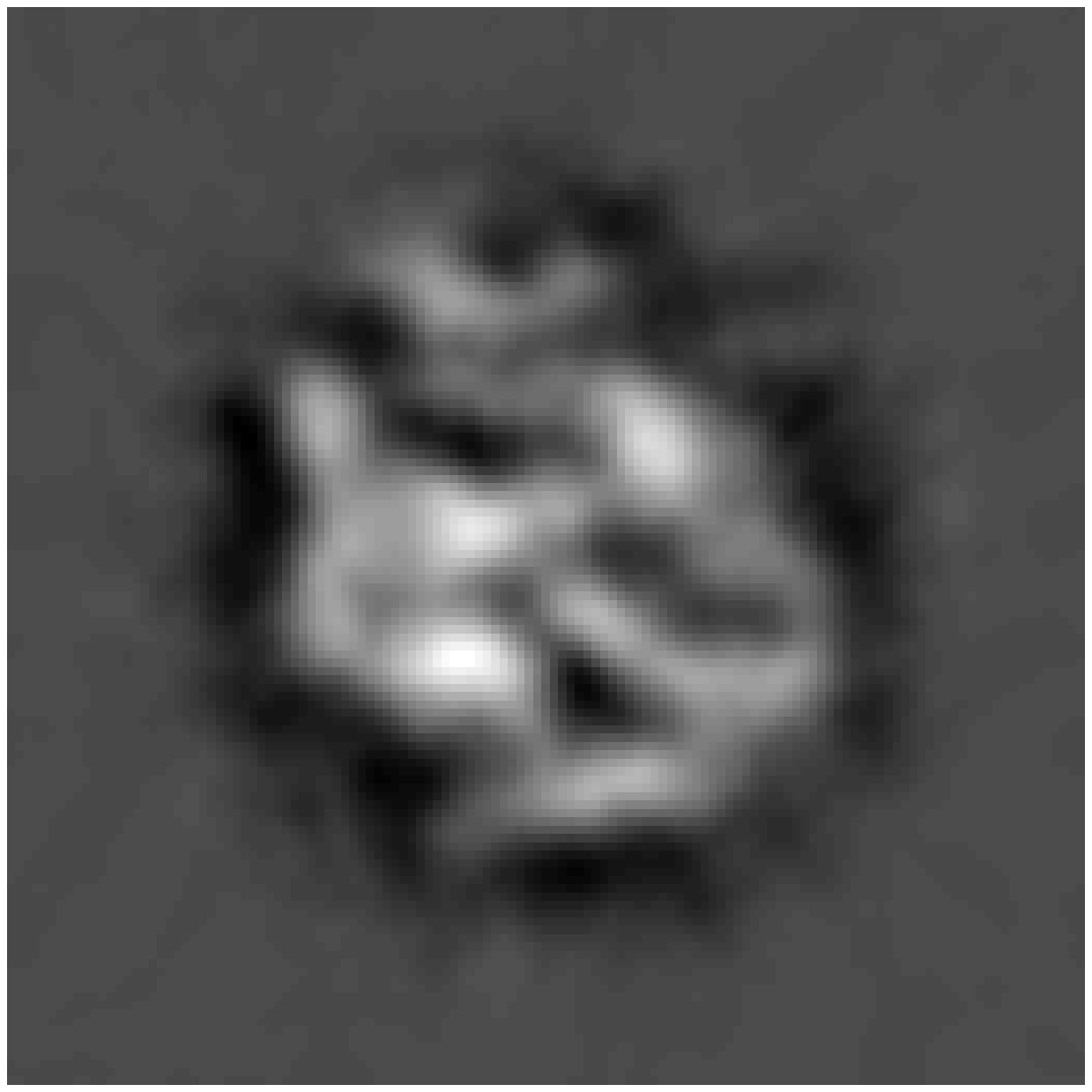}
     & \includegraphics[width=0.2\textwidth]{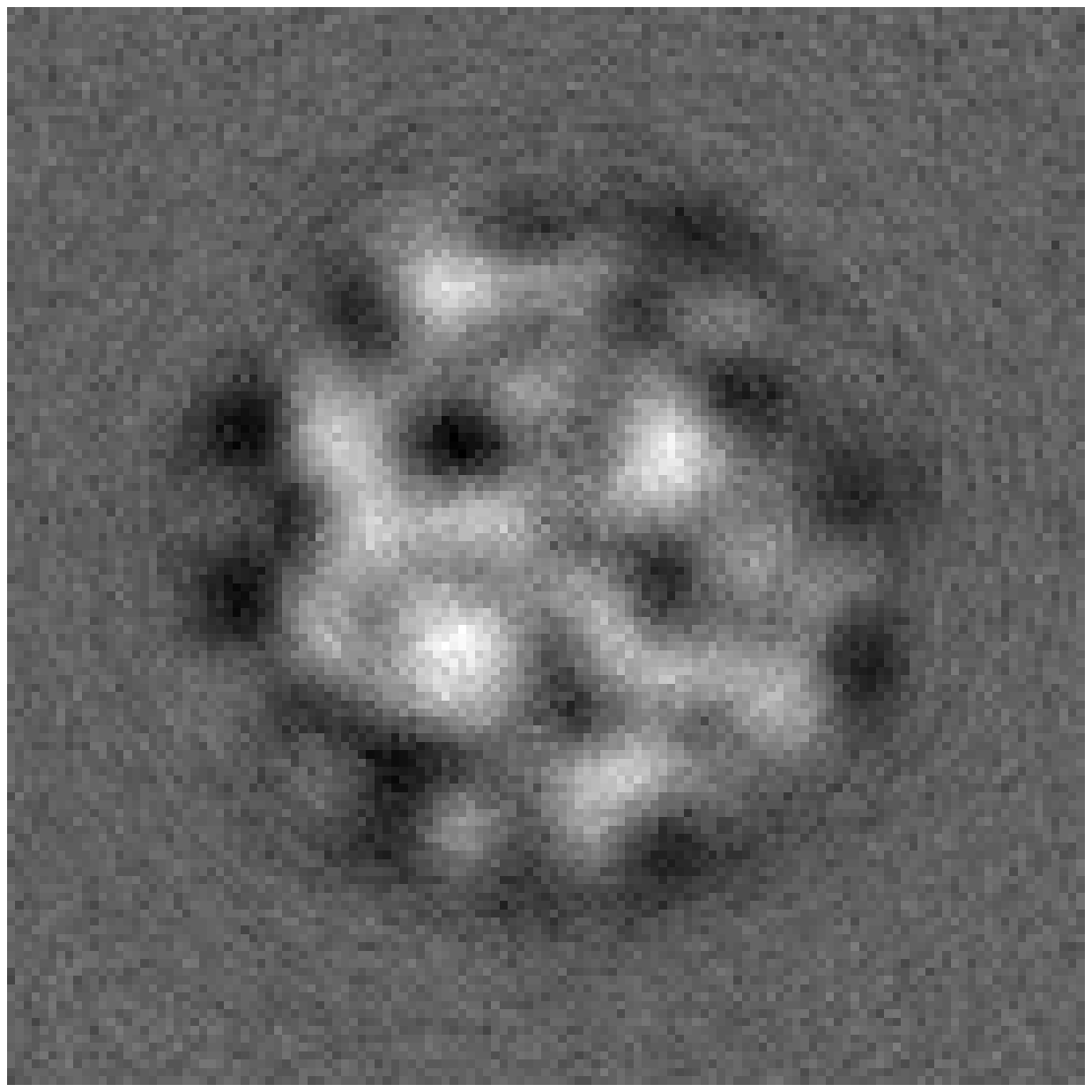} \\\\\\
       \includegraphics[width=0.2\textwidth]{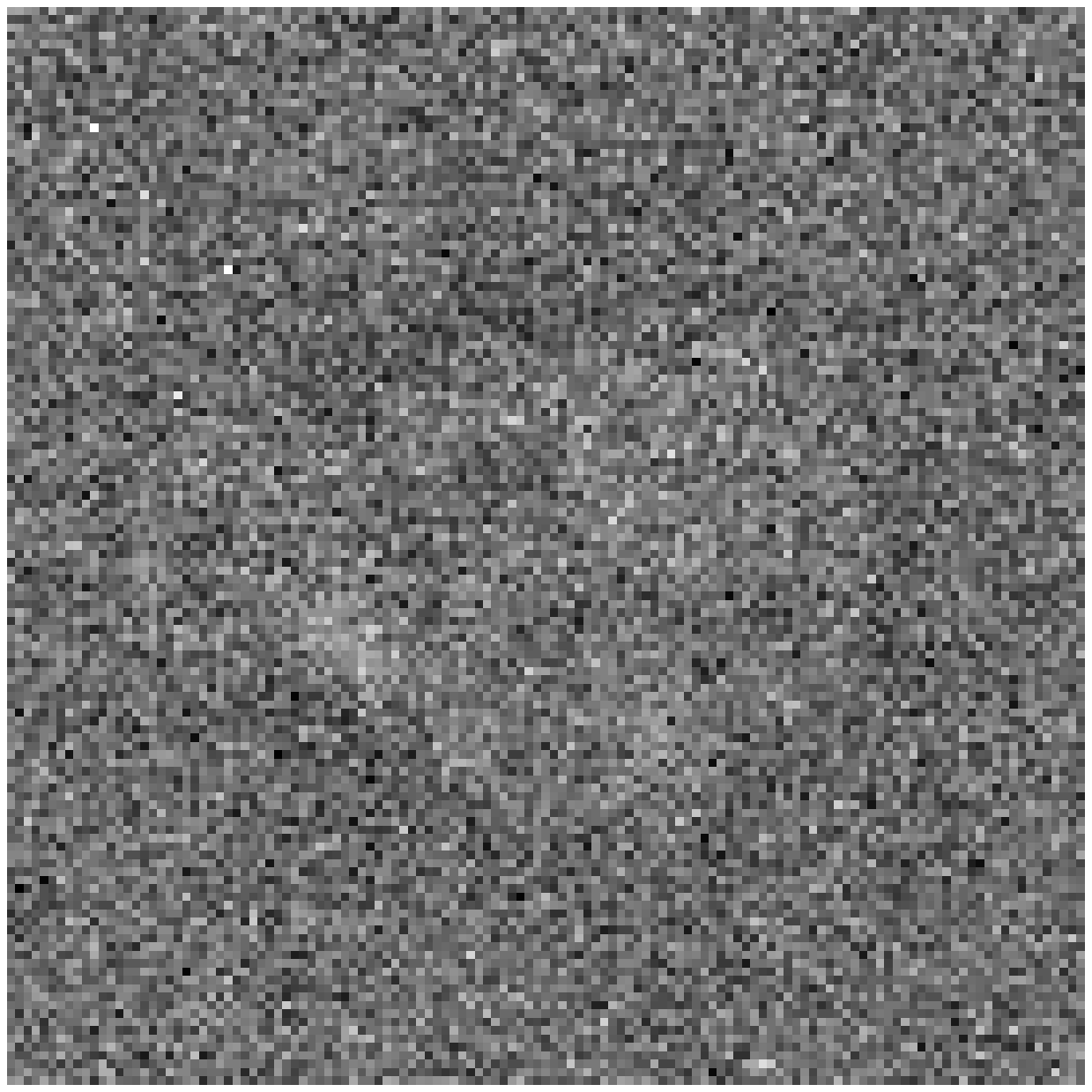}
     & \includegraphics[width=0.2\textwidth]{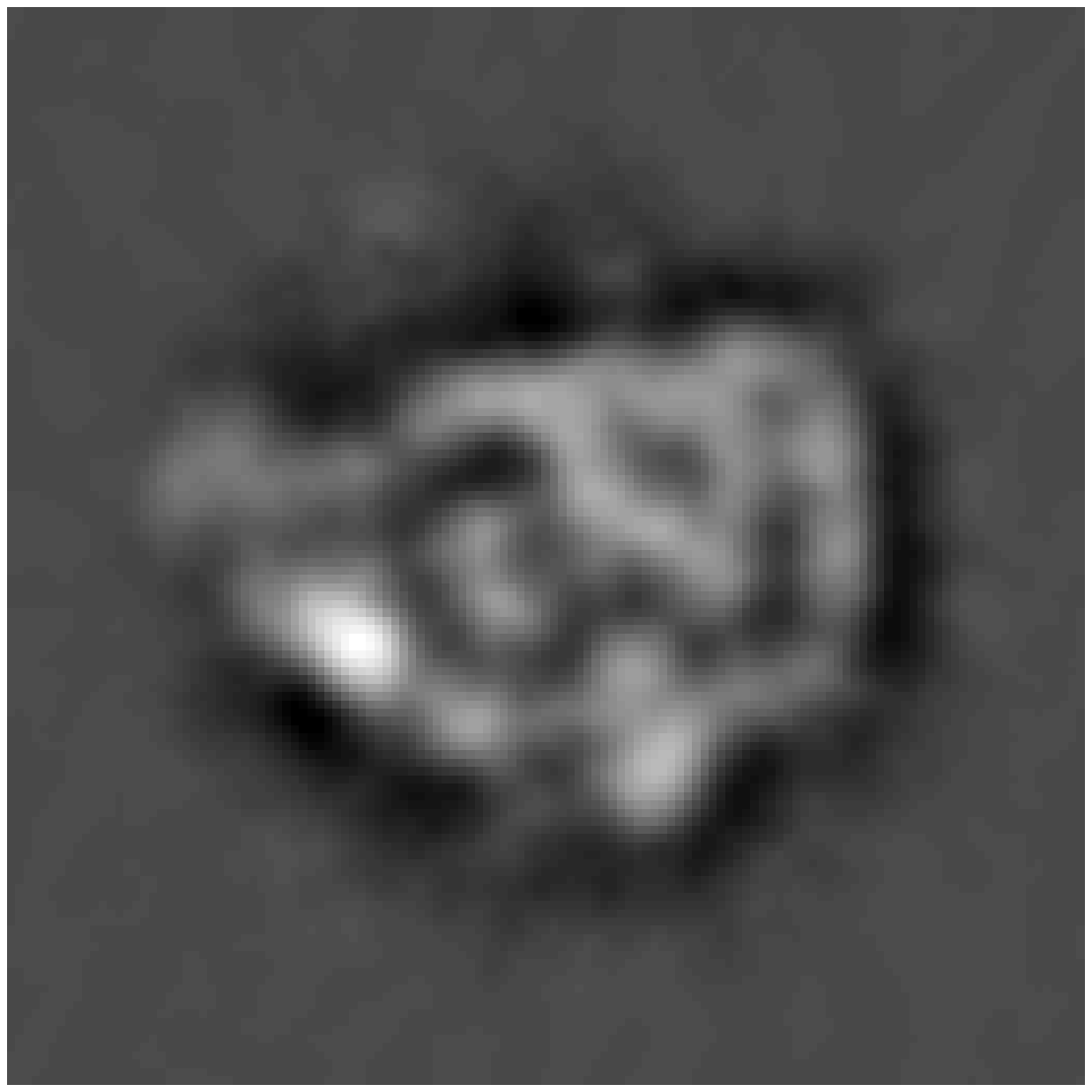}
     & \includegraphics[width=0.2\textwidth]{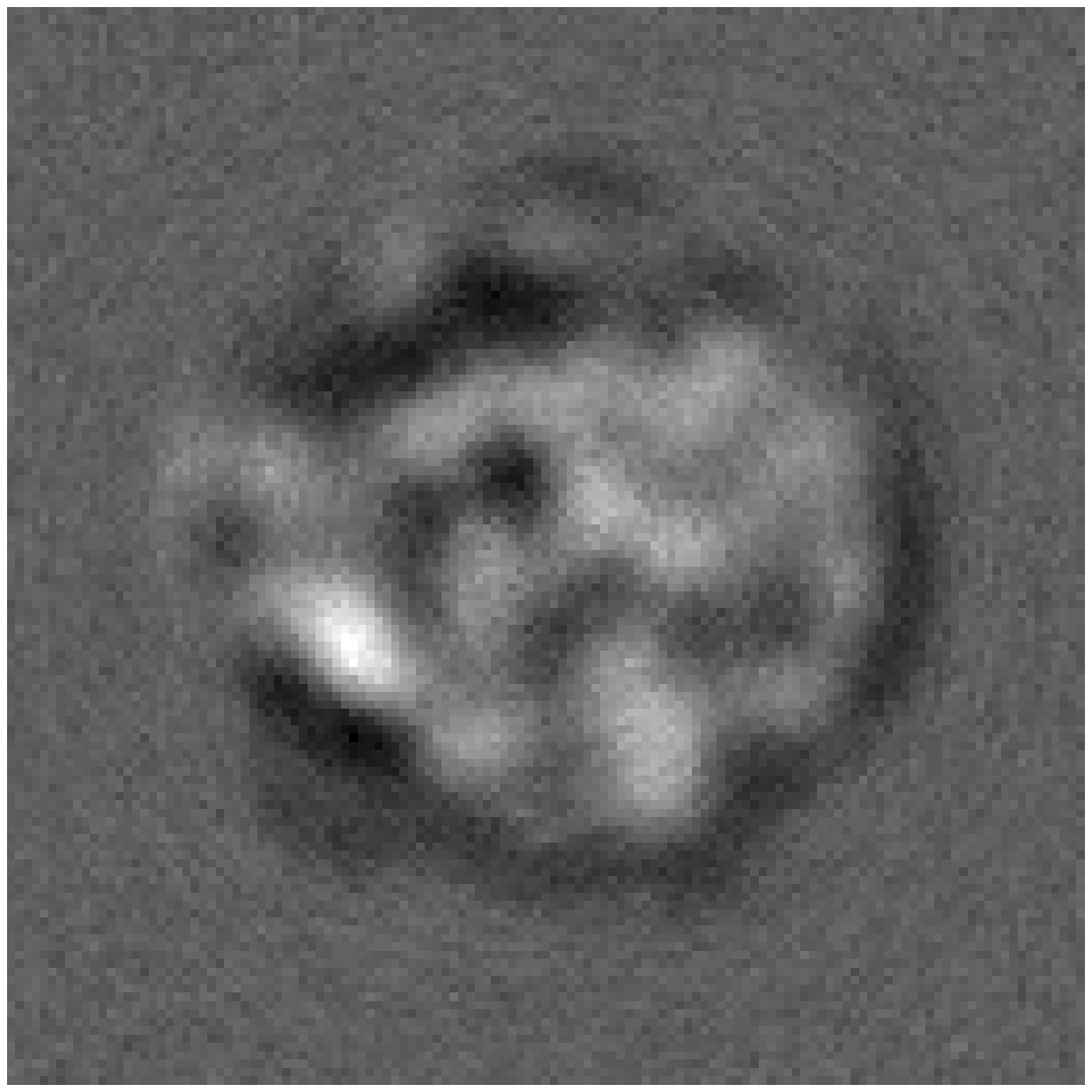} \\\\\\
       \includegraphics[width=0.2\textwidth]{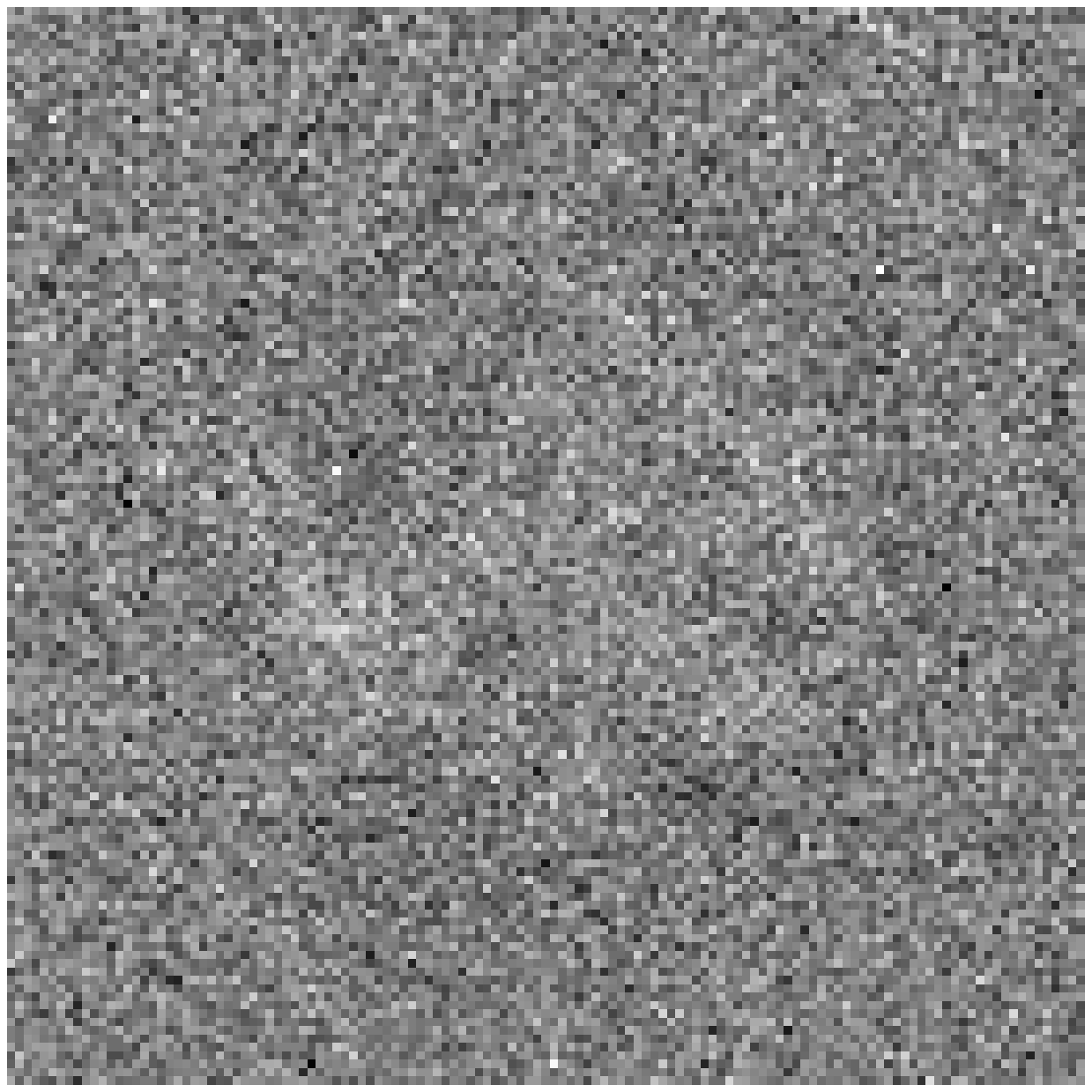}
     & \includegraphics[width=0.2\textwidth]{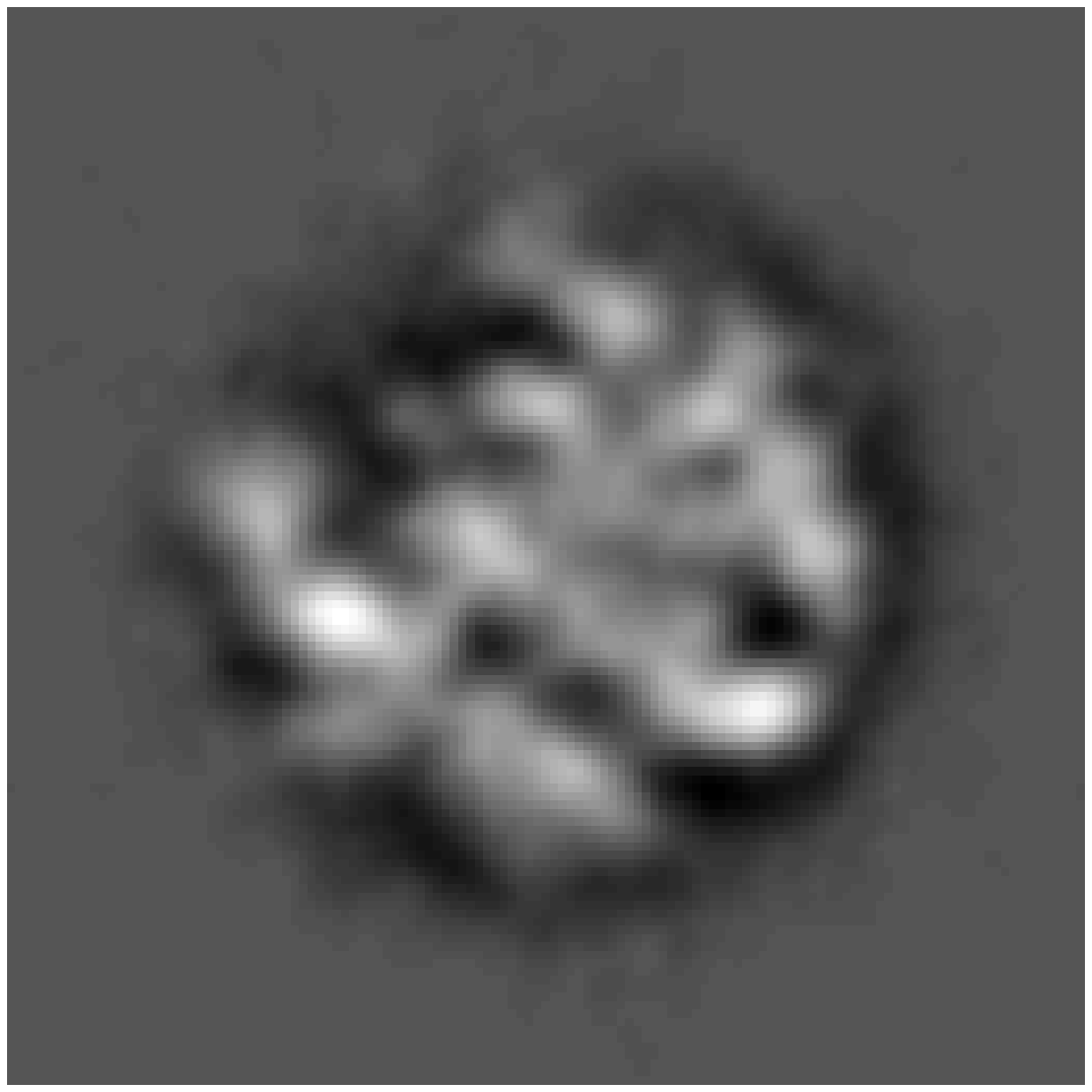}
     & \includegraphics[width=0.2\textwidth]{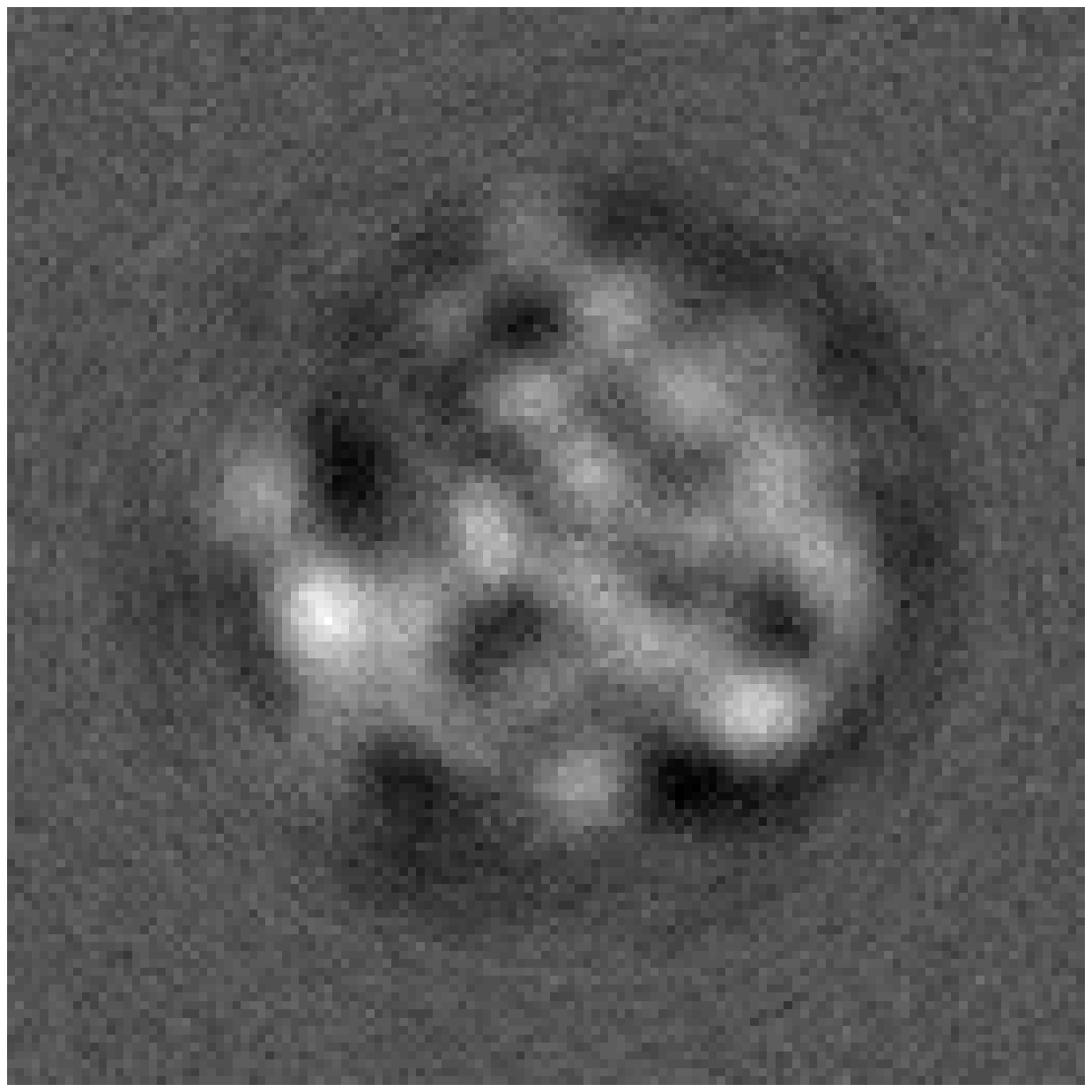} \\\\\\
       \includegraphics[width=0.2\textwidth]{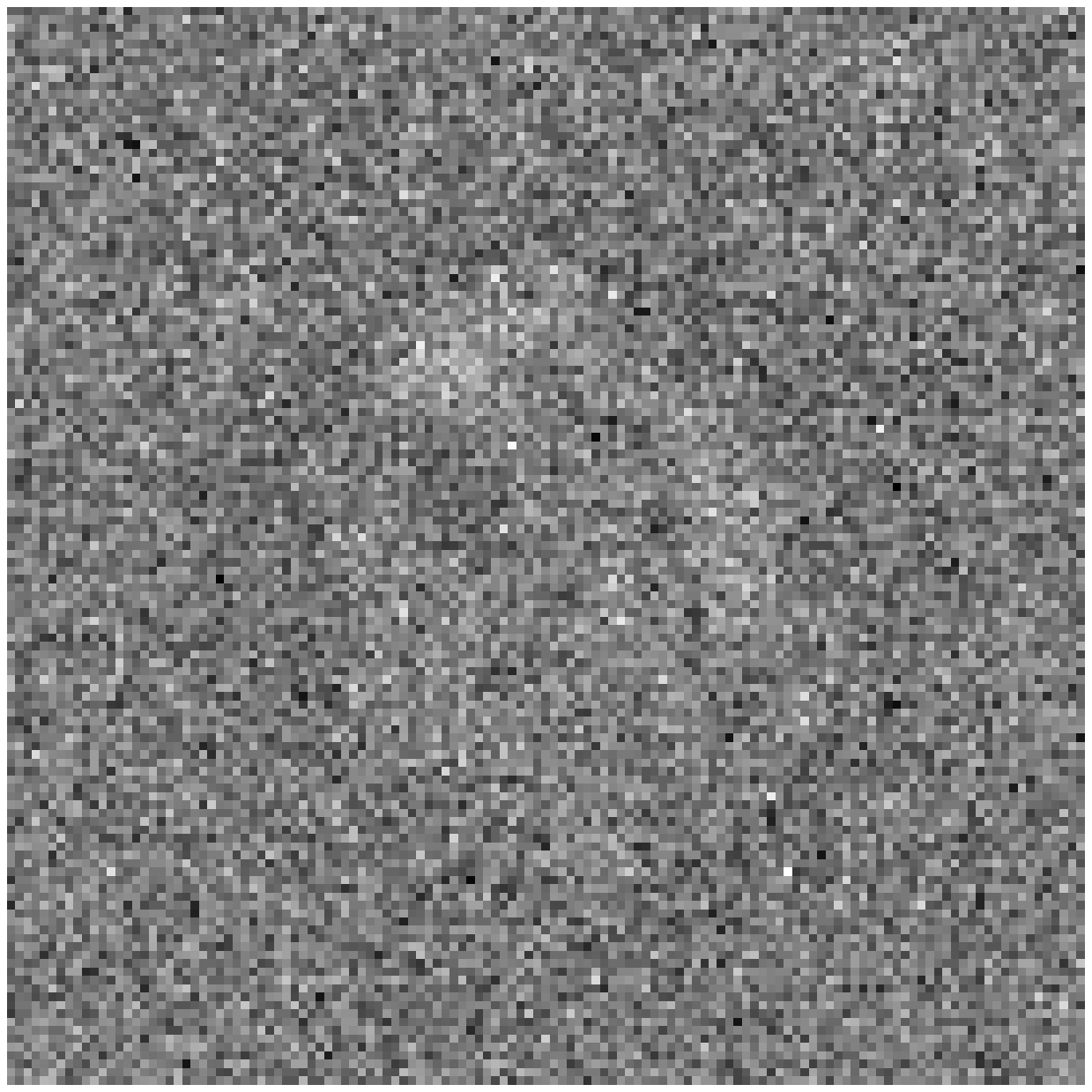}
     & \includegraphics[width=0.2\textwidth]{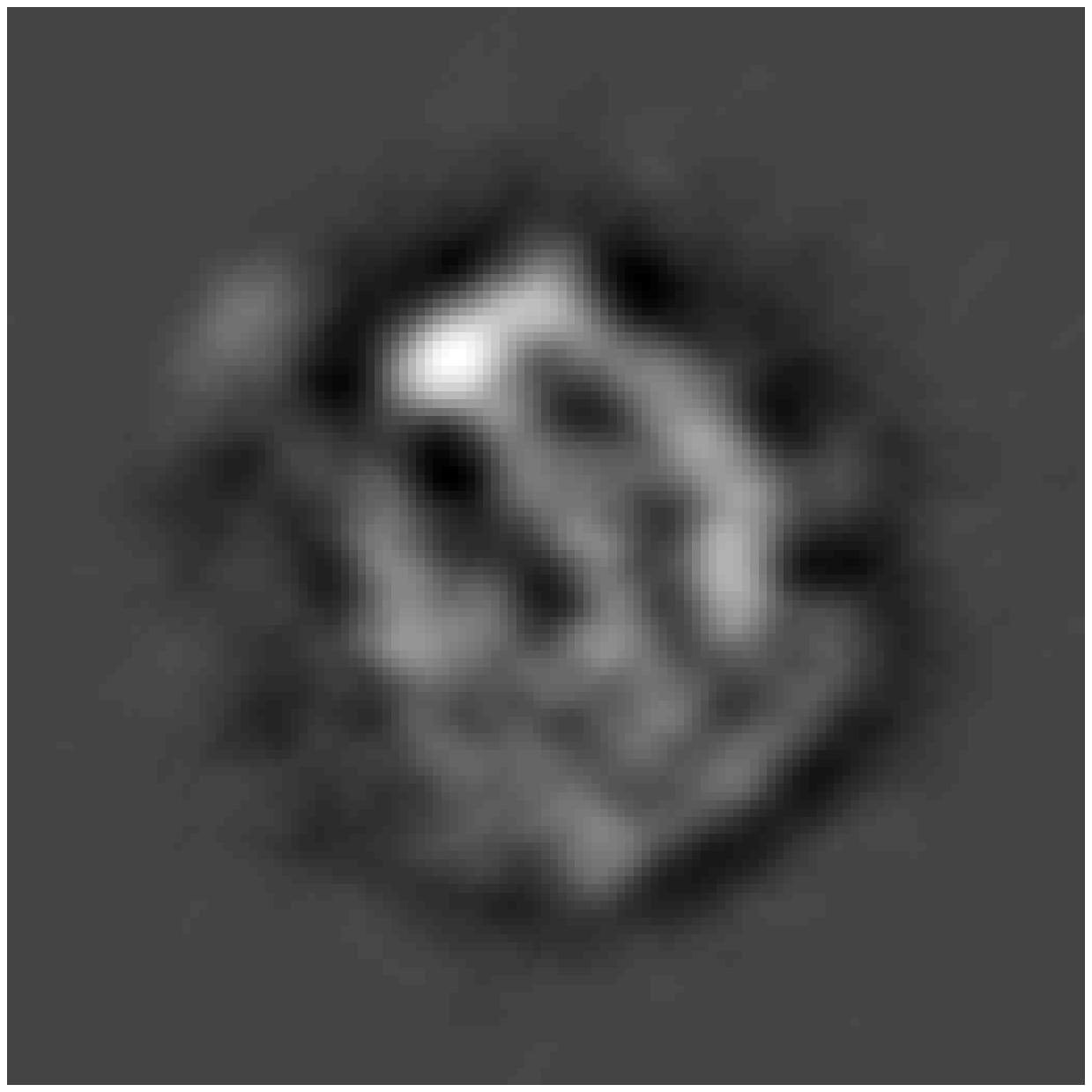}
     & \includegraphics[width=0.2\textwidth]{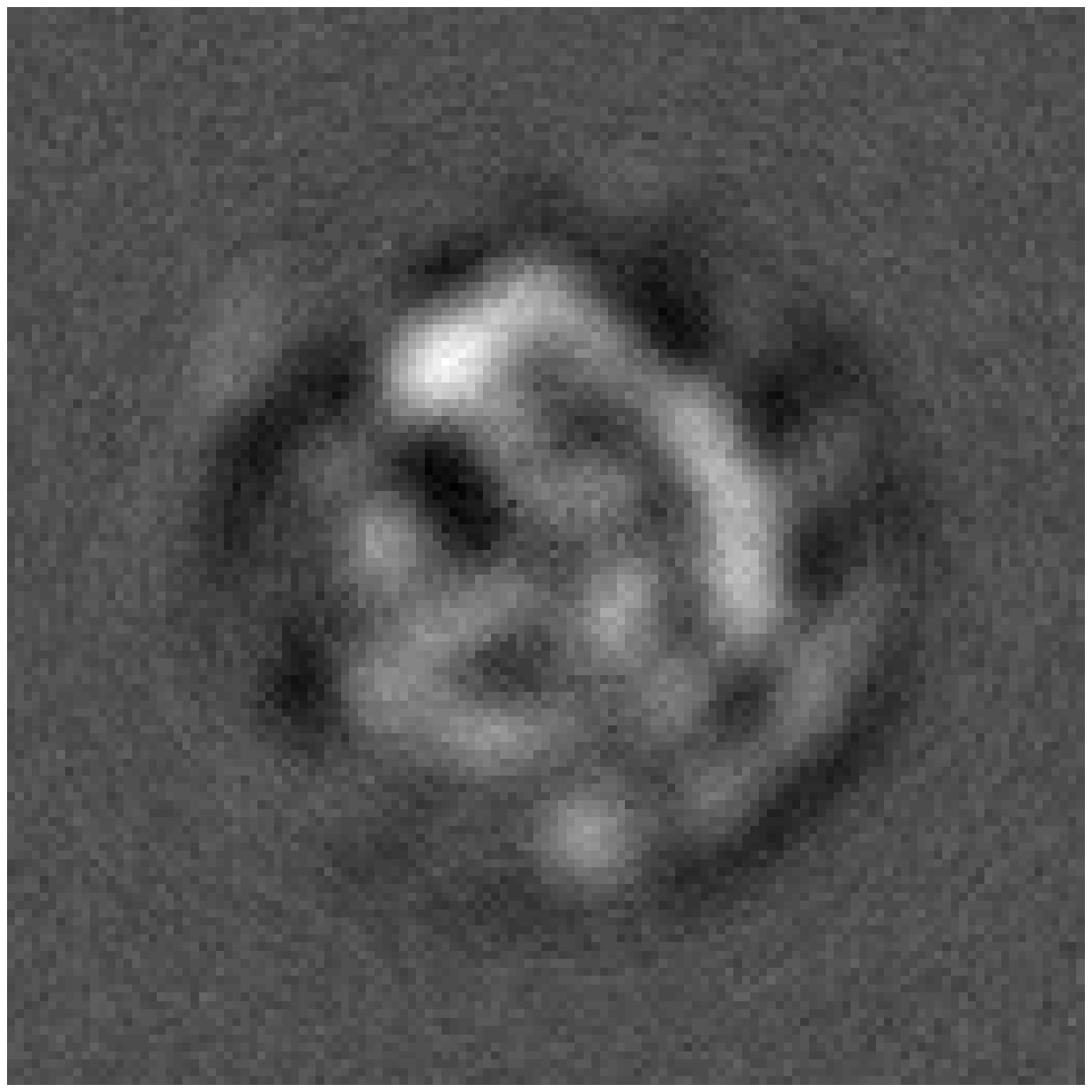} \\\\\\
\end{tabular}
\\\vspace{.25em}
\footnotesize{
{\sc Fig. 3c.} {\it Noisy, clean, and denoised images for the E. coli data set.}
}
\end{center}
\end{figure}

\section{Conclusion}
\label{conclusion}

The present article describes techniques
for the principal component analysis of data sets that are too large
to be stored in random-access memory (RAM),
and illustrates the performance of the methods on data from various sources,
including standard test sets, numerical simulations, and physical measurements.
Several of our data sets stored on disk were so large that
less than a hundredth of any of them could fit in our computer's RAM;
nevertheless, the scheme always succeeded.
Theorems, their rigorous proofs, and their numerical validations
all demonstrate that the algorithm of the present paper produces
nearly optimal spectral-norm accuracy.
Moreover, similar results are available for the Frobenius/Hilbert-Schmidt norm.
Finally, the core steps of the procedures parallelize easily;
with the advent of widespread multicore and distributed processing,
exciting opportunities for further development and deployment abound.

\section*{Appendix}
In this appendix, we describe a method for estimating
the spectral norm $\|D\|_2$ of a matrix $D$.
This procedure is particularly useful for checking whether
an algorithm has produced a good approximation to a matrix
(for this purpose, we choose $D$ to be the difference between the matrix
being approximated and its approximation).
The procedure is a version of the classic power method,
and so requires the application of $D$ and $D^\T$ to vectors,
but does not use $D$ in any other way.
Though the method is classical, its probabilistic analysis summarized below was
introduced fairly recently in~\cite{dixon} and~\cite{kuczynski-wozniakowski}
(see also Section~3.4 of~\cite{woolfe-liberty-rokhlin-tygert}).

Suppose that $m$ and $n$ are positive integers,
and $D$ is a real $m \times n$ matrix.
We define $\omega^{(1)}$, $\omega^{(2)}$, $\omega^{(3)}$, \dots\
to be real $n \times 1$ column vectors
with independent and identically distributed entries,
each distributed as a Gaussian random variable of zero mean and unit variance.
For any positive integers $j$ and $k$, we define
\begin{equation}
p_{j,k}(D) = \max_{1 \le q \le k}
             \sqrt{\frac{\|(D^\T \, D)^j \, \omega^{(q)}\|_2}
                  {\|(D^\T \, D)^{j-1} \, \omega^{(q)}\|_2}},
\end{equation}
which is the best estimate of the spectral norm of $D$ produced by
$j$ steps of the power method, started with $k$ independent random vectors
(see, for example,~\cite{kuczynski-wozniakowski}).
Naturally, when computing $p_{j,k}(D)$,
we do not form $D^\T \, D$ explicitly,
but instead apply $D$ and $D^\T$ successively to vectors.

Needless to say, $p_{j,k}(D) \le \|D\|_2$ for any positive $j$ and $k$.
A somewhat involved analysis shows that the probability that
\begin{equation}
p_{j,k}(D) \ge \|D\|_2/2
\end{equation}
is greater than
\begin{equation}
\label{high_prob}
1 - \left( \frac{2n}{(2j-1) \cdot 16^{^j}} \right)^{k/2}.
\end{equation}
The probability in~(\ref{high_prob}) tends to 1 very quickly
as $j$ increases. Thus, even for fairly small $j$,
the estimate $p_{j,k}(D)$ of the value of $\|D\|_2$ is accurate
to within a factor of two, with very high probability;
we used $j=6$ for all numerical examples in this paper.
We used the procedure of this appendix to estimate the spectral norm
in~(\ref{sort_of_svd}), choosing $D = A-U\,\Sigma\,V^\T$,
where $A$, $U$, $\Sigma$, and $V$ are the matrices from~(\ref{sort_of_svd}).
We set $k$ for $p_{j,k}(D)$ to be equal to the rank
of the approximation $U\,\Sigma\,V^\T$ being constructed.

For more information, see~\cite{dixon}, \cite{kuczynski-wozniakowski},
or Section~3.4 of~\cite{woolfe-liberty-rokhlin-tygert}.

\section*{Acknowledgements}

We would like to thank the mathematics departments of UCLA and Yale,
especially for their support during the development of this paper
and its methods.
Nathan Halko and Per-Gunnar Martinsson were supported in part
by NSF grants DMS0748488 and DMS0610097.
Yoel Shkolnisky was supported in part
by Israel Science Foundation grant 485/10.
Mark Tygert was supported in part by an Alfred P. Sloan Research Fellowship.
Portions of the research in this paper use the FERET database
of facial images collected under the FERET program,
sponsored by the DOD Counterdrug Technology Development Program Office.


\clearpage



\begin{thebibliography}{10}

\bibitem{LSA}
{\sc S. Deerwester, S. T. Dumais, G. W. Furnas, T. K. Landauer,
  and R. Harshman},
  {\it Indexing by latent semantic analysis}, J. Amer. Soc. Inform. Sci.,
  41 (1990), pp. 391--407.

\bibitem{dixon}
{\sc J. D. Dixon},
  {\it Estimating extremal eigenvalues and condition numbers of matrices},
  SIAM J. Numer. Anal., 20 (1983), pp. 812--814.

\bibitem{frank}
{\sc J. Frank},
  {\it Three-dimensional electron microscopy of macromolecular assemblies:
  Visualization of biological molecules in their native state},
  Oxford University Press, Oxford, UK, 2006.

\bibitem{golub-van_loan}
{\sc G. H. Golub and C. F. {Van L}oan},
  {\it Matrix Computations}, 3rd ed.,
  Johns Hopkins University Press, Baltimore, Maryland, 1996.

\bibitem{halko-martinsson-tropp}
{\sc N. Halko, P.-G. Martinsson, and J. Tropp},
  {\it Finding structure with randomness:
  Probabilistic algorithms for constructing approximate matrix decompositions},
  SIAM Review, 53 (2011), issue 2.

\bibitem{kuczynski-wozniakowski}
{\sc J. Kuczy\'nski and H. Wo\'zniakowski},
  {\it Estimating the largest eigenvalue by the power and Lanczos algorithms
  with a random start}, SIAM J. Matrix Anal. Appl., 13 (1992), pp. 1094--1122.

\bibitem{liberty-woolfe-martinsson-rokhlin-tygert}
{\sc E. Liberty, F. Woolfe, P.-G. Martinsson, V. Rokhlin, and M. Tygert},
  {\it Randomized algorithms for the low-rank approximation of matrices},
  Proc. Natl. Acad. Sci. USA, 104 (2007), pp. 20167--20172.

\bibitem{martinsson-szlam-tygert}
{\sc P.-G. Martinsson, A. Szlam, and M. Tygert},
  {\it Normalized power iterations for the computation of SVD},
  Proceedings of the Neural and Information Processing Systems (NIPS) Workshop
  on Low-Rank Methods for Large-Scale Machine Learning,
  Vancouver, Canada (2011),
  available at http://www.math.ucla.edu/$\sim$aszlam/npisvdnipsshort.pdf.

\bibitem{moon-phillips}
{\sc H. Moon and P. J. Phillips},
  {\it Computational and performance aspects of {PCA}-based face-recognition
  algorithms},
  Perception, 30 (2001), pp. 303--321.

\bibitem{feret1}
{\sc P. J. Phillips, H. Moon, S. A. Rizvi, and P. J. Rauss},
  {\it The FERET evaluation methodology for face recognition algorithms},
  IEEE Trans. Pattern Anal. Machine Intelligence, 22 (2000), pp. 1090--1104.

\bibitem{feret2}
{\sc P. J. Phillips, H. Wechsler, J. Huang, and P. J. Rauss},
  {\it The FERET database and evaluation procedure for face recognition
  algorithms}, J. Image Vision Comput., 16 (1998), pp. 295--306.

\bibitem{ponce-singer}
{\sc C. Ponce and A. Singer},
  {\it Computing steerable principal components of a large set of images
  and their rotations},
  Technical report, Princeton Applied and Computational Mathematics, 2010.
  Available at
  http://math.princeton.edu/$\sim$amits/publications/LargeSetPCA.pdf.

\bibitem{rokhlin-szlam-tygert}
{\sc V. Rokhlin, A. Szlam, and M. Tygert},
  {\it A randomized algorithm for principal component analysis},
  SIAM J. Matrix Anal. Appl., 31 (2009), pp. 1100--1124.

\bibitem{woolfe-liberty-rokhlin-tygert}
{\sc F. Woolfe, E. Liberty, V. Rokhlin, and M. Tygert},
  {\it A fast randomized algorithm for the approximation of matrices},
  Appl. Comput. Harmon. Anal., 25 (2008), pp. 335--366.

\end{thebibliography}
\end{document}